\documentclass[useAMS,usenatbib]{mn2e}
\usepackage{graphics}
\usepackage{times}
\usepackage{mn2e-breakabs} 
\newcommand{\hompc}{\,h\,{\rm Mpc}^{-1}}
\newcommand{\mpcoh}{\,h^{-1}\,{\rm Mpc}}

\newcommand{\apj}{ApJ}
\newcommand{\apjl}{ApJ Lett.}
\newcommand{\apjs}{ApJ Suppl.}
\newcommand{\aj}{Astron. J.}
\newcommand{\mnras}{MNRAS}

\newcommand{\physrep}{Physics Reports}
\newcommand{\prd}{Phys. Rev. D}

\begin{document}

\title[Cosmology from the SDSS DR7 LRG Clustering] {Cosmological
  Constraints from the Clustering of the Sloan Digital Sky Survey DR7 Luminous Red
  Galaxies}

\author[Beth A.\ Reid et al.]{
  \parbox{\textwidth}{
    Beth A.\ Reid$^{1,2}$\thanks{E-mail: beth.ann.reid@gmail.com},
    Will J.\ Percival$^{3}$,
    Daniel J.\ Eisenstein$^{4}$, Licia Verde$^{1,5}$, David N. Spergel$^{2,6}$, Ramin A.\ Skibba$^{7}$, Neta A.\ Bahcall$^{2}$, Tamas Budavari$^{8}$, Masataka Fukugita$^{9}$, J.\ Richard Gott$^{2}$, James E.\ Gunn$^{2}$, \v{Z}eljko Ivezi\'{c}$^{10}$, Gillian R.\ Knapp$^{2}$, Richard G.\ Kron$^{11,12}$, Robert H.\ Lupton$^{2}$, Timothy A.\ McKay$^{13}$, Avery Meiksin$^{14}$, Robert C.\ Nichol$^{3}$, Adrian C.\ Pope$^{15}$, David J.\ Schlegel$^{16}$, Donald P.\ Schneider$^{17}$, Michael A.\ Strauss$^{2}$, Chris Stoughton$^{18}$, Alexander S.\ Szalay$^{8}$, Max Tegmark$^{19}$, David H.\ Weinberg$^{20}$, Donald G.\ York$^{11,21}$, Idit Zehavi$^{22}$
  }
  \vspace*{4pt} \\
  $^{1}$ Institute of Space Sciences (CSIC-IEEC), UAB, Barcelona 08193, Spain and \\
    Institute for Sciences of the Cosmos (ICC), University of Barcelona, Barcelona 08028, Spain \\
  $^{2}$ Department of Astrophysical Sciences, Princeton University, 
  Princeton, NJ 08544, USA \\
  $^{3}$ Institute of Cosmology and Gravitation, University of
  Portsmouth, Portsmouth, P01 2EG, UK \\
  $^{4}$ Steward Observatory, University of Arizona, 933
  N. Cherry Ave., Tucson, AZ 85121, USA \\
  $^{5}$ ICREA (Institucio Catalana de Recerca i Estudis Avancats) \\
  $^{6}$ Princeton Center for Theoretical Science, Princeton University,
  Jadwin Hall, Princeton NJ 08542, USA \\
  $^{7}$ Max-Planck-Institute for Astronomy, K\"{o}nigstuhl 17, D-69117 Heidelberg, Germany \\
  $^{8}$ Department of Physics and Astronomy, 
  The Johns Hopkins University, 3701 San Martin Drive, Baltimore, MD 21218, USA \\
  $^{9}$ Institute for Cosmic Ray Research, University of Tokyo, Kashiwa 277-8582, Japan \\
  $^{10}$ Department of Astronomy, University of Washington, Box 351580, Seattle, WA 98195, USA \\
  $^{11}$ Department of Astronomy and Astrophysics, The University of Chicago, 5640 South Ellis Avenue, Chicago, IL 60615, USA \\
  $^{12}$ Fermi National Accelerator Laboratory, P.O. Box 500, Batavia, IL 60510, USA \\
  $^{13}$ Departments of Physics and Astronomy, University of Michigan, Ann Arbor, MI, 48109, USA \\
  $^{14}$ SUPA; Institute for Astronomy, University of Edinburgh, Royal Observatory, Blackford Hill, Edinburgh, EH9 3HJ, UK \\
  $^{15}$ Los Alamos National Laboratory, PO Box 1663, Los Alamos, NM 87545, USA \\
  $^{16}$ Lawrence Berkeley National Lab, 1 Cyclotron Road, MS 50R5032, Berkeley, CA 94720, USA \\
  $^{17}$ Department of Astronomy and Astrophysics, The Pennsylvania State University, University Park, PA 16802, USA \\
  $^{18}$ Fermilab PO Box 500, Batavia, IL  60510, USA \\
  $^{19}$  Department of Physics, Massachusetts Institute of Technology, Cambridge, Massachusetts, 02139, USA \\
  $^{20}$ Department of Astronomy, The Ohio State University, 140 West, 18th Avenue, Columbus, OH 43210, USA \\
  $^{21}$ The Enrico Fermi Institute, The University of Chicago, 5640 South Ellis Avenue, Chicago, IL 60615, USA \\
  $^{22}$  Department of Astronomy, Case Western Reserve University, Cleveland, OH 44106, USA
} 
\date{\today} 
\maketitle
\begin{abstract}
We present the power spectrum of the reconstructed halo density field
  derived from a sample of Luminous Red Galaxies (LRGs) from the Sloan
  Digital Sky Survey Seventh Data Release (DR7).  The halo power spectrum has a direct connection to the underlying dark matter power for $k \leq 0.2 \; h$ Mpc$^{-1}$, well into the quasi-linear regime.
This enables us to use a factor of $\sim 8$
more modes in the cosmological analysis than an analysis with $k_{max} = 0.1 \; h$  Mpc$^{-1}$,
as was adopted in the SDSS team analysis of the DR4 LRG sample \citep{tegmark/etal:2006}.
The observed halo power spectrum for $0.02 < k < 0.2 \; h$  Mpc$^{-1}$ is well-fit by our model: $\chi^2 = 39.6$ for 40 degrees of freedom for the best-fitting $\Lambda$CDM model.  We find $\Omega_m h^2 (n_s/0.96)^{0.13} = 0.141^{+0.009}_{-0.012}$ 
for a power law primordial power spectrum with spectral index $n_s$ and $\Omega_b h^2 = 0.02265$ fixed,
consistent with CMB measurements.  The halo power spectrum also constrains the ratio of the comoving sound horizon at the baryon-drag epoch to an effective distance to $z=0.35$: $r_s/D_V(0.35) = 0.1097^{+0.0039}_{-0.0042}$.
   Combining the halo power spectrum measurement  with the WMAP 5 year
  results, for the flat $\Lambda$CDM model we find 
  $\Omega_m = 0.289 \pm 0.019$ and $H_0 = 69.4 \pm 1.6$ km s$^{-1}$ Mpc$^{-1}$.
  Allowing for massive neutrinos in $\Lambda$CDM, we find
  $\sum m_{\nu} < 0.62$ eV at the 95\% confidence level.
  If we instead consider the effective number of relativistic species $N_{eff}$
  as a free parameter, we find $N_{eff} = 4.8^{+1.8}_{-1.7}$. 
  Combining also with the \citet{kowalski/etal:2008} 
  supernova sample, we find
 $\Omega_{tot} = 1.011 \pm 0.009$ and $w = -0.99 \pm 0.11$ 
 for an open cosmology
  with constant dark energy equation of state $w$.
    The power spectrum and a module to calculate the likelihoods 
    is publicly available at {\tt http://lambda.gsfc.nasa.gov/toolbox/lrgdr/}.
\end{abstract}


\section{Introduction}
The past decade has seen a dramatic increase in the quantity and quality
of cosmological data, from the discovery of cosmological acceleration 
using supernovae \citep{riess/etal:1998, perlmutter/etal:1999} 
to the precise mapping of the cosmic microwave background (CMB)
with the Wilkinson Microwave Anisotropy Probe \citep{page/etal:2003b, nolta/etal:2009} 
to the detection of the imprint of baryon acoustic oscillations (BAO) in the early 
universe on galaxy clustering \citep{eisenstein/etal:2005, cole/etal:2005}.  
Combining the most recent of these three cosmological probes, 
\citet{komatsu/etal:2009} detect no significant deviation 
from the minimal flat $\Lambda$CDM cosmological
model with adiabatic, power law primordial fluctuations, and constrain that model's parameters to within a few percent.

The broad shape of the power spectrum of density fluctuations in the evolved universe
provides a probe of cosmological parameters that is highly complementary to the CMB
and to probes of the expansion history (e.g., supernovae, BAO).  The last decade has
also seen a dramatic increase in the scope of galaxy redshift surveys.  The PSCz 
\citep{saunders/etal:2000} contains $\sim 15000$ IRAS galaxies out to $z=0.1$,
the 2dF Galaxy Redshift Survey(2dFGRS; \citealt{colless/etal:2001, colless/etal:2003}) collected 221,414
galaxy redshifts with median redshift 0.11, and the Sloan Digital Sky Survey (SDSS; \citealt{york/etal:2000}) is now complete
with 929,555 galaxy spectra \citep{abazajian/etal:2009} including both main galaxies ($\left<z\right> \sim 0.1$; \citealt{strauss/etal:2002})
and Luminous Red Galaxies (LRGs; $z \sim 0.35$; \citealt{eisenstein/etal:2001}).
To harness the improvement in statistical power available now from these surveys requires 
stringent understanding of modeling uncertainties.  
The three major components of this uncertainty are the non-linear 
gravitational evolution of the matter density field (e.g., \citealt{zeldovich:1970, davis/groth/peebles:1977, davis/peebles:1977}), the relationship 
between the galaxy and underlying matter density fields (``galaxy bias'', e.g., \citealt{kaiser:1984, rees:1985, cole/kaiser:1989}), 
and redshift space distortions (e.g., \citealt{kaiser:1987, davis/peebles:1983} and \citealt{hamilton:1998} for a review).

Several major advances have enabled previous analyses of 
2dFGRS and SDSS to begin to address these complications.    
Progress in $N$-body simulations (e.g., \citealt{heitmann/etal:2008}), 
analytical methods (see \citealt{carlson/white/padmanabhan:2009} for
an overview and comparison of many recent methods), 
and combinations thereof (e.g., \citealt{smith/etal:2003,eisenstein/seo/white:2007})
have allowed significant progress in the study of the non-linear 
real space matter power spectrum.  Recent power spectrum analyses 
have accounted for the luminosity dependence of a scale independent 
galaxy bias \citep{tegmark/etal:2004:la,cole/etal:2005}, which can 
introduce an artificial tilt in $P(k)$ in surveys which are not volume-limited 
\citep{percival/verde/peacock:2004}.  \citet{cresswell/percival:2009} have recently 
examined the scale dependence of galaxy bias as a function of luminosity and color.
\citet{tegmark/etal:2004:la} 
applied a matrix-based method using pseudo-Karhunen-Lo\`{e}ve 
eigenmodes to measure three power spectra from the SDSS galaxy distribution, 
allowing a quantification of 
 the clustering anisotropy and a more accurate reconstruction of the real-space 
power spectrum than can be obtained from the 
angle-averaged redshift space power spectrum.  
Non-linear redshift space distortions, caused in part by the virialized motions
of galaxies in their host dark matter haloes, create features known as 
Fingers-of-God (FOGs) along the line of sight in the redshift space 
galaxy density field \citep{davis/peebles:1983, gramann/cen/gott:1994}.  
Both \citet{tegmark/etal:2004:la} and \citet{cole/etal:2005} apply 
cluster-collapsing algorithms to mitigate the effects of FOGs 
before computing power spectra.  
Previous analyses have fit galaxy power spectra to linear 
\citep{percival/etal:2001,percival/etal:2007} or non-linear matter models
\citep{spergel/etal:2003, tegmark/etal:2004}, but did not attempt to model
the scale dependence of the galaxy bias.
  \citet{cole/etal:2005} 
introduced a phenomenological model to account for both 
matter non-linearity and the 
non-trivial relation between the galaxy power spectrum $P_{gal}(k)$ and matter power spectrum:
\begin{equation}
\label{qeqn}
P_{gal}(k) = \frac{1+Qk^2}{1+Ak} P_{lin}(k) \,,
\end{equation}
where $P_{lin}$
denotes the underlying linear matter power spectrum.
For the 2dFGRS analysis, 
\citet{cole/etal:2005} fit $A$ using mock galaxy catalogues and 
derive expected central values of $Q$.  In the fit to the observed galaxy power spectrum, 
they allow $Q$ to 
vary up to twice the expected value, which is supported by 
halo model calculations of the cosmological dependence 
of the galaxy $P(k)$.  This approach appears to work well
 for the case of 2dFGRS galaxies because it was calibrated on 
 mock catalogues designed to match the properties of this galaxy population;
  however, its application to the 
 LRG sample in \citet{tegmark/etal:2006}, 
 where the best-fitting $Q$ was much larger than for 2dFGRS galaxies, is questionable 
 (see \citealt{reid/spergel/bode:2008} and \citealt{yoo/etal:2009}, but also \citealt{sanchez/cole:2008}).

In this paper we focus our efforts on accurately modeling the relationship
 between the galaxy and matter density fields for the SDSS LRG sample.  
 Several authors have studied this relation using the small and intermediate 
 scale clustering in the SDSS LRG sample 
 \citep{masjedi/etal:2006, zehavi/etal:2005a, kulkarni/etal:2007,wake/etal:2008,zheng/etal:2008,reid/spergel:2009} 
 and galaxy-galaxy lensing \citep{mandelbaum/etal:2006}.
The LRG selection algorithm in the SDSS \citep{eisenstein/etal:2001}
 was designed to provide a homogenous galaxy sample
probing a large volume with a number density, $\bar{n}_{LRG}$, which
maximizes the effective survey volume $V_{eff}(k)$ on the large scales of 
interest, $k \sim 0.1 h$ Mpc$^{-1}$.  $V_{eff}$ is given by 
\citep{feldman/kaiser/peacock:1994, tegmark:1997}:
\begin{equation}
V_{eff}(k) = \int d^3 r \left[\frac{n({\bf r}) P(k)}{1+n({\bf r}) P(k)}\right]^2\,,
\end{equation}
where $P(k)$ denotes the measured galaxy power spectrum,
$\bar{n}({\bf r})$ the average galaxy number density in the sample at position ${\bf r}$, and the
integral is over the survey volume.
The total error on $P(k)$ is minimized (i.e., $V_{eff}$ is maximized)
when $\bar{n} P \sim 1$, which optimally balances cosmic variance and shot 
noise for a fixed number of galaxies.
The LRG sample 
has proven its statistical power through the detection of the BAO \citep{eisenstein/etal:2005,
percival/etal:2007}.  
However, parameterizing the LRG power spectrum
with a heuristic model for the non-linearity (Eqn.~\ref{qeqn}) and marginalizing
over fitting parameters
limits our ability to extract the full cosmological
information available from the power spectrum shape and
can introduce systematic biases
\citep{sanchez/cole:2008,dunkley/etal:2009,verde/peiris:2008,
reid/spergel/bode:2008}.

On sufficiently large scales, we expect galaxies to be linearly biased
with respect to the underlying matter density field \citep{mo/white:1996, scherrer/weinberg:2008}.
However, 
an often overlooked consequence of a sample with $\bar{n}_{LRG}
P_{LRG} \sim 1$ is that errors in the treatment of the shot noise can
introduce significant changes in the measured shape of $P_{LRG}(k)$ and 
can be interpreted 
as a scale dependent galaxy bias.  
In the halo model picture, the LRGs occupy massive dark matter haloes, which
themselves may not be Poisson tracers of the underlying matter density
field, as they form at the high peaks of the initial Gaussian density
distribution (e.g., \citealt{bardeen/etal:1986}).
  Moreover, an
additional shot noise-like term is generated when multiple LRGs
occupy individual dark matter haloes \citep{peacock/smith:2000, cooray/sheth:2002}.  Our
approach is to first eliminate the one-halo contribution to the power
spectrum by identifying groups of galaxies occupying the same dark
matter halo, and then to calibrate the relation between the power
spectrum of the reconstructed halo density field, $P_{halo}(k,{\bf p})$, and
the underlying matter power spectrum, $P_{DM}(k)$, using the $N$-body
simulation results presented in \citet{reid/spergel/bode:2008}.  As a
result, the effects of non-linear redshift space distortions caused by
pairs of galaxies occupying the same halo are diminished.  However, a
further complication is that LRGs occupy the massive end of the halo
mass function, and velocities of isolated LRGs within their host haloes could still
 be quite large.  The details of the relation between LRGs
and the underlying matter distribution can then have a significant
impact on the non-linear corrections to the power spectrum.

The DR7 LRG sample has sufficient statistical power that the
details of the relation between LRGs and the underlying matter density field
become important and need to be reliably modeled before attempting a 
cosmological interpretation of the data.  This paper offers three sequential key improvements to the 
modeling of LRG clustering compared with \citet{eisenstein/etal:2005} and \citet{tegmark/etal:2006}:
\begin{itemize}
\item We reconstruct the underlying halo density field traced by the LRGs before computing the power spectrum, while \citet{tegmark/etal:2006} apply an aggressive FOG compression algorithm.  The reconstructed halo density field power spectrum deviates from the underlying matter power spectrum by $< 4\%$ at $k=0.2 \hompc$, while the \citet{tegmark/etal:2006} power spectrum differs by $\sim 40\%$ at $k=0.2 \hompc$ \citep{reid/spergel/bode:2008}.
\item We produce a large set of mock LRG catalogues drawn from $N$-body simulations of sufficient resolution to trace a halo mass range relevant to LRGs without significant errors
in the small-scale halo clustering and velocity statistics (see Appendix A of
\citealt{reid/spergel/bode:2008}).  We present novel consistency checks between the mock and observed LRG density fields in  
halo-scale higher order clustering, FOG features, and the effective shot noise.
\item We use these tests along with the halo model framework to determine tight bounds on the remaining modeling uncertainties, and marginalize over these in our likelihood calculation.  In contrast, \citet{eisenstein/etal:2005} assume no uncertainty in their model LRG correlation function, and \citet{tegmark/etal:2006} marginalize over $Q$ in Eqn.~\ref{qeqn} with only an extremely weak prior on $Q$.
\end{itemize}
This paper represents a first attempt
to analyse a galaxy redshift survey with a model that accounts
for the non-linear galaxy bias and its uncertainty; other
approaches that utilize the galaxy distribution rather than the 
halo density field are in development \citep{yoo/etal:2009}.

\begin{table*}
\begin{center}
\begin{tabular}{lll}
  $P(k)$ & Definition & Reference \\
  \hline
  $\hat{P}_{LRG}(k)$   & 
    measured angle averaged redshift-space power spectrum of the LRGs & - \\
  $\hat{P}_{halo}(k)$  & 
    measured power spectrum of reconstructed halo density field &  - \\
  $P_{lin}(k)$        & 
    linear power spectrum computed by CAMB & \citet{lewis/challinor/lasenby:2000}\\
  $P_{DM}(k)$         & 
    theoretical real-space non-linear power spectrum of dark matter & - \\
  $P_{nw}(k)$         & 
    theoretical linear power spectrum without BAO (``no wiggles'') & 
    \citet{eisenstein/hu:1998}\\
  $P_{damp}(k)$      &  
    theoretical linear power spectrum with damped BAO (Eqn.~\ref{eq:Pdamp}) & 
    \citet{eisenstein/seo/white:2007}\\
  $P_{halo}(k,{\bf p})$ & 
    model for the reconstructed halo power spectrum for cosmological parameters ${\bf p}$ & 
    \citet{reid/spergel/bode:2008} \\
  $P_{halo,win}(k,{\bf p})$ & 
  $P_{halo}(k,{\bf p})$ convolved with survey window function (Eqn.~\ref{defwindow}) & \citet{percival/etal:2007} \\
    & and directly compared with $\hat{P}_{halo}(k)$ in the likelihood calculation (Eqn.~\ref{likeeqn}) & \\

\end{tabular}
\caption{\label{table:powspec} Definitions of power spectra referred to throughout the paper.}
\end{center}
\end{table*}

In this paper we present and analyse a measurement of the power spectrum of the
reconstructed halo density field from the SDSS DR7 LRG sample.  
DR7 represents a factor of $\sim 2$
increase in effective volume over the analyses presented
in \citet{eisenstein/etal:2005} and \citet{tegmark/etal:2006}, and covers a coherent  
region of the sky.  
Section~\ref{data} describes the measurement of the 
reconstructed halo density field power spectrum, $\hat{P}_{halo}(k)$, along with the window
and covariance matrices used in our likelihood analysis.
Section~\ref{modelsec} describes the details
of our model for the reconstructed halo power spectrum, $P_{halo}(k, {\bf p})$.  In Section \ref{quantnuis} we 
summarize the tests we have performed for 
various systematics in our modeling of the relation
between the galaxy and dark matter density field.  We quantify the
expected level of uncertainty through two nuisance parameters and
present several consistency checks between the model and observed
reconstructed halo density field.  
In Section~\ref{cosmoconstraints} we discuss the cosmological constraints
from $\hat{P}_{halo}(k)$ alone as well as in combination with WMAP5
 \citep{dunkley/etal:2009} and the Union supernova dataset
  \citep{kowalski/etal:2008}.  Section~\ref{compare} compares our findings 
  with the results of previous analyses 
of galaxy clustering, and Section~\ref{conc} summarizes our conclusions.

In a companion paper (\citealt{percival/etal:prep}; hereafter, P09) we measure and analyse
BAO in the SDSS DR7 sample, of which the LRG sample considered here is
a subset. BAO are detected in seven redshift shells, leading to a
2.7\% distance measure at redshift $z=0.275$, and a measurement of the
gradient of the distance-redshift relation, this quantified by
the distance ratio between $z=0.35$ and $z=0.2$.  We show in 
Section~\ref{cosmoconstraints} that the results from
these measurements are in agreement
with our combined results from BAO and the shape of the power
spectrum calculated using just the LRGs.  The results from these different
analyses will be correlated because of the overlapping data used, so
they should not be combined in cosmological analyses. The best data
set to be used will depend on the cosmological model to be tested.
While the inclusion of 2dFGRS and main SDSS galaxies in P09 provides
a higher significance detection of the BAO, we show in Section \ref{shapepayoff} 
that the full power spectrum information provides tighter constraints on 
both massive neutrinos and the number of relativistic species.

  Throughout the paper we make use of two specific
cosmological models.  The simulation set described in
\citet{reid/spergel/bode:2008} and used to calibrate the model $P_{halo}(k,{\bf p})$
adopts the WMAP5 recommended $\Lambda$CDM values: ($\Omega_m$, $\Omega_b$,
$\Omega_{\Lambda}$, $n_s$, $\sigma_8$, $h$) = (0.2792, 0.0462, 0.7208,
0.960, 0.817, 0.701).  We refer to this model throughout the paper 
as our `fiducial cosmological model.'
  To convert redshifts to distances in the
computation of the $\hat{P}_{halo}(k)$, we adopt a flat $\Lambda$CDM cosmology with $\Omega_m = 0.25$
and $\Omega_{\Lambda} = 0.75$.  Throughout we refer to the power spectrum of
several different density fields and several theoretical spectra.
Table~\ref{table:powspec} summarizes their definitions.

\section{DATA}  \label{data}

\subsection{LRG sample}  \label{sampleselection}
The SDSS \citep{york/etal:2000} is the largest galaxy survey ever
produced; it used a 2.5m telescope \citep{gunn/etal:2006} to obtain
imaging data in 5 passbands $u$, $g$, $r$, $i$ and $z$
\citep{fukugita/etal:1996,gunn/etal:2006}. The images were reduced
\citep{stoughton/etal:2002,pier/etal:2003,ivezic/etal:2004} and
calibrated \citep{hogg/etal:2001,smith/etal:2002,tucker/etal:2006,padmanabhan/etal:2008b},
and galaxies were selected for follow-up spectroscopy. The second phase
of the SDSS, known as SDSS-II, has recently finished, and the DR7
\citep{abazajian/etal:2009} sample has recently been made public. The
SDSS project is now continuing with SDSS-III where the extragalactic
component, the Baryon Oscillation Spectroscopic Survey (BOSS;
\citealt{schlegel/white/eisenstein:2009}), has a different galaxy
targeting algorithm. DR7 therefore represents the final data set that
will be released with the original targeting and galaxy selection
\citep{eisenstein/etal:2001,strauss/etal:2002}.

In this paper we analyse a subsample containing $110\,576$ Luminous
Red Galaxies (LRGs: \citealt{eisenstein/etal:2001}), which were
selected from the SDSS imaging based on $g$, $r$ and $i$ colours, to
give approximately $15$ galaxies per square degree. The SDSS also
targeted a magnitude limited sample of galaxies for spectroscopic
follow-up \citep{strauss/etal:2002}. The LRGs extend this main galaxy
sample to $z\simeq0.5$, covering a greater volume. Our DR7 sample
covers $7931$\,deg$^2$ (including a 7190\,deg$^2$ contiguous region in
the North Galactic Cap), with an effective volume of
$V_{eff}=0.26$\,Gpc$^3$h$^{-3}$, calculated with a model power
spectrum amplitude of $10^4$\,$h^{-3}$Mpc$^3$. This power spectrum
amplitude is approximately correct for the LRGs at $k\sim0.15\hompc$.  For
comparison, the effective volume of the sample used by
\citet{eisenstein/etal:2005} was $V_{eff}=0.13$\,Gpc$^3$h$^{-3}$, and
$V_{eff}=0.16$\,Gpc$^3$h$^{-3}$ in \citet{tegmark/etal:2006}; this
work represents a factor of $\sim$2 increase in sample size over these
analyses.  The sample is the same as that used in P09, and its
construction follows that of \citet{percival/etal:2007}, albeit with a
few improvements.

We use SDSS Galactic extinction-corrected Petrosian magnitudes
calibrated using the ``\"{u}bercalibration'' method
\citep{padmanabhan/etal:2008b}.  However, we find that the power
spectrum does not change significantly when one adopts the old
standard calibration instead \citep{tucker/etal:2006}. Luminosities
are K-corrected using the methodology of
\citet{blanton/etal:2003,blanton03b}. We remove LRGs that are not
intrinsically luminous by applying a cut ${\rm M}_{^{0.1}r}<-21.8$,
where ${\rm M}_{^{0.1}r}$ is our estimate of the absolute magnitude in
the $r$-band for a galaxy at $z=0.1$.

Spectroscopic LRG targets were selected using two color-magnitude cuts
\citep{eisenstein/etal:2001}.  The tiling algorithm ensures nearly
complete samples \citep{blanton/etal:2003}.  However, spectroscopic
fiber collisions prohibit simultaneous spectroscopy for objects
separated by $<55''$, leaving $\sim 7\%$ of targeted objects without
redshifts \citep{masjedi/etal:2006}.  We correct for this effect as in
\citet{percival/etal:2007}: for an LRG lacking a spectrum but 55''
from an LRG with a redshift, we assign both galaxies the measured
redshift.  If the LRG lacking a redshift neighbors only a galaxy from
the low redshift SDSS main sample, we do not assign it a redshift.
These galaxies are assumed to be randomly distributed, and simply
contribute to the analysis by altering the completeness, the fraction
of targeted galaxies with good redshifts, in a particular region.  The
impact of the fiber collision correction is addressed in
Appendix \ref{cicstats} and Appendix \ref{cicvsnotcompare}.

\begin{figure}
  \centering
  \resizebox{0.9\columnwidth}{!}{\includegraphics{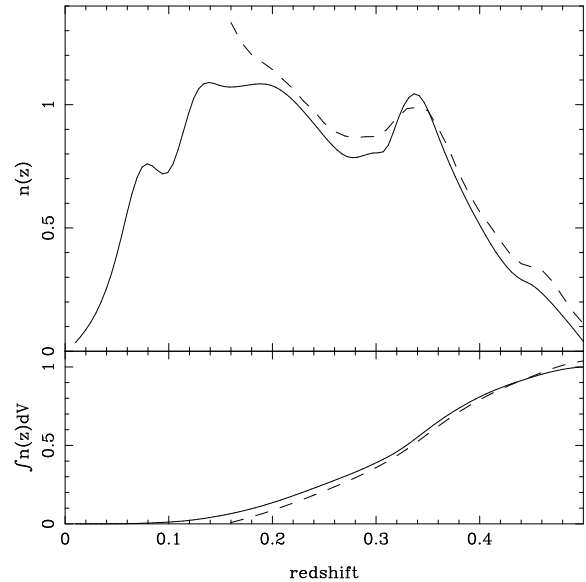}}
  \caption{\label{fig:nzcompare} Fits to the redshift distributions
    for the LRG selection used in this work (solid curves) and the
    \citet{zehavi/etal:2005a} $-23.2 < M_g < -21.2$ sample used in
    \citet{tegmark/etal:2006} (dashed curves).  {\em Upper panel}:
    $n(z)$ vs redshift in units of $10^{-4}$ ($h^{-1}$ Mpc)$^{-3}$
    {\em Lower panel}: $N(<z) = \int dz n(z) dV/dz$ (arbitrary overall
    normalization).}
\end{figure}

Fig.~\ref{fig:nzcompare} compares the number density as a function of
redshift for the LRG selection used in this paper
\citep{percival/etal:2007} and the one used in
\citet{tegmark/etal:2006} and presented in \citet{zehavi/etal:2005a}.
The main differences are that our selection includes a small number of
galaxies at $z < 0.15$, and our cut on the intrinsic luminosity of the
LRGs slightly reduces the number density of galaxies at high $z$. The
different selections produce a similar number of galaxies per unit
volume, and we expect no difference between the samples on the large
scale structure statistics of interest here.

\subsection{Recovering the halo density field}  \label{preprocess}
In real space, the impact of more than one LRG per halo on the large scale power spectrum can be accurately modeled as an additional shot noise term \citep{cooray/sheth:2002,reid/spergel/bode:2008}.  However, this picture is much more complicated in redshift-space because of the velocity dispersion of the LRGs shifts them along the line of sight by $\sim 9$ $h^{-1}$ Mpc \citep{reid/spergel/bode:2008}, and the distribution of intrahalo velocities has long tails.  This shifting causes power to be shuffled between scales and causes even the largest scale modes along the line of sight to be damped by these FOG features \citep{davis/peebles:1983,peacock/dodds:1994,seljak:2001}.  We substantially reduce the impact of these effects by using the power spectrum of the reconstructed halo density field.

We follow the Counts-In-Cylinders (CiC)
technique in \citet{reid/spergel/bode:2008} to identify LRGs occupying
the same halo and thereby estimate the halo density field.  Two
galaxies are considered neighbors when their transverse comoving
separation satisfies $\Delta r_{\perp} \leq 0.8$ $h^{-1}$ Mpc and
their redshifts satisfy $\Delta z/(1+z) \leq \Delta v_p/c = 0.006$
($\Delta v_p = 1800$ km s$^{-1}$). A cylinder should be a good
approximation to the density contours of satellites surrounding
central galaxies in redshift space, as long as the satellite velocity
is uncorrelated with its distance from the halo centre and the
relative velocity dominates the separation of central and satellite
objects along the line of sight.  Galaxies are then grouped with their
neighbors by a Friends-of-Friends (FoF) algorithm.  The reconstructed
halo density field is defined by the superposition of the centres of
mass of the CiC groups. We refer to the power spectrum of the
reconstructed halo density field as $\hat{P}_{halo}(k)$; it is our best estimate 
of the power spectrum of the haloes traced by the LRGs.  
For comparison we also compute the power
spectrum without applying any cluster-collapsing algorithm, $\hat{P}_{LRG}(k)$.

Our reconstructed halo density field contains $104\,337$ haloes derived from $110\,576$ LRGs.

\subsection{Calculating power spectra, window functions and
  covariances}

In this paper we focus on using the angle-averaged power spectrum to
derive constraints on the underlying linear theory power spectrum.  On
linear scales the redshift space power spectrum is proportional to the
real space power spectrum \citep{kaiser:1987,hamilton:1998}.  Our halo
density field reconstruction mitigates the effects of FOGs from
objects occupying the same halo. Though we do not explore it here, we
expect that our halo density field reconstruction will be useful to an
analysis of redshift-space anisotropies (e.g.,
\citealt{hatton/cole:1999}).

The methodology for calculating the power spectrum of the
reconstructed halo density field, $\hat{P}_{halo}(k)$, is based on the
Fourier method of \citet{feldman/kaiser/peacock:1994}. The halo
density is calculated by throwing away all but the brightest galaxy
where we have located a set of galaxies within a single halo. This
field is converted to an over-density field by placing the haloes on
a grid and subtracting an unclustered ``random catalogue'', which
matches the halo selection.  To calculate this random catalogue, we
fit the redshift distributions of the halo sample with a spline model
\citep{press92} (shown in Fig.~\ref{fig:nzcompare}), and the angular
mask was determined using a routine based on a {\sc HEALPIX}
\citep{gorski05} equal-area pixelization of the sphere as in
\citep{percival/etal:2007}. This procedure allows for the variation in radial
selection seen at $z>0.38$, which is caused by the spectroscopic features of
the LRGs moving across the wavebands used in the target selection. The
haloes and randoms are weighted using a luminosity-dependent bias model
that normalizes the fluctuations to the amplitude of $L_*$ galaxies
\citep{percival/verde/peacock:2004}.  To do this we assume that each
galaxy used to locate a halo is biased with a linear deterministic
bias model, and that this bias depends on ${\rm M}_{^{0.1}r}$
according to \citet{tegmark/etal:2004:la} and \citet{zehavi05}, where ${\rm
  M}_{^{0.1}r}$ is the Galactic extinction and K-corrected $r$-band
absolute galaxy magnitude. This procedure is similar to that adopted
by P09.

\begin{figure}
  \centering
  \resizebox{0.9\columnwidth}{!}{\includegraphics{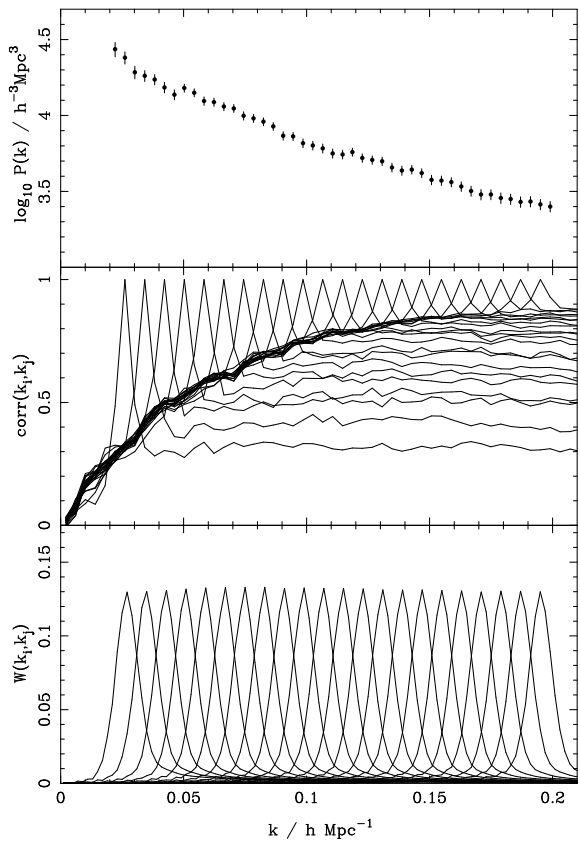}}
  \caption{\label{fig:windowandcorr} {\em Top panel:} Measured
    $\hat{P}_{halo}(k)$ bandpowers.  Error bars indicate
    $\sqrt{C_{ii}}$ (Eq.~\ref{covdefn}).  {\em Middle panel}:
    Correlations between data values calculated using Log-normal
    catalogues, assuming our fiducial cosmological model. {\em Bottom
      panel:} The normalized window function for each of our binned
    power spectrum values with $0.02<k<0.2 \hompc$. Each curve shows
    the relative contribution from the underlying power spectrum as a
    function of $k$ to the measured power spectrum data. The
    normalisation is such that the area under each curve is unity. For
    clarity we only plot curves for every other band power.}
\end{figure}

The power spectrum was calculated using a $1024^3$ grid in a series of
cubic boxes. A box of length $4000\mpcoh$ was used initially, but we
then sequentially divide the box length in half and apply periodic
boundary conditions to map galaxies that lie outside the box. For each
box and power spectrum calculation, we include modes that lie between $1/4$
and $1/2$ the Nyquist frequency (similar to the method described by
\citealt{cole/etal:2005}), and correct for the smoothing effect of the
cloud-in-cell assignment used to locate galaxies on the grid
\citep[e.g.][chap.~5]{he81}. The power spectrum is then spherically
averaged, leaving an estimate of the ``redshift-space'' power. The
upper panel of Fig.~\ref{fig:windowandcorr} shows the shot-noise
subtracted bandpowers measured from the halo density field, calculated
in bands linearly separated by $\Delta k=0.004\hompc$. This spacing is
sufficient to retain all of the cosmological information.

The calculation of the likelihood for a cosmological model given the
measured bandpowers $\hat{P}_{halo}(k)$ requires three additional
components determined by the survey geometry and the properties of the
galaxy sample: the covariance matrix of measured bandpowers $C_{ij}$,
the window function $W(k_i,k_n)$, and the model power spectrum as a
function of the underlying cosmological parameters, $P_{halo}(k,{\bf
  p})$. The calculation of model power spectra is considered in
Section~\ref{modelsec}.

The covariance matrix and corresponding correlation
coefficients between bandpowers $i$ and $j$ are defined as
\begin{eqnarray}
  C_{ij} & = & \langle\hat{P}_{halo}(k_i)\hat{P}_{halo}(k_j)\rangle
    - \langle\hat{P}_{halo}(k_i)\rangle\langle\hat{P}_{halo}(k_j)\rangle
    \label{covdefn} \\
    corr(k_i,k_j) & = & \frac{C_{ij}}{\sqrt{C_{ii}C_{jj}}} \label{corrdefn}
\end{eqnarray}
The covariance matrix was calculated from $10^4$ Log-Normal (LN)
catalogues \citep{coles91,cole/etal:2005}. Catalogues were calculated on a
$(512)^3$ grid with box length $4000\mpcoh$ as in P09, where LN
catalogues were similarly used to estimate covariance matrices. Unlike
$N$-body simulations, these mock catalogues do not model the growth of
structure, but instead return a density field with a log-normal
distribution, similar to that seen in the real data. The window
functions for these catalogues were matched to that of the halo
catalogue. The input power spectrum was a cubic spline fit matched to
the data power spectra, multiplied by a damped $\Lambda$CDM BAO model
calculated using {\sc CAMB} \citep{lewis/challinor/lasenby:2000}. The recovered LN power
spectra were clipped at 5$\sigma$ to remove extreme outliers which
contribute less than 0.05\% of the simulated power spectra, and are
clearly non-Gaussian. This covariance matrix calculation matches the
procedure adopted by P09. The middle panel of
Fig.~\ref{fig:windowandcorr} shows the correlations expected between
band-powers calculated using this procedure.

As described in \citet{cole/etal:2005}, the window function can be
expressed as a matrix relating the theory power spectrum
for cosmological parameters ${\bf p}$ and evaluated at wavenumbers $k_n$,
 $P_{halo}(k_n, {\bf p})$, to the central wavenumbers of the observed
bandpowers $k_i$:
\begin{equation}  \label{defwindow}
  P_{halo,win}(k_i,{\bf p}) =
    \sum_nW(k_i,k_n)P_{halo}(k_n,{\bf p}) - W(k_i,0).
\end{equation}
The term $W(k_i,0)$ arises because we estimate the average halo
density from the sample, and is related to the integral constraint in
the correlation function \citep{percival/etal:2007}. The window
function allows for the mode-coupling induced by the survey geometry.
Window functions for the measured power spectrum (Eqn.~15 of \citealt{percival/verde/peacock:2004}) were calculated as
described in \citet{percival/etal:2001}, \citet{cole/etal:2005}, and
\citet{percival/etal:2007}: an unclustered random catalogue with the
same selection function as that of the haloes was Fourier transformed
using the same procedure adopted for our halo overdensity field
described above. The shot noise was subtracted, and the power spectrum
for this catalogue was spherically averaged, and then fitted with a
cubic spline, giving a model for $W(k_i,k_n)$.  For ease of use this
is translated into a matrix by splitting input and output power
spectra into band powers as in Eqn.~\ref{defwindow}.

The window functions $W(k_i,k_j)$ and the corresponding correlation
coefficients for every other bandpower are shown in the lower panel of
Fig.~\ref{fig:windowandcorr}. In addition to the window coupling for
nearby wavenumbers, there is a beat-coupling to survey-scale modes
\citep{hamilton/rimes/scoccimarro:2006, reid/spergel/bode:2008}.  That
is, density fluctuations on the scale of the survey couple to the
modes we can measure from the survey.  However, this effect
predominantly changes only the amplitude of $\hat{P}_{halo}(k)$, which
is marginalized over through the bias parameter $b_0^2$ in
Eqn.~\ref{eq:nuis} below.  Fig.~\ref{fig:windowandcorr} can be
compared with Fig.~10 in \citet{percival/etal:2007}, where the windows
and correlations were presented for the SDSS DR5 data. For the DR5
plot, variations in the amplitude were removed leaving only the
small-$k$ difference couplings.  The power spectrum, window functions,
and inverse covariance matrix are electronically available with the
likelihood code we publicly release (see
Section~\ref{cosmoconstraints}).

\subsection{$\hat{P}_{halo}(k)$ likelihood}
\label{nongaussianbao}
We assume that the likelihood distribution of the power spectrum band
powers is close to a standard multi-variate Gaussian; by the central
limit theorem, this should be a good approximation in the limit of 
many modes per band. The final
expression for the likelihood for cosmology ${\bf p}$ is then
\begin{equation} \label{likeeqn}
  -2 \ln \; L({\bf p}) = \chi^2({\bf p}) 
    = \sum_{ij} \Delta_i C_{ij}^{-1} \Delta_j,
\end{equation}
where $\Delta_i\equiv\left[(\hat{P}_{halo}(k_i) - P_{halo,win}(k_i,{\bf p})\right]$.

A single comoving distance-redshift relation $\chi_{fid}(z)$, 
that of a flat, $\Omega_m = 0.25$
cosmology, is assumed to assign positions to the galaxies in our
sample before computing $\hat{P}_{halo}(k)$.  Rather than recomputing
$\hat{P}_{halo}(k)$ for each comoving distance-redshift relation to be tested, 
\citet{percival/etal:2007} and P09
account for this when evaluating the likelihood of other cosmological
models by altering the window function.  
$D_V(z, {\bf p})$ \citep{eisenstein/etal:2005} 
quantifies the model dependence of the conversion between ({\bf ra}, {\bf dec},
z) and comoving spatial coordinates when galaxy pairs are distributed isotropically:
\begin{equation}  \label{dveqn}
  D_V(z) = \left[(1+z)^2 D_A(z)^2 \frac{cz}{H(z)}\right]^{1/3},
\end{equation}
where $D_A(z)$ is the physical angular diameter distance.  
Following \citet{tegmark/etal:2006} we partially correct for the
discrepancy between the fiducial model $\chi_{fid}(z)$ and the $\chi(z)$
of the model to be tested by introducing a single dilation of scale.
To first order, changes in the cosmological
distance--redshift model alter the scale of the measured power
spectrum through $D_V(z)$, so we 
 introduce a scale parameter that depends on this quantity,
\begin{equation}   \label{asclcorr}
  a_{scl}(z) = \frac{D_V(z)}{D_V^{\rm fiducial}(z)}.
\end{equation}
Strictly, we should allow for variations in $a_{scl}$ across the redshift range of the
survey, as in P09. However, to first
approximation we can simply allow for a single scale change at an
effective redshift for the survey $z_{eff}$.
When comparing $\hat{P}_{halo}(k)$, computed using $\chi_{fid}(z)$,
with a model comoving distance-redshift relation $\chi(z, {\bf p})$, in practice we use
\footnote{This correction was incorrectly applied in previous versions of
{\sc cosmomc}, and is corrected in the code we release.  This 
correction is primarly important for constraining the BAO scale rather than the
turnover scale, and so previous analyses with {\sc cosmomc}
should be minimally affected.}
\begin{equation}
\Delta_i = \left(\hat{P}_{halo}(k_i) - P_{halo,win}(k_i/a_{scl}, {\bf p})\right).
\end{equation}
In Appendix \ref{dvapprox} we verify that this approximation is valid for our 
sample with $z_{eff} = 0.313$.  

In our cosmological analysis, 
we include modes up to $k_{max} = 0.2 \hompc$, where the
model power spectrum deviates from the input linear power spectrum by
$<15\%$.  We also impose a conservative
lower bound at $k_{min} = 0.02$, above which galactic extinction
corrections (see the analysis in \citealt{percival/etal:2007}), galaxy
number density modeling, and window function errors should be
negligible.

P09 present a detailed analysis demonstrating that the
BAO contribution to the likelihood surface is non-Gaussian; this is in large
part due to the relatively low signal-to-noise ratio of the BAO signature in our sample.
Therefore, to match expected and recovered confidence intervals, 
P09 find that the 
covariance matrix of the LRG-only sample must be inflated by a factor $1.1^2 = 1.21$.  
Though our likelihood surface
incorporates constraints from the shape of the power spectrum, for
which the original covariance matrix should be accurate, we conservatively
multiply the entire covariance matrix by this factor required for the BAO
constraints throughout the analysis.  Therefore our constraints likely
slightly underestimate the true constraints available from the data.  
This factor is already included in the electronic version we release with the
full likelihood code.

\section{MODELING THE HALO POWER SPECTRUM} \label{modelsec}

We consider three effects that cause the shape of $P_{halo}(k, {\bf p})$ to
deviate from the linear power spectrum, $P_{lin}(k, {\bf p})$, for cosmological
parameters ${\bf p}$.  We will assume
that these modifications of the linear power spectrum can be treated
independently.  These effects are the damping of the BAO, 
the change in the broad shape of the power spectrum because of 
non-linear structure formation, and the bias because we observe 
galaxies in haloes in redshift space rather than the real space matter distribution. We also need to consider the 
evolution of these effects with redshift.

\citet{reid/spergel/bode:2008} construct a large set 
of mock LRG catalogues based on $N$-body simulations evaluated 
at a single cosmological model ${\bf p}_{fid}$.  We use these catalogues to 
calibrate the model halo power spectrum, and make detailed 
comparisons between the observed and 
mock density fields in Appendix \ref{quantnuisdetails}.

\subsection{BAO damping}
\label{baodamping}
The primary effect of non-linear structure formation and peculiar
velocities on the BAOs is to damp them at large $k$.  
\citet{eisenstein/seo/white:2007} showed that this can be
accurately modelled as a Gaussian smoothing, where
\begin{equation}  \label{eq:Pdamp}
  P_{\rm damp}(k,{\bf p},\sigma) 
  = P_{\rm lin}(k,{\bf p}) e^{-\frac{k^2\sigma^2}{2}}
  + P_{\rm nw}(k,{\bf p}) \left(1-e^{-\frac{k^2\sigma^2}{2}}\right). 
\end{equation}
Here $P_{\rm lin}(k, {\bf p})$ is the linear matter power spectrum
 computed by CAMB \citep{lewis/challinor/lasenby:2000} and shown in the
 upper left panel of Fig.~\ref{fig:LRGtheory} for our fiducial cosmological model. 
$P_{\rm nw}(k, {\bf p})$, defined by Eqn. 29 of
\citet{eisenstein/hu:1998}, is a smooth version of $P_{\rm lin}(k,
{\bf p})$ with the baryon oscillations removed.  The upper right panel of
Fig.~\ref{fig:LRGtheory} shows the ratio $P_{\rm lin}(k)/P_{\rm
  nw}(k)$ for our fiducial cosmology.  The amplitude of the damping
is set by $\sigma$ and depends on the cosmological parameters, whether
the power spectrum is in real or redshift space, and whether we are
considering the matter or a tracer like the LRGs.
We fix $\sigma_{halo}$, i.e., the value of $\sigma$ appropriate for the reconstructed halo density field, using fits to the
reconstructed halo density field power spectrum in the mock LRG
catalogues presented in \citet{reid/spergel/bode:2008} and shown here
in Fig.~\ref{fig:fidfits}.  
We performed tests in the $\Lambda$CDM case which
demonstrate that cosmological constraints are not altered when
$\sigma$ is allowed to vary with cosmology {\bf p} according to the
dependence given in \citet{eisenstein/seo/white:2007}, and in Appendix \ref{pnwapprox}
we show that using a spline fit to $P_{\rm lin}$ instead of the \citet{eisenstein/hu:1998}
formula for $P_{\rm nw}$ does not affect the likelihood surface in the region of interest.

\subsection{Non-linear structure growth}
\begin{figure}
  \centering
  \resizebox{0.9\columnwidth}{!}{\includegraphics{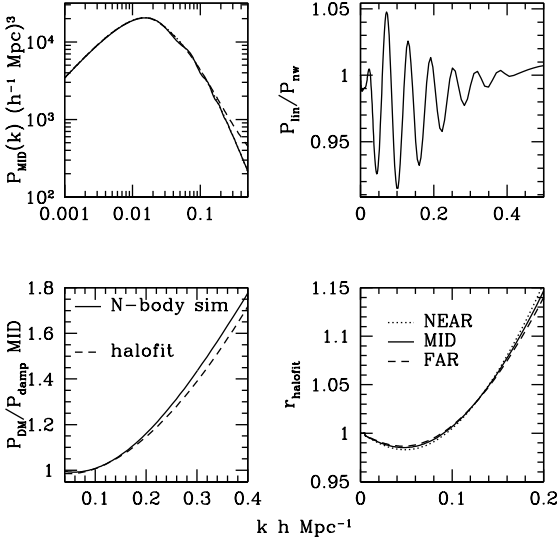}}
  \caption{\label{fig:LRGtheory} {\it Upper left panel}: Power spectra
    for the fiducial cosmology.  The solid curve is $P_{\rm lin}(k)$
     and the dashed curve is
    $P_{\rm nw}(k) r_{\rm halofit}$, the non-linear power spectrum from {\sc halofit}
    using $P_{\rm nw}(k)$ as the input.  {\it Upper right panel}:
    $P_{\rm lin}(k)/P_{\rm nw}(k)$.  {\it Bottom left panel}: 
    $P_{DM}(k)/P_{damp}(k, \sigma_m)$ measured in $N$-body
    simulation snapshots at $z_{MID}$, reported in
    \citet{reid/spergel/bode:2008}, compared with the smooth
    correction predicted by {\sc halofit}, $r_{\rm halofit}$.
      {\it Bottom right panel}: $r_{\rm halofit}$
    at \{$z_{NEAR}$, $z_{MID}$, $z_{FAR}$\} = \{0.235, 0.342, 0.421\}.}
\end{figure}

As the small perturbations in the early universe evolve, 
gravitational instability drives the density field non-linear, 
and power on small scales is enhanced as structures form.  
{\sc halofit} \citep{smith/etal:2003} provides an analytic formalism to estimate the real space non-linear matter power as a function of the underlying linear matter power spectrum.  
While Eqn.~\ref{eq:Pdamp} accounts for the effects of non-linear growth of structure
on the BAO features in $P_{halo}(k,{\bf p})$, {\sc halofit} provides a more 
accurate fit to the smooth component of the non-linear growth in the 
quasi-linear regime ($k \leq 0.2$) when evaluated 
with an input spectrum $P_{nw}(k,{\bf p})$ rather than the linear matter power spectrum containing BAO wiggles:
\begin{eqnarray}
  r_{halofit}(k, {\bf p}) 
    & \equiv & \frac{P_{\rm halofit,nw}(k,{\bf p})}{P_{\rm nw}(k,{\bf p})} \label{eq:halofit} \\
  P_{DM,halofit}(k,{\bf p}) & = & P_{damp}(k,{\bf p},\sigma_m) r_{halofit}(k,{\bf p}). \label{mattermodel}
\end{eqnarray}
Eqn.~\ref{mattermodel} is our modified {\sc halofit} model real space
power spectrum, using Eqn.~\ref{eq:Pdamp} to account for BAO
damping and {\sc halofit} for the smooth component.
The bottom left panel of Fig.~\ref{fig:LRGtheory} shows that 
$P_{DM}(k)/P_{damp}(k,\sigma_m)$ and $r_{\rm halofit}$ 
agree at the $\sim 1.5\%$ level for $k \leq 0.2$ 
in our fiducial cosmology.  Since we normalize the final model $P_{halo}(k, {\bf p})$
using our mock catalogues at the fiducial cosmology ${\bf p}_{fid}$, 
in practice {\sc halofit} only provides the cosmological dependence 
of the non-linear correction to the matter power spectrum:
\begin{equation}  \label{eq:DM_model}
  r_{DM,damp}(k, {\bf p}) 
    = \frac{r_{halofit}(k, {\bf p})}{r_{halofit}(k,{\bf p}_{\rm fid})}
    \frac{P_{DM}(k,{\bf p}_{\rm fid})}
      {P_{\rm damp}(k,{\bf p}_{\rm fid},\sigma_{DM})}.
\end{equation}
$r_{DM,damp}(k,{\bf p})$ is our model for the ratio of the
 non-linear matter power spectrum to the damped 
 linear power spectrum.  The normalization of $r_{DM,damp}$ 
 accounts for the small offset between the $N$-body and {\sc halofit} results
  in Fig.~\ref{fig:LRGtheory} at the fiducial cosmology.
  In the space of cosmologies consistent with the data, the small cosmology-dependence of this correction is primarily
 through $\sigma_8$.  In Section \ref{lrgonlycosmodep} we find that the LRG-only 
 likelihood surface is independent
 of the assumed value of $\sigma_8$ over the range 0.7 to 0.9.
 
\subsection{Halo bias}
\label{modelhalobias}
In our likelihood calculation 
we marginalize over the overall amplitude of $\hat{P}_{halo}(k)$,
so in this Section we are concerned only with the scale dependence
of the relation between the reconstructed halo and matter power
spectra.  \citet{smith/scoccimarro/sheth:2007} show that the scale dependence of 
halo bias in real space is large for the most massive haloes, but should be 
rather weak for the halo mass range which host the majority of the LRGs; 
\citet{matsubara:2008} demonstrates this analytically in redshift space
 in the quasi-linear regime.
Indeed, \citet{reid/spergel/bode:2008} find that the power spectrum of
the (redshift space) reconstructed halo density field is nearly linearly biased with
respect to the underlying real space matter power spectrum
for $k < 0.2 \hompc$ and our fiducial $\Lambda$CDM model, and we
assume this should remain approximately true in the narrow range of cosmologies
consistent with the data. For the fiducial cosmology, we can
use our simulations to calibrate the relation between the halo and matter spectra:
\begin{equation}  \label{eq:bias}
  r_{halo, DM}(k, {\bf p}_{\rm fid}) 
  = \frac{P_{\rm halo}(k,{\bf p}_{\rm fid})/P_{damp}(k,{\bf p},\sigma_{halo})}
  {P_{\rm DM}(k,{\bf p}_{\rm fid})/P_{damp}(k,{\bf p},\sigma_{m})}.
\end{equation}
This is our model for the smooth component of the bias between 
the halo and dark matter power spectra.  To account for any dependence of
$r_{halo, DM}(k, {\bf p}_{\rm fid})$ on the cosmological model and other 
remaining modeling uncertainties, we introduce a smooth multiplicative correction to the
final model $P_{halo}(k,{\bf p})$ containing three nuisance parameters 
$b_0$, $a_1$ and $a_2$:
\begin{equation}  
F_{nuis}(k) = b_0^2 \left(1+a_1 \left(\frac{k}{k_{\star}}\right) + a_2 \left(\frac{k}{k_{\star}}\right)^2\right) \label{eq:nuis}
\end{equation}
where we set $k_{\star} = 0.2 \hompc$.
The parameter $b_0$ is the effective bias of the LRGs at the effective sample redshift, $z_{eff}$, relative to $L^{\star}$ galaxies (Eqn.~18 of \citealt{percival/verde/peacock:2004}).  In Section \ref{quantnuis} we will use consistency checks between the observed and mock catalogue galaxy density fields as well as the halo model framework to establish the allowed region of $a_1-a_2$ parameter space.  An allowed trapezoidal region in $a_1-a_2$ space is completely specified through two parameters, $u_{0.1}$ and $u_{0.2}$.  These two parameters specify the maximum absolute deviation allowed by $F_{nuis}(k)/b_0^2$ away from 1 for $k \leq 0.1$ ($u_{0.1}$) and $0.1 \leq k \leq 0.2$ ($u_{0.2}$).  When evaluating the likelihood of a particular cosmological model we marginalize analytically over $b_0$ using a flat prior on $b_0^2 \geq 0$, and we marginalize numerically over the allowed $a_1-a_2$ region with a flat prior in this region.  We discuss the impact of these priors on the cosmological constraints in Appendix \ref{nuiseffects}.

\subsection{Model fits and evolution with redshift}
\label{modelz}
Our final model halo power spectrum at fixed redshift treats
each of the three non-linear effects independently: Eqn.~\ref{eq:Pdamp} converts the linear power spectrum to the damped linear power spectrum, $r_{DM,damp}$
converts the damped linear power spectrum to the real space
non-linear matter power spectrum, $r_{halo,DM}$ converts the real 
space non-linear matter power spectrum to the redshift space reconstructed
halo density field power spectrum 
(assuming this relation is cosmology independent), and
$F_{nuis}(k)$ allows for smooth deviations
from our model due to modeling errors, uncertainties,
and unaccounted cosmological parameter dependencies:
\begin{eqnarray} 
P_{halo}(k,{\bf p}) = P_{damp}(k,{\bf p}) r_{DM,damp}(k,{\bf p}) \times \nonumber\\
 r_{halo,DM}(k,{\bf p_{fid}}) F_{nuis}(k). \label{finalmodel}
\end{eqnarray}
For this multiplicative model, 
the $P_{\rm DM}(k,{\bf p}_{\rm fid})/P_{damp}(k,{\bf p},\sigma_{m})$
terms from Eqns.~\ref{eq:DM_model} and ~\ref{eq:bias} cancel, 
so calibration of the model only requires 
fits to $\sigma_{\rm halo}$ and $P_{halo}(k,{\bf p}_{fid})/P_{damp}(k,{\bf 
p}_{fid}, \sigma_{\rm halo})$ using the mock catalogues.

\begin{figure}
  \centering
  \resizebox{0.9\columnwidth}{!}{\includegraphics{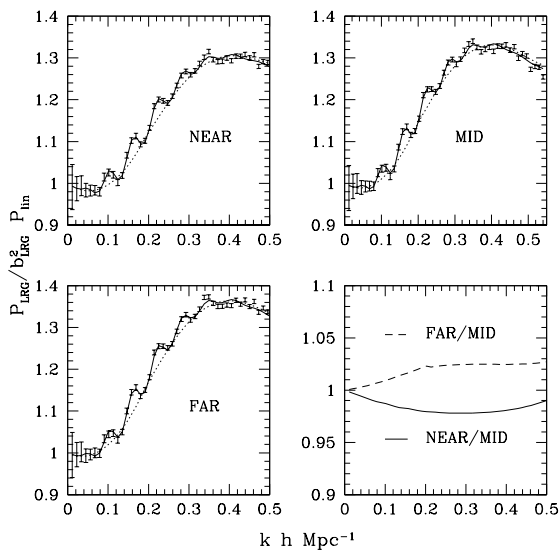}}
  \caption{\label{fig:fidfits} BAO-damping times polynomial fits to
    $P_{halo}(k,{\bf p}_{fid})/P_{lin}(k,{\bf p}_{fid})$ for our mock
    NEAR, MID, and FAR LRG reconstructed halo density field subsamples
    in \citet{reid/spergel/bode:2008}; \{$z_{NEAR}$, $z_{MID}$, $z_{FAR}$\} = \{0.235, 0.342, 0.421\}.  The smooth component of these
    fits (dashed curves) enter our model $P_{halo}(k, {\bf p})$ through
    Eqns.~(\ref{eq:bias}), while the amplitude
    of the BAO suppression $\sigma^2_{halo}$ enters in
    Eqn.~\ref{eq:Pdamp}.  {\it Lower right panel}: Ratio of the
    shape of the smooth components for the NEAR and FAR redshift subsamples to the MID redshift subsample.}
\end{figure}

The model in Eqn.~\ref{finalmodel} is strictly only valid at a single redshift.  
In order to match our model to the observed redshift distribution of the LRGs and
their associated haloes, we use the mock halo catalogues constructed in
\citet{reid/spergel/bode:2008} at three redshift snapshots. These are
centered on the NEAR ($z_{NEAR} = 0.235$), MID ($z_{MID} = 0.342$),
and FAR ($z_{FAR} = 0.421$) LRG subsamples of
\citet{tegmark/etal:2006}.  
  Fig.~\ref{fig:fidfits} shows our fits to $P_{halo}(k,{\bf
  p}_{fid})/P_{lin}(k,{\bf p}_{fid})$ 
  for each redshift snapshot.
We first fit for $\sigma_{halo}$ in Eqn.~\ref{eq:Pdamp} using
 our LRG mock catalogue results $P_{halo}(k,{\bf
  p}_{fid})$.  We include modes between $k=0 \hompc$ and $k=0.2 \hompc$
  in the fit and marginalize over an arbitrary fourth order polynomial
  to account for the smooth deviations from $P_{damp}$ with $k$.
 We find $\sigma_{halo, \;
  NEAR} = 9.3\; h^{-1}$ Mpc, $\sigma_{halo, \; MID} = 9.2\; h^{-1}$
Mpc, and $\sigma_{halo, \; FAR} = 9.2 \; h^{-1}$ Mpc.  These numbers
are roughly consistent with the results presented in
\citet{eisenstein/seo/white:2007}, and are somewhat degenerate with
the smooth polynomial correction.  

After fixing these values for $\sigma_{halo}$, we calibrate the smooth
component of the model, $r_{DM,damp}(k,{\bf p}_{fid}) r_{halo,DM}(k,{\bf p}_{fid})$.
For $k \leq 0.2$ we fit
$P_{halo}(k,{\bf p}_{fid})/P_{\rm damp}(k,{\bf
  p}_{fid},\sigma_{halo})$ to a second order
polynomial, and a fourth order polynomial for $k \leq 0.5$.  This component
of the fit is shown in the first three panels of Fig.~\ref{fig:fidfits} by the dotted curves,
while the solid lines show the full fit to $P_{halo}(k,{\bf
  p}_{fid})/P_{lin}(k,{\bf p}_{fid})$.  Both the BAO-damping and smooth
  increase in power with $k$ are well described by our fits out to $k=0.5 \hompc$.

Our final model for the reconstructed halo power spectrum is a weighted sum over
our model $P_{halo}(k, {\bf p})$ (Eqn.~\ref{finalmodel}) from each of the NEAR, MID,
and FAR redshift slices fit in Fig.~\ref{fig:fidfits}:
\begin{equation}
  P_{halo}(k, {\bf p}) =\sum_{i=NEAR,MID,FAR} w_i P_{halo}(k, {\bf p}, z_{i})\,,
\end{equation}
where $w_i$ specifies the weight of each redshift subsample.
The lower right panel of Fig.~\ref{fig:LRGtheory} shows that the
smooth correction for the non-linear matter power spectrum varies by
$<1\%$ over the redshift range of the LRGs.  Moreover, the lower right
panel of Fig.~\ref{fig:fidfits} shows that the relative shape of the
power spectrum of the reconstructed halo density field varies by 
$\pm \sim 2.5\%$ between the redshift subsamples, 
so moderate biases in the determination of these weights will
induce negligible changes in the predicted shape $P_{halo}(k, {\bf
  p})$.

In the limit that most pairs of galaxies contributing power to mode
${\bf k}$ come from the same redshift, the fractional contribution to
the power spectrum from a large redshift subsample is
\begin{equation}
\label{pairwgt}
  w(z_{min}, z_{max}) \propto \int_{z_{min}}^{z_{max}} 
    n^2(z) \frac{w^2(z)}{b^2(z)} \frac{dV}{dz} dz\,,
\end{equation}
where $n(z)$, $b(z)$, and $w(z)$ specify the average number density,
bias, and weight of the sample at redshift $z$ 
as defined in \citet{percival/verde/peacock:2004}.
Since the integrand is slowly varying with redshift, this
approximation should be fairly accurate.  We derive weights $w_{NEAR}
= 0.395$, $w_{MID} = 0.355$, and $w_{FAR} = 0.250$.

\subsection{Comparison with fiducial model $P_{halo}(k,{\bf p_{fid}})$}
Our fiducial $P_{halo}(k,{\bf p})$ model is calibrated on simulations with the
WMAP5 recommended parameters \citep{komatsu/etal:2009}: ($\Omega_m,
\Omega_b, \Omega_{\Lambda}, n_s, \sigma_8, h)$ = (0.2792, 0.0462,
0.7208, 0.960, 0.817, 0.701).  For the 45 observed bandpowers satisfying 
$0.02 < k < 0.2 \hompc$, $\chi^2 = 44.0$ if we hold nuisance 
parameters $a_1 = a_2 = 0$ and choose $b_0$ to minimize $\chi^2$;
our fiducial model is therefore sufficiently close to the measured $\hat{P}_{halo}(k)$
to be used to calibrate the cosmology-dependent model.  The best-fitting 
nuisance parameters within the allowed range that we determine in Section \ref{getmaxnuisvals}, $a_1 = 0.172$ and
$a_2 = -0.198$, lower the $\chi^2$ to 40.9 for 42 DOF.  The 
 best-fitting model to the LRG-only likelihood presented in 
 Section \ref{DR7only} is lower by only $\Delta \chi^2 \approx 1.7$ for the same treatment of the three nuisance parameters.

\section{QUANTIFYING MODEL UNCERTAINTIES AND CHECKS FOR SYSTEMATICS}
\label{quantnuis}
While the non-linear evolution of a collisionless dark matter density
field can be accurately studied using $N$-body simulations, there
remain many uncertainties in the mapping between the galaxy and matter
density field.  We first review the generic halo model predictions for 
a galaxy power spectrum, which provide the context for exploring the
uncertainties in the relation between the galaxy and matter density fields.
We summarize the results of Appendix \ref{quantnuisdetails}, which presents our modeling assumptions and
consistency checks between the mock catalogue
and SDSS DR7 LRG density fields that constrain the level of deviation
from our modeling assumptions.  The ultimate goal of this Section is to establish
physically-motivated constraints on the nuisance parameters $a_1$ and
$a_2$ in Eqn.~\ref{eq:nuis} by determining $u_{0.1}$ and $u_{0.2}$ defined in Section \ref{modelhalobias}.  
These nuisance parameter constraints will then be used to compute cosmological
parameter constraints in Section \ref{cosmoconstraints}.

\subsection{Galaxy power spectra in the halo model}
In the simplest picture for a galaxy power spectrum in the halo model, one
considers a separation of the pairs into galaxies occupying the same
dark matter halo, which contribute to $P^{1h}(k)$, and those occupying
different dark matter haloes, which contribute to $P^{2h}(k)$
\citep{cooray/sheth:2002}:
\begin{eqnarray}
  P_{gal}(k) & = & P^{1h}_{gal}(k) + P^{2h}_{gal}(k) \label{onetwopkeqn}\\
  P^{1h}_{gal} & = & \int dM \; n(M) 
  \frac{\left< N_{gal} (N_{gal} - 1 ) | M) \right > }
  {\bar{n}^2_{gal}} \label{onehaloterm}\\
  P^{2h}_{gal}(k) & = & b_{gal}^2 P_{DM}(k).\label{twohaloterm}
\end{eqnarray}
On large scales, treating the haloes as linear tracers of the underlying
matter density field (Eqn.~\ref{twohaloterm}) and ignoring the spatial
extent of haloes in Eqn.~\ref{onehaloterm}  
are good approximations
\citep{reid/spergel/bode:2008}.  Therefore, in real space, the
dominant effect of the inclusion of satellite galaxies is an excess
shot noise given by Eqn.~\ref{onehaloterm}, though they also upweight
highly biased halo pairs and slightly increase $b_{gal}$ as well.
However, in redshift space, satellite galaxies are significantly
displaced along the line of sight from their host haloes by the FOGs, and power
is shuffled between scales, and even the largest scale modes along the
line of sight are damped by the FOG smearing.  There will be residual non-linear
redshift space distortions in the reconstructed halo density field from imperfect reconstruction, 
and potentially from peculiar motion of isolated LRGs in their host haloes as well.

\subsection{Summary of tests for systematics and remaining uncertainties}
In the context of the
halo model, both uncertainty in the distribution of galaxies in groups
as it enters Eqn.~\ref{onehaloterm} and uncertainty in the structure of the 
FOG features will introduce uncertainty in the relation between the reconstructed halo
and matter density fields, and thus their power spectra.  Appendices \ref{haloparam} and \ref{galaxiesinhaloes} discuss the modeling assumptions we have used to derive the \citet{reid/spergel/bode:2008} mock LRG catalogues from $N$-body simulation halo catalogues, and state the expected impact on the relation between the reconstructed halo and matter power spectra.  

Appendix \ref{cicstats} introduces several distinct consistency checks of the uncertainties in Appendices \ref{haloparam} and \ref{galaxiesinhaloes}.  In Section \ref{preprocess} we define the CiC group finder by which we identify haloes.  We demonstrate that this group finder produces group multiplicity functions that are in good agreement between the mock and observed LRG density fields, once fiber collisions are accounted for.  While this agreement demonstrates that our mock catalogues reproduce small scale higher-order clustering statistics and FOG features of the observed density field, this is not a consistency check since the mocks were designed to match these statistics.  We find consistency when we compute a second CiC group multiplicity function allowing a wider separation between pairs perpendicular to the line of sight ($\Delta r_{\perp} = 1.2 \; h^{-1}$ Mpc).  If the observed satellite galaxies were significantly less concentrated than in our mock catalogues, we would detect these galaxies when $\Delta r_{\perp}$ increases from 0.8 $h^{-1}$ Mpc to 1.2 $h^{-1}$ Mpc.  From this comparison we conclude that residual shot noise errors from inaccurate halo density field reconstruction are $\sim 2\%$
of the total shot noise correction and do not dominate our systematic uncertainty.  The second consistency check between the mock and observed LRG catalogues is the distribution of line of sight separations between pairs
of galaxies in the same CiC group (Fig~\ref{fig:FOGhist}).  This check probes the accuracy of our model of the FOG features coming from galaxies occupying the same halo, and the agreement we find indicates that the residual FOG features in the reconstructed observed and mock halo density fields will be in satisfactory agreement.  Appendix \ref{cicvsnotcompare} presents the difference between the power spectra with and without the halo density field reconstruction preprocessing step ($\hat{P}_{halo}(k)$ and $\hat{P}_{LRG}(k)$, respectively).  This difference agrees with the mock catalogues, provided one carefully accounts for the impact of fiber collisions.  In other words, while the treatment of fiber collisions can substantially impact $\hat{P}_{LRG}(k)$, $\hat{P}_{halo}(k)$ is unaffected.  In Appendix \ref{lumweight} we demonstrate that the luminosity weighting used to compute $\hat{P}_{halo}(k)$ but not accounted for in the mock catalogues does not alter the effective shot noise level of $\hat{P}_{halo}(k)$.  Appendix \ref{wmap3cats} presents evidence that the cosmology dependence of the model $P_{halo}(k,{\bf p})$ is sufficiently accurate.  Finally, we note that \citet{lunnan/etal:prep}
have compared the \citet{reid/spergel/bode:2008} mock catalogue genus curve
 with the observed genus curves  \citep{gott/etal:2009}, and find good agreement with no free parameters.

As discussed in detail in Appendix \ref{galaxiesinhaloes}, 
the vast majority of LRGs ($\sim 94\%$) are expected to reside at the centre
of their host dark matter haloes \citep{zheng/etal:2008,reid/spergel/bode:2008}.  The principal modeling uncertainty we identify in Appendix \ref{quantnuisdetails} is
the velocity of these central LRGs within their host haloes; substantial
intrahalo velocities for these galaxies will suppress power in a scale-dependent
manner (Fig.~\ref{fig:veldispPk}).  Note that none of the tests from Appendix \ref{quantnuisdetails} can directly constrain the level of central LRG velocity dispersion.

\subsection{Constraints on $F_{\rm nuis}(k)$}
\label{getmaxnuisvals}
In Section \ref{modelhalobias} we introduced a quadratic
function $F_{nuis}(k)$ to account both for errors in our modeling
at the fiducial cosmology and for any errors in the 
cosmology dependence
of our model.  We parametrized the amplitude of the total modeling uncertainty through $u_{0.1}$ and $u_{0.2}$.  These parameters, which we determine in this subsection, specify the maximum fractional deviation from the model power spectrum at $k=0.1 \hompc$ and $k=0.2 \hompc$, respectively.  We choose these values of $k$ because $k \leq 0.1$ is usually considered safely in the linear regime, while $k=0.2 \hompc$ is the maximum wavenumber we attempt to model.

The dominant uncertainty in our model is in the relation between the
power spectrum of the reconstructed halo density field and the
underlying matter power spectrum, which we describe by Eqn.~\ref{eq:bias}.
At $k=0.1 \hompc$ in the mock catalogues,
the reconstructed halo density field and the redshift space central 
galaxy power spectra agree well below the percent level.  
The total one-halo correction $P^{1h}$ in real
space is 7-10\%.  If we conservatively assume
that the halo reconstruction algorithm incorrectly subtracts the real space
one-halo term by 20\%, then the systematic error at $k=0.1 \hompc$, $u_{0.1}$, is
allowed to be 2\%.  At $k=0.2 \hompc$, the same error would translate to 5\%
in real space, though in redshift space this term is mitigated.  In Appendix \ref{cicvsnotcompare}
we find that the shape difference of $\hat{P}_{halo}(k)$ and $\hat{P}_{LRG}(k)$
 is only 18\% at $k=0.2 \hompc$, and only 8\% after accounting for the shot
 noise introduced by the fiber collision corrections.  If we assume that
our modeling and treatment of the one-halo contribution to the FOGs
are accurate at the $\sim 50\%$ level, we can estimate a conservative
error at $k=0.2 \hompc$ of 5\%.  Therefore, for all the modeling uncertainties besides
central velocity dispersion that we have discussed, $u_{0.1} = 0.02$ and $u_{0.2} = 0.05$
encompass the estimated uncertainties.  These are our `fiducial' nuisance function
constraints.

In Appendix \ref{galaxiesinhaloes} we find that a large amount of central galaxy
misidentification or central--halo velocity bias can reduce the
amplitude of $P_{halo}(k,{\bf p})$ by a smoothly varying function of $k$ at
a level that exceeds these fiducial bounds on $u_{0.1}$ and $u_{0.2}$.
Our approach to mitigating the impact of uncertain central LRG
peculiar velocities is twofold.  First, for all of the analysis in
Section~\ref{cosmoconstraints} we adopt more conservative bounds for the
nuisance function: $u_{0.1} = 0.04$ and $u_{0.2} = 0.10$, which nearly
encompass the change in power spectrum shape in Fig.~\ref{fig:veldispPk} for the
extreme velocity dispersion model.  Furthermore, we calibrate a second
model from the mocks with extreme velocity dispersion,
and in Appendix \ref{nuiseffects} we 
determine the cosmological parameter constraints with this model
to establish the level of remaining systematic uncertainty in our final results.

\section{COSMOLOGICAL CONSTRAINTS}
\label{cosmoconstraints}
In this Section we explore the cosmological constraints derived from  the power spectrum of the reconstructed halo density field,
$\hat{P}_{halo}(k)$.  We first consider constraints obtained from $\hat{P}_{halo}(k)$ alone, and then combine the LRG  likelihood with WMAP5 and the Union Supernova Sample \citep{kowalski/etal:2008} to explore joint
 constraints in several cosmological models.  Throughout, we make use of 
 the COSMOMC package \citep{lewis/bridle:2002} to
compute cosmological constraints using the Markov Chain Monte-Carlo method.  A stand-alone module to compute the $\hat{P}_{halo}(k)$
likelihood is made publicly available.\footnote{http://lambda.gsfc.nasa.gov/toolbox/lrgdr/}
\subsection{Constraints from the halo power spectrum}
\label{DR7only}
\begin{figure}
  \centering
  \resizebox{1.0\columnwidth}{!}{\includegraphics{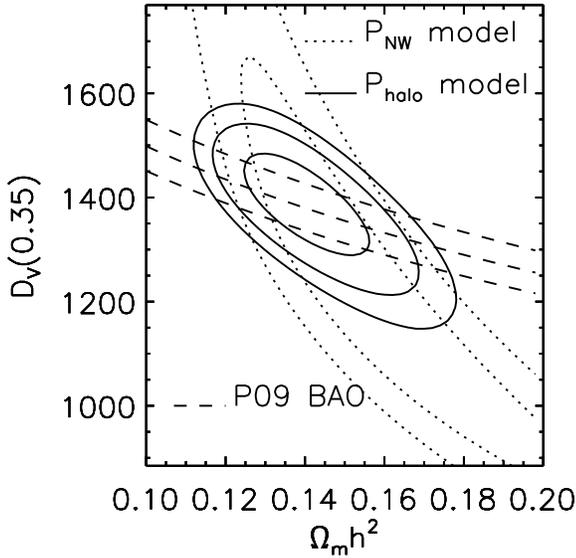}}
  \caption{\label{fig:dr7onlyonepanel} Constraints from the LRG DR7
    $\hat{P}_{halo}(k)$ for a $\Lambda$CDM model with 
    $\Omega_b h^2 = 0.02265$ and $n_s = 0.960$ fixed.  
    The dotted contours show 
    $\Delta \chi^2 = 2.3$ and 6.0 contours for the
    $\hat{P}_{halo}(k)$ fit to a no-wiggles model.
    The solid contours indicate $\Delta \chi^2 = 2.3$, 6.0, and 9.3 contours 
    for $k_{max} =0.2 \hompc$ and our fiducial $P_{halo}(k,{\bf p})$ model.
    The three dashed lines show
    the best-fitting and $\pm 1\sigma$ values $r_s/D_V(0.35) = 0.1097 \pm 0.0036$ from
    P09.}
\end{figure}
\begin{table*}
\begin{center}
\begin{tabular}{llllll}
data/model & $\Omega_m h^2$ & $D_V(0.35)$ (Mpc) & $r_s/D_V(0.35)$ & $\Omega_m h^2 D_V(0.35)$ (Mpc) & $A_{0.35}$ \\
\hline
${\bf k_{max} = 0.2}$ & ${\bf 0.141_{-0.012}^{+0.010}}$ & ${\bf 1380_{-73}^{+61}}$ & ${\bf 0.1097_{-0.0042}^{+0.0039}}$ & ${\bf 194_{-10}^{+10}}$ & ${\bf 0.493_{-0.017}^{+0.017}}$ \\
$k_{max} = 0.15$ & $0.142_{-0.012}^{+0.010}$ & $1354_{-77}^{+64}$ & $0.1118_{-0.0046}^{+0.0043}$ & 
$191_{-11}^{+10}$ & $0.485_{-0.018}^{+0.018}$ \\
$k_{max} = 0.1$ & $0.145_{-0.016}^{+0.014}$ & $1329_{-116}^{+104}$ & $0.1136_{-0.0072}^{+0.0070}$ & $192_{-12}^{+11}$ & $0.480_{-0.024}^{+0.025}$ \\
$k_{max} = 0.2$ weak $F_{nuis}$ & $0.139_{-0.017}^{+0.015}$ & $1384_{-77}^{+64}$ & $0.1099_{-0.0040}^{+0.0039}$ & $192_{-16}^{+17}$ & $0.490_{-0.020}^{+0.020}$ \\
$k_{max} = 0.2$ VD & $0.148_{-0.013}^{+0.011}$ & $1365_{-76}^{+63}$ & $0.1096_{-0.0043}^{+0.0040}$ & $202_{-11}^{+11}$ & $0.499_{-0.018}^{+0.018}$ \\
\hline
$k_{max} = 0.2$ + prior & $0.135_{-0.006}^{+0.004}$ & $1411_{-58}^{+44}$ & $0.1085_{-0.0036}^{+0.0036}$ & $189.9_{-7.5}^{+7.5}$ & $0.493_{-0.016}^{+0.016}$ \\
$k_{max} = 0.15$ + prior & $0.135_{-0.006}^{+0.004}$ & $1387_{-61}^{+48}$ & $0.1104_{-0.0039}^{+0.0040}$ & $186.6_{-7.9}^{+7.9}$ & $0.485_{-0.017}^{+0.017}$ \\
$k_{max} = 0.1$ + prior & $0.134_{-0.007}^{+0.005}$ & $1394_{-81}^{+67}$ & $0.1101_{-0.0053}^{+0.0053}$ & $187.1_{-9.2}^{+9.3}$ & $0.487_{-0.022}^{+0.022}$ \\
$k_{max} = 0.2$ weak $F_{nuis}$ + prior & $0.133_{-0.007}^{+0.005}$ & $1404_{-58}^{+44}$ & $0.1095_{-0.0037}^{+0.0036}$ & $186.1_{-  8.2}^{+  8.4}$ & $0.487_{-0.017}^{+0.017}$ \\
$k_{max} = 0.2$ VD + prior & $0.136_{-0.006}^{+0.004}$ & $1417_{-58}^{+44}$ & $0.1078_{-0.0035}^{+0.0035}$ & $192.9_{-  7.8}^{+  7.4}$ & $0.498_{-0.017}^{+0.016}$ \\
$k_{max} = 0.1$ NW + prior & $0.134_{-0.007}^{+0.005}$ & $1436_{-150}^{+143}$ & $0.1076_{-0.011}^{+0.010}$ & $192_{- 17}^{+ 17}$ & $0.500_{-0.045}^{+0.047}$\\
$k_{max} = 0.2$ NW + prior & $0.134_{-0.007}^{+0.005}$ & $1463_{-142}^{+134}$ & $0.1054_{-0.0095}^{+0.0092}$ & $196_{- 15}^{+ 15}$ & $0.510_{-0.042}^{+0.044}$ \\
\end{tabular}
  \caption{\label{table:lrgdr7only} One-dimensional constraints from the LRG $\hat{P}_{halo}(k)$ likelihood, or in combination with the WMAP5 $\Omega_m h^2$ constraint $\Omega_m h^2 = 0.1326 \pm 0.0063$ (`+ prior', below the line).  We vary the $k_{max}$ (units of $\hompc$) included in the fit, the nuisance function constraints (fiducial vs weak $F_{nuis}$), velocity dispersion in the model (fiducial vs. `VD'), and whether the BAO features are included in the model (fiducial vs `NW').  All constraints have assumed the $\Lambda$CDM relation between $\Omega_m$, $H_0$, and $D_V$, $\Omega_b h^2 = 0.02265$, $n_s = 0.96$, and $\sigma_8 = 0.817$.  In the last column we show $A_{0.35} \equiv \sqrt{\Omega_m H_0^2} D_V(0.35)/0.35c$ \citep{eisenstein/etal:2005}.  Models with weak $F_{nuis}$ constraints or central galaxy velocity dispersion are discussed in Appendix \ref{nuiseffects}.  The $k_{max} = 0.2 \hompc$ constraints highlighted in bold are our main results, and the other cases are shown for comparison.}
\end{center}
\end{table*}
In this subsection we examine the cosmological constraints derived from
the $\hat{P}_{halo}(k)$ alone and in combination with
a prior on $\Omega_m h^2$ from WMAP5.  
In the model $P_{halo}(k,{\bf p})$, the scale factor $a_{scl}$ in Eqn.~\ref{asclcorr} 
is evaluated at $z_{eff} = 0.313$.  For comparison with other works, we scale our
constraint on $D_V(0.313)$ using the fiducial distance-redshift relation,
for which $D_V(0.35)/D_V(z_{eff}) = 1.106$; the variation of this ratio
with cosmological parameters is negligible.  Following \citet{eisenstein/etal:2005},
we consider two free parameters $\Omega_m h^2$ and $D_V(0.35)$.
In this subsection we hold 
$\Omega_b h^2 = 0.02265$, $n_s = 0.960$, and $\sigma_8 = 0.817$
 fixed at their values in the fiducial cosmological model,  and
assume a flat $\Lambda$CDM model;
 in \S \ref{lrgonlycosmodep} we relax these assumptions.

For the 45 bandpowers satisfying
$0.02 < k < 0.2 \hompc$, $\chi^2$ is minimized when 
$D_V(0.35) = 1396$ and $\Omega_m h^2 = 0.136$ with best-fitting nuisance parameters
$a_1 = 0.160$ and $a_2 = -0.181$: $\chi^2 =
39.6$ for 40 degrees of freedom.
Thus the assumed model
power spectrum and covariance matrix provide a reasonable
fit to the observed spectrum.  
In a $\Lambda$CDM model, this point corresponds to  $h = 0.67$
 and $\Omega_m = 0.30$.  Fig~\ref{fig:dr7onlyonepanel} shows $\chi^2$
 contours in the $\Omega_m h^2-D_V(0.35)$ parameter space, while Table~\ref{table:lrgdr7only}
 reports marginalized one-dimensional constraints for several combinations of these parameters.

The information in $\hat{P}_{halo}(k)$ can be roughly divided into broad-shape information and information from the BAO scale. Since in this subsection $n_s$ is fixed, the shape information is the location of the turnover in the power spectrum set by matter-radiation equality, which constrains $\Omega_m h^2 D_V$; 
information from the BAO scale constrains $r_s/D_V$.  Here, $r_s$
is the sound horizon at the baryon-drag epoch, which we evaluate using Eqn. 6 
of \citet{eisenstein/hu:1998}.  These two scales correspond to constraints
 on $h\Omega_m^{0.93}$ and $h \Omega_m^{-0.37}$ respectively, in a 
 $\Lambda$CDM cosmology \citep{tegmark/etal:2006}.

To isolate information from the power spectrum turnover and exclude that of
the BAO scale,
we alter our model so that $P_{\rm damp}(k,{\bf p}) = P_{\rm
  nw}(k,{\bf p})$ in Eqn.~\ref{eq:Pdamp}.  The dashed lines in
Fig.~\ref{fig:dr7onlyonepanel} show the constraints
when using this `no wiggles' model.
Most of the available shape
 information comes from large scales with $k<0.1\hompc$; we demonstrate
 this in Table~\ref{table:lrgdr7only} by fitting the $P_{\rm damp}(k,{\bf p}) = P_{\rm
  nw}(k,{\bf p})$ model with the $\Omega_m h^2$ prior
   to the data up to $k_{max} = 0.1 \hompc$ and $k_{max} = 0.2 \hompc$.  
   The number of independent modes is proportional to 
   $(k_{max}^3-k_{min}^3)$; thus between 
   $k=0.1 \hompc$ and $k=0.2 \hompc$ there are about 7 times 
   more modes than between $k_{min}$ and $0.1 \hompc$.  
   Nevertheless, the constraint on $\Omega_m h^2 D_V(0.35)$ 
   only improves by $\approx 10\%$  with the inclusion of modes 
   between $k_{max} = 0.1 \hompc$ and $k_{max} = 0.2 \hompc$ 
   and does not shift appreciably.   
   This also indicates that our modeling in the 
   quasi-linear regime $0.1 < k < 0.2 \hompc$ 
    does not bias or substantially improve this constraint.
   
  If we reintroduce the BAO features in the model $P_{halo}(k,{\bf p})$, then the
degeneracy between $D_V(0.35)$ and $\Omega_m h^2$ is partially broken (solid contours in Fig.~\ref{fig:dr7onlyonepanel}), and the
constraints grow tighter as we include additional modes.  This is understandable as the region $0.1<k<0.2 \hompc$ includes the location of the second BAO.  
The constraints on both $r_s/D_V(0.35)$ and $\Omega_m h^2 D_V(0.35)$
listed in Table~\ref{table:lrgdr7only} 
improve with $k_{max}$.  The mean value of $\Omega_m h^2 D_V(0.35)$
is consistent with what we find
using the $P_{\rm damp}(k,{\bf p}) = P_{\rm nw}(k,{\bf p})$ model 
with the WMAP5 $\Omega_m h^2$ prior, 
and does not shift substantially with increasing $k_{max}$.
Because the BAO features break the degeneracy between
$\Omega_m h^2$ and $D_V(0.35)$, the LRG $\hat{P}_{halo}(k)$ provides
an independent constraint on $\Omega_m h^2$.  For $n_s = 0.96$,
we find $\Omega_m h^2 = 0.141^{+0.010}_{-0.012}$, which is consistent
with the WMAP5 constraint, $\Omega_m h^2 = 0.1326 \pm 0.0063$, 
but with a 70\% larger error.

Fig.~\ref{fig:dr7onlyonepanel} shows that
the LRG-only constraints derived with $k_{max} = 0.2 \hompc$ are consistent
 with the intersection of the power spectrum shape constraint (dotted lines)
  combined with constraints on
  $r_s/D_V(0.35)$ from P09: the best-fitting and $\pm 1\sigma$ lines, $0.1097 \pm 0.0036$ are shown as dashed lines. 
  Note that these are one parameter $1 \sigma$ errors.
 Table~\ref{table:lrgdr7only} shows excellent agreement
  for this quantity for the LRG-only constraints, with $r_s/D_V(0.35) = 0.1097^{+0.0039}_{-0.0042}$
   for $k_{max} = 0.2 \hompc$.  This agreement reinforces the argument in
Appendix \ref{dvapprox} that our neglect of the model dependence of the
window function does not introduce significant bias in the $D_V(0.35)$
constraint.  Moreover, this constraint does not change
   if we adopt very weak constraints on the nuisance function, $|F_{nuis}(0.1 \hompc)|/b_0^2 < 0.2$ and $|F_{nuis}(0.2 \hompc)|/b_0^2 < 0.5$, 
   or use the extreme central galaxy velocity dispersion model instead.  
   We show in Appendix \ref{nuiseffects} that the largest known source of systematic 
   uncertainty, the central galaxy velocity dispersion, impacts the cosmological parameter constraints at well below the statistical errors, and can be safely neglected for this analysis.  We also demonstrate that our results are robust to the treatment of the nuisance parameters $a_1$ and $a_2$.
   
We estimate the significance of the detection of the BAO
feature as the difference between the best-fitting $\chi^2$ for the
fiducial and no wiggles models when $a_1, a_2$, and $b_o^2$ are
chosen to minimize $\chi^2$; we find $\Delta \chi^2_{BAO} = 8.9$.
The resulting constraint on $r_s/D_V(0.35)$ is much tighter than is available
from the shape information alone.  To see this result, in Table~\ref{table:lrgdr7only}
we combine the LRG $\hat{P}_{halo}(k)$ likelihood with a WMAP5 prior
on $\Omega_m h^2$.  The constraint from the shape alone, obtained by
fitting the no wiggles model, gives a constraint on $r_s/D_V$ that is
consistent with the constraint from the model including BAOs, 
but with a factor of $\sim 2.3$ larger errors.
Finally, we note that P09 estimate the total BAO detection significance
to be $\Delta \chi^2 = 13.1$; it is substantially larger than the value we find
due to the inclusion of lower redshift galaxies
from both the SDSS main sample and 2dFGRS.

Finally, Table~\ref{table:lrgdr7only} also reports our constraint on $A_{0.35}$ \citep{eisenstein/etal:2005}:
\begin{equation}
A_{0.35} \equiv \sqrt{\Omega_m H_0^2} \frac{D_V(0.35)}{0.35c}.
\end{equation}
This parameter is tightly constrained by the $\hat{P}_{halo}(k)$ measurement and is independent of $H_0$.

\subsection{Dependence of LRG-only constraints on the cosmological model}
\label{lrgonlycosmodep}
\begin{figure}
  \centering
  \resizebox{0.4\textwidth}{!}{\includegraphics{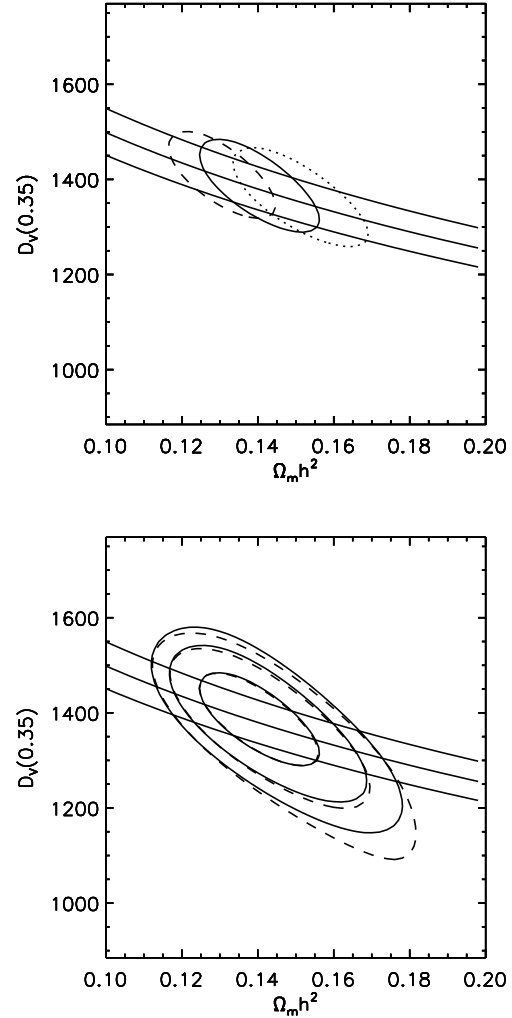}}
  \caption{\label{fig:cosmodepend} {\em Upper panel}: Change in $\Delta \chi^2 = 2.3$ contour as $n_s$ is varied, with all other parameters as in Fig.~\ref{fig:dr7onlyonepanel}.  $n_s = 1.02$ (dashed), $n_s = 0.96$ (solid), and $n_s = 0.90$ (dotted).  The degeneracy is well-described as $\Omega_m h^2 (n_s/0.96)^{1.2} = 0.14$.  {\em Lower panel}: The impact of assuming a $\Lambda$CDM relation between $\Omega_m$, $h$, and $D_V$ (solid contours) compared with applying the non-linear corrections at $h=0.7$ and assuming no relation between $\Omega_m$, $h$, and $D_V(0.35)$ (dashed contours).  As in Fig~\ref{fig:dr7onlyonepanel} the lines show the constraints for constant $r_s/D_V(0.35)$ from P09.}
\end{figure}

In Section \ref{DR7only} the cosmological parameters 
$\Omega_b h^2$, $n_s$, and $\sigma_8$ were fixed at their WMAP5 recommended values.
For our purposes, $r_s$ changes negligibly as a function of $\Omega_b h^2$ since this parameter is so tightly constrained by CMB data.
The parameters $\Omega_m h^2$ and $n_s$ both affect the linear power spectrum
and are degenerate in shifting the contours along  the constant $r_s/D_V$ direction, 
as illustrated in the
upper panel of Figure~\ref{fig:cosmodepend}.  This degeneracy is well described
as $\Omega_m h^2 (n_s/0.96)^{1.2} = 0.141$.

In Figure~\ref{fig:dr7onlyonepanel}, we have 
assumed the $\Lambda$CDM relation between $\Omega_m$, $h$, and $D_V$.
This determines the scale at which to apply the non-linear corrections,
which are at fixed $k$ values in units of $h \; {\rm Mpc}^{-1}$.  In
the bottom panel of Figure~\ref{fig:cosmodepend} we show that this assumption
is not restrictive.  The dashed curve fixes $h=0.7$ and assumes no 
relation between $h$ and $D_V$, which also depends on $\Omega_k$ and $w$.
Varying $\sigma_8$ by $\pm 0.1$, which enters the {\sc halofit} calculation
of the smooth component of the non-linear matter power spectrum 
in Eqn.~\ref{eq:halofit}, changes the contours in 
Figure~\ref{fig:dr7onlyonepanel} negligibly.

\subsection{Combined constraints with WMAP5 and Union SN}
\label{cosmowithwmap}
\begin{figure}
  \centering
  \resizebox{0.9\columnwidth}{!}{\includegraphics{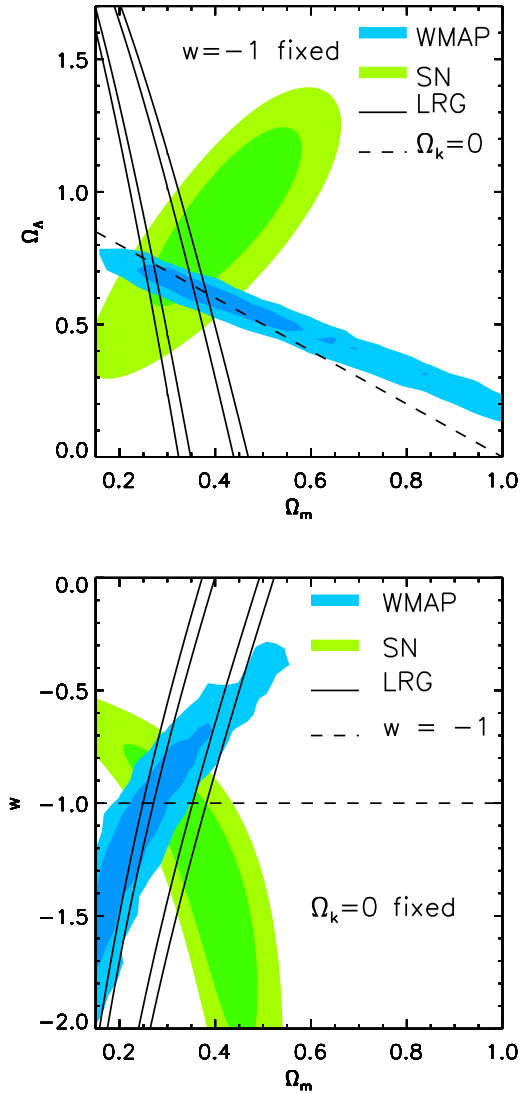}}
  \caption{\label{fig:prettypic} WMAP5, Union supernova sample, and the LRG $\hat{P}_{halo}(k)$ $A_{0.35}$ constraint on the geometry of the universe.  {\em Upper panel:} curvature varies and $w=-1$ is fixed.  The dashed line shows a flat universe, $\Omega_m + \Omega_\Lambda = 1$.  {\em Lower panel:} $w$ varies (assumed independent of redshift), and a flat universe is assumed.  The dashed line indicates a cosmological constant, $w=-1$.  WMAP5 and Union supernova contours are MCMC results, while for $\hat{P}_{halo}(k)$, we approximate $\Delta \chi^2 = 2.3$ and $\Delta \chi^2 = 6$ contours by showing $A_{0.35} \pm \sqrt{2.3} \sigma_{A_{0.35}}$ and $A_{0.35} \pm \sqrt{6.0} \sigma_{A_{0.35}}$ from the constraints in the top row of Table~\ref{table:lrgdr7only}.}
\end{figure}
As probes of the redshift-distance relation,
the three cosmological datasets we use in this Section are highly
complementary for constraining the geometry of the universe and 
the equation of state of dark energy:
WMAP5 effectively constrains the
distance to the surface of last scatter and $\Omega_m h^2$, supernova
data constrains angular diameter distance ratios up to $z \sim 1$,
and $\hat{P}_{halo}(k)$ sets joint constraints on $r_s/D_V(0.35)$ and $\Omega_m h^2 (n_s/0.96)$.  In
Fig.~\ref{fig:prettypic} we show the intersection of these constraints for
two models assuming a power law primordial power spectrum and no  
massive neutrinos.  The blue bands indicate the WMAP5 constraints and the green
bands show constraints using the Union Supernova Sample \citep{kowalski/etal:2008}.
For the $\hat{P}_{halo}(k)$, we show the constraint on $A_{0.35}$ (open bands), which has assumed $n_s = 0.96$ and $\Omega_b h^2 = 0.02265$, and is independent of $H_0$.  In the upper panel, we have assumed $w=-1$ and allow
curvature to vary.  The three independent constraints intersect near $\Omega_m = 0.3$ and a flat 
universe (dashed line).  In the lower panel, we assume flatness but allow $w$ to vary; again the contours intersect near $\Omega_m = 0.3$ and $w=-1$, a cosmological constant.

In this Section we combine these probes using the Markov Chain
Monte Carlo (MCMC) method to obtain constraints on four 
cosmological models: a flat universe with a cosmological constant 
($\Lambda$CDM), a $\Lambda$CDM universe with curvature 
(o$\Lambda$CDM), a flat universe with a dark energy component
 with constant equation of state w (wCDM), and a wCDM 
 universe with curvature (owCDM).  In each model we combine 
 the constraints from $\hat{P}_{halo}(k)$ with the WMAP5
 results \citep{dunkley/etal:2009}.  In the last model, we also
 present constraints in combination with both WMAP5 and 
 the Union Supernova Sample \citep{kowalski/etal:2008}.
 Marginalized one-dimensional parameter constraints are 
 presented in Table~\ref{table:lrgowcdm}.

\begin{figure*}
  \centering
  \resizebox{0.7\textwidth}{!}
 {\includegraphics{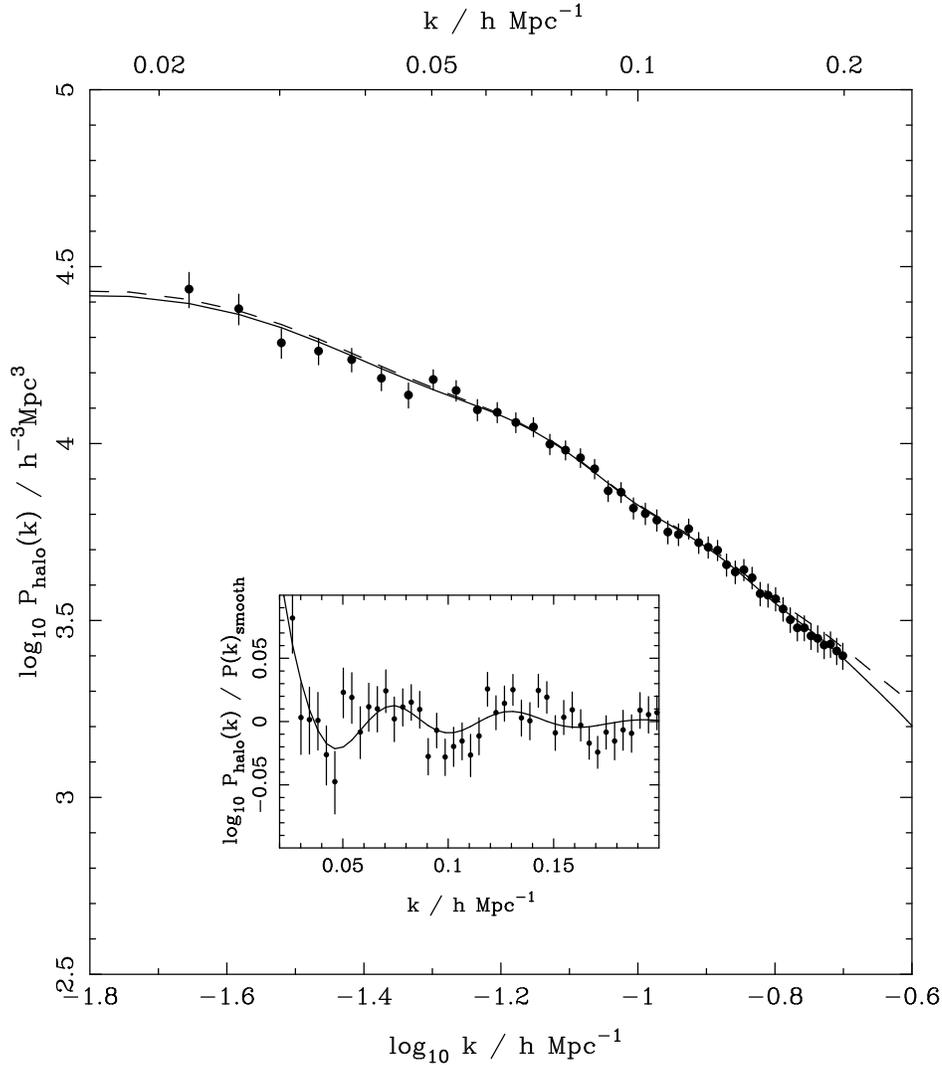}}
  \caption{\label{fig:obstofid}  Points with errors show our measurement
   of $\hat{P}_{halo}(k)$.  We show $\sqrt{C_{ii}}$ as error bars; recall
    that the points are positively correlated.  We plot the best-fitting WMAP5$+$LRG
    $\Lambda$CDM model ($\Omega_m,
\Omega_b, \Omega_{\Lambda}, n_s, \sigma_8, h)$ = 
(0.291, 0.0474, 0.709, 0.960, 0.820, 0.690) with best-fitting nuisance parameters
     $a_1 = 0.172$ and $a_2 = -0.198$ (solid curve), for which $\chi^2 = 40.0$; 
     the dashed line shows the same model but with $a_1 = a_2 = 0$, 
     for which $\chi^2 = 43.3$.  The BAO inset shows the same data
      and model divided by a spline fit to the smooth component, 
      $P_{smooth}$, as in Fig. 4 of P09.  In Section \ref{DR7only} we find the significance of the BAO detection in the $\hat{P}_{halo}(k)$ measurement is $\Delta \chi^2 = 8.9$.}
\end{figure*}

The best-fitting $\Lambda$CDM fit to the WMAP5$+$LRG
likelihoods is ($\Omega_m, \Omega_b, \Omega_{\Lambda},
 n_s, \sigma_8, h)$ = (0.291, 0.0474, 0.709, 0.960, 0.820, 0.690)
  with best-fitting nuisance parameters 
  $a_1 = 0.172$ and $a_2 = -0.198$.  This model has 
  $\chi^2_{LRG} = 40.0$ when fitting to 45 bandpowers,
  and is shown with the data in Fig.~\ref{fig:obstofid}.
  In this model adding the information from $\hat{P}_{halo}$
  breaks the partial degeneracy between $\Omega_m$ and $H_0$
  in the WMAP5 data and reduces the uncertainties in each 
  by a factor of $\sim 1.6$ compared to WMAP5 alone: $\Omega_m = 0.289 \pm 0.019$
  and $H_0 = 69.4 \pm 1.6$ km s$^{-1}$ Mpc$^{-1}$ ($\Omega_m = 0.258 \pm 0.03$
  and $H_0 = 71.9^{+2.6}_{-2.7}$ km s$^{-1}$ Mpc$^{-1}$ for WMAP5).  The constraint on $\sigma_8$
  also tightens by 30\% because of the $\sigma_8-\Omega_m h^2$
  partial degeneracy in the WMAP5 data.  Note that since we marginalize
  over the galaxy bias, we have no constraint on $\sigma_8$ directly from the LRGs.
  
\begin{figure*}
  \centering
  \resizebox{0.7\textwidth}{!}{\includegraphics{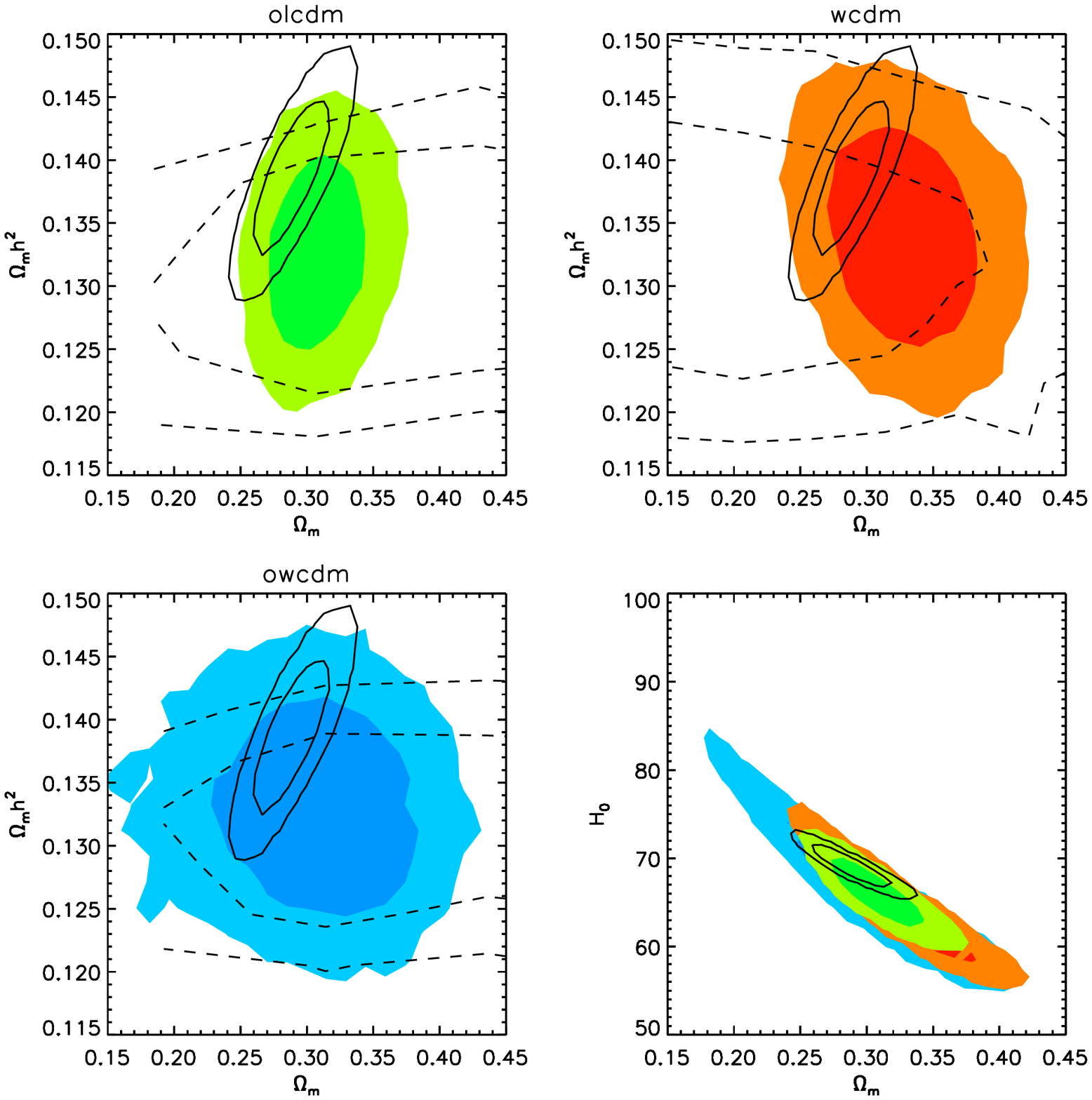}}
  \caption{\label{fig:lrgwmapgrid} WMAP5$+$LRG constraints on
   $\Omega_m h^2$, $\Omega_m$, and $H_0$ for $\Lambda$CDM 
   (solid black contours), o$\Lambda$CDM (shaded green contours), 
   wCDM (shaded red contours), and owCDM (shaded blue contours) models.  
   The first three panels show WMAP5-only constraints (dashed contours)
   and WMAP5$+$LRG constraints (colored contours) in the $\Omega_m h^2$-
   $\Omega_m$ plane as the model is varied.  In the lower right we 
   show all constraints from WMAP5$+$LRG for all four models
   in the $\Omega_m-h$ plane, which lie within the tight
   $\Omega_m h^2 \approx 0.133$ WMAP5-only constraints.}
\end{figure*}

In Fig.~\ref{fig:lrgwmapgrid} we show the effect of opening the 
cosmological parameter space to include curvature and a constant
dark energy equation of state $w$.  Solid contours show the 
$\Lambda$CDM constraint in each panel for comparison.
The dashes show WMAP5-only constraints.  Without the $\Lambda$CDM
assumption, WMAP5 cannot constrain $\Omega_m$ and $H_0$ 
separately from $\Omega_m h^2$.  In each of these models,
the inclusion of the $\hat{P}_{halo}(k)$ information can break the degeneracy
through the BAO constraint on $r_s/D_V$.  Table~\ref{table:lrgowcdm} 
shows that the cold dark matter density, $\Omega_c h^2$, constraint improves by $\sim 15\%$ compared to the
WMAP5-only constraint ($\sim \pm 0.0063$) due to the power spectrum
shape information in the non-$\Lambda$CDM models.  
Moreover, the $r_s/D_V(0.35)$ constraint does not
deviate substantially from the $\hat{P}_{halo}(k)$+$\Omega_m h^2$ prior
constraint presented in Table~\ref{table:lrgdr7only}.  In the context
of power-law initial conditions, $\hat{P}_{halo}(k)$ information
does not improve constraints on the spectral index $n_s$.

Allowing curvature relaxes the constraints on $\Omega_m$ and $H_0$ 
to the WMAP5-only $\Lambda$CDM errors on these parameters, while
tightly constraining $\Omega_{\rm tot} = 1 - \Omega_k$ to $1.0114_{-0.0076}^{+0.0077}$ 
($-0.027 < \Omega_k < 0.003$ with 95\% confidence).
If instead we assume flatness but allow the dark energy equation of state
as an additional parameter $w$ (assumed constant), $w$ is constrained to
$-0.79 \pm 0.15$.  Since the effective LRG sample redshift is
$z_{eff} = 0.313$, allowing $w$ to deviate from $-1$ significantly degrades the $z=0$
constraints, $\Omega_m$ and $H_0$.

\begin{figure}
  \centering
  \resizebox{0.9\columnwidth}{!}{\includegraphics{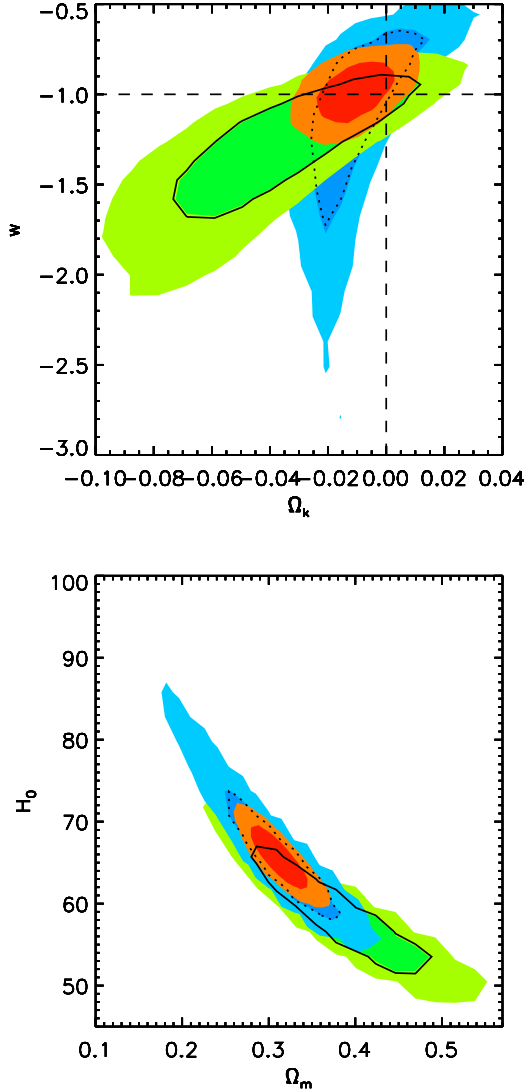}}
  \caption{\label{fig:omkwSN} For the owCDM model we compare the
  constraints from WMAP5$+$LRG (blue contours), WMAP5$+$SN
  (green contours), and WMAP5$+$LRG+$SN$ (red contours).  In the upper panel the vertical line indicates a flat universe ($\Omega_k = 0$), and the horizontal line indicates a cosmological constant ($w=-1$).  In the both panels we overplot the WMAP5$+$SN 68\% contour (solid black) and WMAP5$+$LRG (dotted black) for ease of comparison.}
\end{figure}

When both $\Omega_k$ and $w$ vary, there remains a large degeneracy
between $\Omega_m$, $H_0$, and $w$.  Curvature is still tightly constrained
and consistent with flatness at the percent level: 
$\Omega_{\rm tot} = 1.009 \pm 0.012$.  Figure~\ref{fig:omkwSN}
demonstrates that supernovae can break the degeneracy in this model.  The combination
of all three data sets simultaneously constrains $\Omega_k$ within 0.009 and
$w$ to 11\%, while still improving constraints on $\Omega_m$ and $H_0$ compared
with WMAP5 alone in the $\Lambda$CDM model.  Allowing $\Omega_k \neq 0$ and/or $w \neq -1$
all act to increase $\Omega_m$ and decrease $H_0$ compared with the $\Lambda$CDM
model.  The upper panel of Figure~\ref{fig:omkwSN} shows that
the $\Lambda$CDM model is only $\sim 1\sigma$ away from the best fit.  
The full set of constraints on all parameters is reported in Table ~\ref{table:lrgowcdm}.

\begin{table*}
\begin{center}
\begin{tabular}{llllll}
  parameter & $\Lambda$CDM & o$\Lambda$CDM & wCDM & 
  owCDM & owCDM$+$SN\\
  \hline
$\Omega_m$ & 0.289 $\pm$ 0.019 & 0.309 $\pm$ 0.025 & 0.328 $\pm$ 0.037 & 0.306 $\pm$ 0.050 & 0.312 $\pm$ 0.022  \\
$H_0$ & 69.4 $\pm$ 1.6 & 66.0 $\pm$ 2.7 & 64.3 $\pm$ 4.1 & $66.7^{+5.9}_{-5.6}$ & 65.6 $\pm$ 2.5  \\
$D_V(0.35)$& 1349 $\pm$ 23 & 1415 $\pm$ 49 & 1398 $\pm$ 45 & 1424 $\pm$ 49 & 1418 $\pm$ 49 \\
$r_s/D_V(0.35)$& 0.1125 $\pm$ 0.0023 & 0.1084 $\pm$ 0.0034 & 0.1094 $\pm$ 0.0032 & $0.1078^{+0.0033}_{-0.0034}$ & 0.1081 $\pm$ 0.0034 \\
$\Omega_k$ &  - & $-0.0114^{+0.0076}_{-0.0077}$ &  - & -0.009 $\pm$ 0.012 & -0.0109 $\pm$ 0.0088  \\
$w$ &  - &  - & -0.79 $\pm$ 0.15 & -1.06 $\pm$ 0.38 & -0.99 $\pm$ 0.11  \\
$\Omega_\Lambda$ & 0.711 $\pm$ 0.019 & 0.703 $\pm$ 0.021 & 0.672 $\pm$ 0.037 & $0.703^{+0.057}_{-0.058}$ & 0.699 $\pm$ 0.020  \\
Age (Gyr) & 13.73 $\pm$ 0.13 & 14.25 $\pm$ 0.37 & 13.87 $\pm$ 0.17 & 14.27 $\pm$ 0.52 & 14.24 $\pm$ 0.40  \\
$\Omega_{\rm tot}$ & - & $1.0114_{-0.0076}^{+0.0077}$ & - & 1.009 $\pm$ 0.012 & 1.0109 $\pm$ 0.0088 \\
$100 \Omega_b h^2$ & 2.272 $\pm$ 0.058 & 2.274 $\pm$ 0.059 & $2.293^{+0.062}_{-0.063}$ & $2.279^{+0.066}_{-0.065}$ & $2.276^{+0.060}_{-0.059}$  \\
$\Omega_c h^2$ & $0.1161^{+0.0039}_{-0.0038}$ & 0.1110 $\pm$ 0.0052 & $0.1112^{+0.0056}_{-0.0057}$ & $0.1103^{+0.0055}_{-0.0054}$ & $0.1110^{+0.0051}_{-0.0052}$  \\
$\tau$ & 0.084 $\pm$ 0.016 & 0.089 $\pm$ 0.017 & 0.088 $\pm$ 0.017 & 0.088 $\pm$ 0.017 & 0.088 $\pm$ 0.017  \\
$n_s$ & 0.961 $\pm$ 0.013 & 0.962 $\pm$ 0.014 & 0.969 $\pm$ 0.015 & 0.965 $\pm$ 0.016 & 0.964 $\pm$ 0.014  \\
$\ln(10^{10} A_{05})$ & $3.080^{+0.036}_{-0.037}$ & 3.068 $\pm$ 0.040 & $3.071^{+0.040}_{-0.039}$ & 3.064 $\pm$ 0.041 & 3.068 $\pm$ 0.039  \\
$\sigma_8$ & 0.824 $\pm$ 0.025 & 0.796 $\pm$ 0.032 & 0.735 $\pm$ 0.073 & 0.79 $\pm$ 0.11 & $0.790^{+0.045}_{-0.046}$  \\
\end{tabular}
  \caption{\label{table:lrgowcdm} Marginalized
    one-dimensional constraints (68\%) for WMAP5$+$LRG for flat $\Lambda$CDM, $\Lambda$CDM with curvature (o$\Lambda$CDM), flat wCDM (wCDM), wCDM with curvature (owCDM), and wCDM with curvature and including constraints from the Union Supernova sample.  Here $\tau$ is the optical depth to reionization, $n_s$ is the scalar spectral index, and $A_{05}$ is the amplitude of curvature perturbations at $k = 0.05/$Mpc; these parameters are constrained directly by the CMB only.}
\end{center}
\end{table*}

\subsection{Additional constraints from the broad $\hat{P}_{halo}(k)$ shape}
\label{shapepayoff}
\begin{figure}
  \centering
  \resizebox{0.9\columnwidth}{!}{\includegraphics{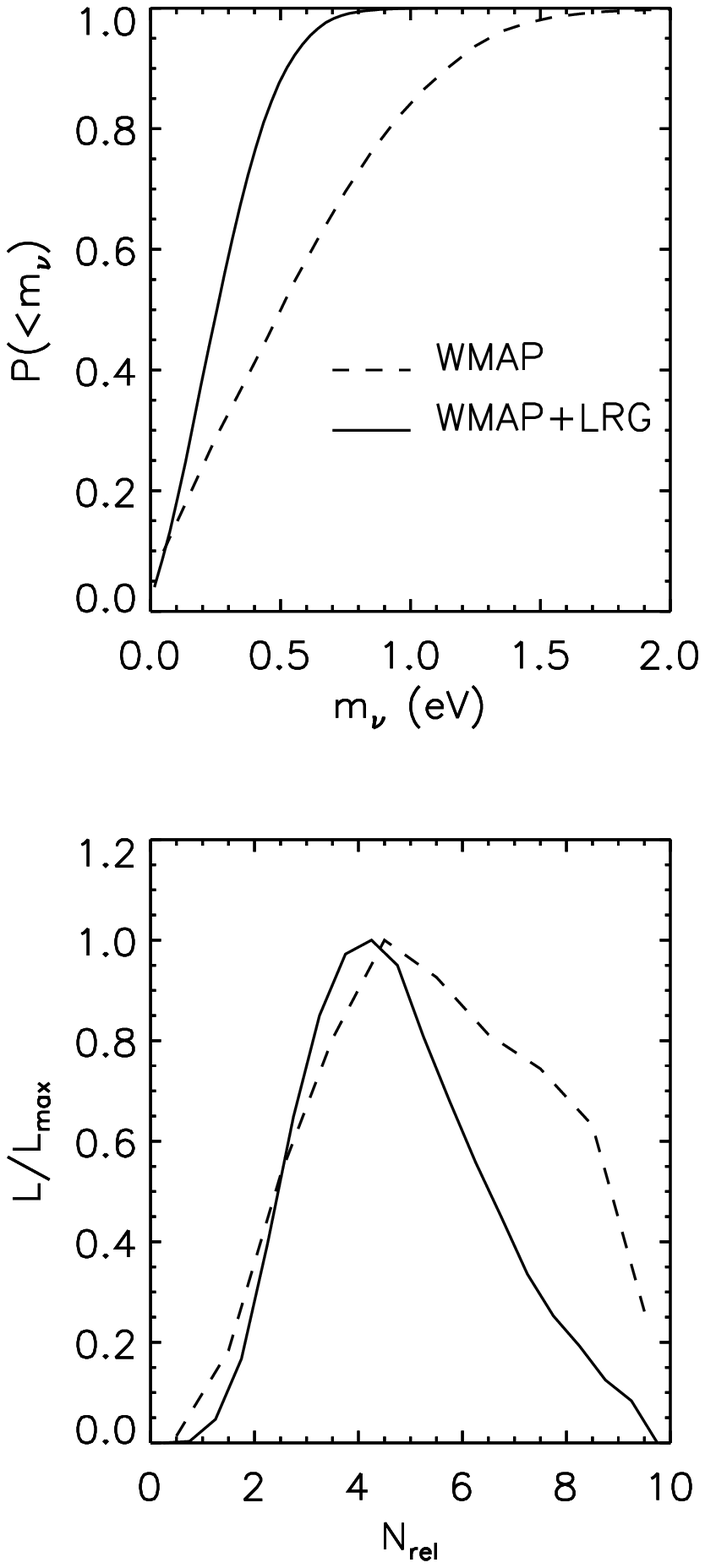}}
  \caption{\label{fig:shapeneutrinos} $\hat{P}_{halo}(k)$ improves constraints on neutrinos in the $\Lambda$CDM model through both the BAO scale and the broad power spectrum shape constraints.  We show the one dimensional cumulative probability for WMAP alone (dashed) and WMAP+$\hat{P}_{halo}(k)$ (solid) for the neutrino mass
(upper panel) and the one dimensional likelihood for the effective number of relativistic species $N_{eff}$ (lower panel).}
\end{figure}

For the models considered thus far, 
we have shown that gains in cosmological parameter
constraints from adding constraints on the broad shape
of $\hat{P}_{halo}(k)$ to WMAP5 results are moderate: 
$\sim 15\%$ improvement in $\Omega_c h^2$ for all the models
considered in Table~\ref{table:lrgowcdm}.
On the other hand, when the constraints on 
$\Omega_b h^2$ and $\Omega_c h^2$ from WMAP5
are used, our constraint on the BAO scale 
provides a much more precise determination
of $D_V$ at the effective redshift of the survey than the shape information alone.

In more extended models than we have thus far considered, we may
expect the additional shape information to allow tighter constraints.   
The cosmological  parameters most closely constrained by the broad 
$P(k)$ shape are those which affect the shape directly or which affect 
parameters degenerate with the shape: these are expected to be 
the power spectrum spectral slope $n_s$, its running $dn_s/d\ln k$, 
neutrino mass $m_{\nu}$, and the number of relativistic species $N_{eff}$.  
Thus far in our analysis, we have assumed $dn_s/d\ln k = 0$, $m_{\nu} = 0$, and 
$N_{eff} = 3.04$.

One intuitively expects the measurement of $\hat{P}_{halo}(k)$ to improve 
constraints on the primordial power spectrum.  In a $\Lambda$CDM
model where both running of the spectral index and tensors are allowed,
WMAP5 still places relatively tight constraints on the primordial power spectrum:
 $n_s = 1.087^{+0.072}_{-0.073}$ and $d\ln n_s/d\ln k = -0.05 \pm 0.03$. 
The measurement reported in this paper probes at most 
$\Delta \ln k \sim 2$ and covers a range corresponding to $\ell \sim 300-3000$;
this range overlaps CMB measurements but extends to smaller scales.  Over this $k$-range and for this model, WMAP5 constrains the $P(k)$ shape to vary by $\sim 8\%$ from variations in the primordial power spectrum. Due to the uncertainties in the relation between the galaxy  and underlying matter density fields, our nuisance parameters alone allow
$P_{halo}(k,{\bf p})$ to vary by up to $10-14\%$ over this region.  Therefore we do not
expect significant gains on $n_s$ or $d\ln n_s/d\ln k$ from our measurement.

The effect of massive neutrinos in the CMB power spectrum is to 
increase the height of the high $\ell$ acoustic peaks: 
free streaming neutrinos smooth out perturbations,
 thus boosting acoustic oscillations. In the matter power spectrum 
 instead, neutrino free streaming gives a scale-dependent suppression
 of power on the scales that large scale structure measurements currently probe \citep{lesgourgues/pastor:2006}.
  This makes these two observables 
 highly complementary in constraining neutrino masses with cosmology.

We start by  comparing the constraints from WMAP5+$\hat{P}_{halo}(k)$
and WMAP5+BAO (using the P09 BAO likelihood) in the
$\Lambda$CDM model with three degenerate massive neutrino species.  
While WMAP5 alone finds $\sum m_{\nu} < 1.3$ eV with 95\% confidence, 
WMAP5+$\hat{P}_{halo}(k)$ yields $\sum m_{\nu} < 0.62$ eV, which is
a significant improvement over $\sum m_{\nu} < 0.73$ eV (WMAP5+BAO).  The
upper panel of Fig.~\ref{fig:shapeneutrinos} compares the likelihood for $m_{\nu}$ 
for WMAP5 data alone (dashed) and in combination with $\hat{P}_{halo}(k)$.

A change in the number of relativistic species in the early universe
 changes the epoch of matter-radiation equality and thus shifts the CMB acoustic peaks. 
The CMB constrains the redshift of matter-radiation equality through the ratio
of the third to first peak heights \citep{komatsu/etal:2009}.  If the effective number of relativistic
species $N_{eff}$ is allowed to vary, this constraint defines a degeneracy between $\Omega_c h^2$
and $N_{eff}$ \citep{dunkley/etal:2009}.  
Note that the physical quantity that is being constrained is the physical energy 
density in relativistic particles. In the standard model this is given by photons 
and neutrinos but $N_{\nu}$ should really be considered an ``effective'' number 
of relativistic neutrino species: $N_{eff}=3.04$ for standard neutrinos. 
Departures from this number can be interpreted also in terms of decay of 
dark matter particles, quintessence, exotic models, and additional 
hypothetical relativistic particles such as a light majoron or 
a sterile neutrino.  

In the $\Lambda$CDM model, which specifies a rigid
relation between the angular diameter distance at last
scattering measured by the CMB and low redshift
distance scales,
the degeneracy between 
$N_{eff}$ and $\Omega_c h^2$ can be broken by a low 
redshift distance constraint such as the BAO.
However, $N_{eff}$ will also impact the matter power spectrum,
which probes the horizon size at matter-radiation equality 
(e.g., \citealt{eisenstein/hu:1998}).
Therefore, $\hat{P}_{halo}(k)$ is an excellent
probe of $N_{eff}$: WMAP5+$\hat{P}_{halo}(k)$
finds $N_{eff} = 4.8^{+1.8}_{-1.7}$, while WMAP5+BAO
yields $N_{eff} \approx 6.0 \pm 2.5$.   For comparison, \citet{komatsu/etal:2009}
find $N_{eff} = 4.4 \pm 1.5$ when combining WMAP, BAO, supernovae,
and the Hubble Space Telescope key project \citep{freedman/etal:2001}.  The
lower panel of Fig.~\ref{fig:shapeneutrinos} compares the likelihood for $N_{eff}$ 
for WMAP5 data alone with a prior $N_{eff} \leq 10$ (dashed) and in combination with $\hat{P}_{halo}(k)$; no prior on $N_{eff}$ is needed in this case.

\section{Comparison with Other Analyses}
\label{compare}
\subsection{Comparison with previous galaxy clustering results}
There have been several previous analyses of the clustering of the SDSS LRG spectroscopic sample.  \citet{eisenstein/etal:2005} use the correlation function of the DR3 SDSS LRG sample to derive constraints on $\Omega_m h^2 = 0.133(n_s/0.96)^{-1.2} \pm 0.011$ and $D_V(0.35) = 1381 \pm 64$ Mpc, where we have adjusted their constraints to match our assumed values of $\Omega_b h^2$ and $n_s$; recall that these constraints are not independent.  Comparison with their Figure 7 indicates that our model is slightly more than $1\sigma$ away from their best fit.  Our analysis prefers larger $\Omega_m h^2$ and lower $r_s/D_V$.  In interpreting this comparison one should consider the differences in modeling and the fact that we have a factor of $\sim 2$ larger volume.  Given this larger volume, naively we would expect an improvement on the constraints by a factor of $\sim \sqrt{2}$.  Comparison with Table~\ref{table:lrgdr7only} shows that our LRG-only constraints on $\Omega_m h^2$ and $D_V$ have approximately the same uncertainty as \citet{eisenstein/etal:2005}.  This is partly because we conservatively increased our covariance matrix by a factor of $1.21$ to account for the non-Gaussianity in the BAO contribution to the likelihood surface (see Section~\ref{nongaussianbao} discussion).  However, this increase will artificially weaken the constraint from the shape.  Marginalization over the two nuisance parameters $a_1$ and $a_2$ to account for our uncertainty in the $P_{halo}(k,{\bf p})$ as well as our conservative cut at $k_{min}$ also slightly weaken the constraint from the power spectrum shape.

\citet{tegmark/etal:2006} report cosmological constraints from a somewhat larger LRG sample (SDSS DR4)  and combine their results with WMAP3 data.  To compare LRG-only constraints, we use the value derived from the \citet{tegmark/etal:2006} power spectrum in \citet{sanchez/cole:2008}: $\Omega_m h = 0.173 \pm 0.017$ for $n_s = 1.0$ and $h=0.72$.  For a $\Lambda$CDM model scaled to $n_s = 1.0$, our LRG-only constraints yield $\Omega_m h = 0.200^{+0.012}_{-0.011}$.  Restricting our analysis to $k_{max} = 0.1 \hompc$ to match \citet{tegmark/etal:2006}, we find $\Omega_m h = 0.195 \pm 0.013$.  Besides the increase in sample volume, the discrepancy between these results could be due to differences in the FOG compression and the degeneracy between their nuisance parameter $Q$ (see Eqn.~\ref{qeqn}) and cosmological parameters.  A detailed comparison of our modeling approaches is given in \citet{reid/spergel/bode:2008}.  
Note that \citet{sanchez/etal:2009} have also recently completed an analysis of the LRG correlation
function, but they do not present a constraint from their shape measurement with which we can compare.

Our results agree with analyses of photometric LRG samples.  \citet{padmanabhan/etal:2007} find $\Omega_m = 0.30 \pm 0.03$ for $h=0.7$ and $n_s = 1$, and \citet{blake/etal:2007} find $\Omega_m h = 0.195 \pm 0.023$ for $h=0.75$ and $n_s = 1$.  Our constraint is also consistent with determinations from other galaxy samples.  For the 2dFGRS sample, \citet{cole/etal:2005} find $\Omega_m h = 0.168 \pm 0.016$ for fixed $n_s = 1.0$ and $h=0.72$; allowing a 10\% Gaussian uncertainty in $h$ yields $\Omega_m h = 0.174 \pm 0.019$, which is within $1\sigma$ of our LRG-only constraint.  Our results are also in good agreement with the SDSS main sample: \citet{tegmark/etal:2004a:la} find $\Omega_m h = 0.201 \pm 0.017$, again with fixed $n_s = 1.0$ and $h=0.72$.

\subsection{Comparison with P09}
\label{compareP09}
The P09 constraints overlap significantly
with our analysis.  We showed in Section \ref{DR7only} that our
LRG-only constraint on $r_s/D_V(0.35)$ is in very good agreement
with the determination in P09.  When
combined with the WMAP5 constraint on $\Omega_m h^2$,
our use of the shape information in $\hat{P}_{halo}(k)$ allows
$\sim 10\%$ improvement on $r_s/D_V(0.35)$.  Moreover,
the shape information
provides a tighter constraint on $\Omega_c h^2$.
  However, the P09 inclusion
of SDSS main and 2dFGRS galaxies allows an additional constraint
on $r_s/D_V(0.2)$, which generally makes the P09 constraints on 
$\Omega_m$ and $H_0$ tighter.  Our constraints on $\Omega_k$ and $w$
are comparable to P09.  Across the models we have studied,
 WMAP5+$\hat{P}_{halo}(k)$ constraints yield lower values
of $H_0$ than the P09 results.  This is driven
by the P09 $r_s/D_V(0.2)$ constraint, which pulls the overall distance scale
slightly lower compared to $r_s/D_V(0.35)$ alone, but does not signal
any inconsistency between these analyses.  

\subsection{Comparison with \citet{riess/etal:2009} $H_0$}
\citet{riess/etal:2009} recently released a new determination of the Hubble
constant using a differential distance ladder: $H_0 = 74.2 \pm 3.6$ km s$^{-1}$ Mpc$^{-1}$.
This value is consistent at the $\sim 1\sigma$ level
 with the WMAP5+$\hat{P}_{halo}(k)$ result 
 for the $\Lambda$CDM model, $H_0 = 69.4 \pm 1.6$ km s$^{-1}$ Mpc$^{-1}$.  
 Table~\ref{table:lrgowcdm} shows that if we allow $\Omega_k \neq 0$
 and/or $w \neq -1$, the mean value of $H_0$ decreases to $\sim 64-67$ km s$^{-1}$ Mpc$^{-1}$.  Therefore,
 combining the \citet{riess/etal:2009} measurement with our constraints should reduce the uncertainties further and push the best-fitting model closer to $\Lambda$CDM.  P09 present constraints including the \citet{riess/etal:2009} $H_0$ constraint for the owCDM model; the impact should be similar when using $\hat{P}_{halo}(k)$ rather than the P09 BAO constraints.

\section{CONCLUSIONS}
\label{conc}
In this paper we have presented the power spectrum of the reconstructed halo density field derived from a sample of Luminous Red Galaxies (LRGs) from the Sloan Digital Sky Survey DR7.
The size of LRG DR7 sample has sufficient statistical power that the details of the relation between LRGs and the underlying linear density field become important and need to be reliably modeled before attempting a cosmological interpretation of the data.
Here, we have adopted the method of \citet{reid/spergel/bode:2008}, which applies a pre-processing step to the measured galaxy density field to reconstruct the halo density field before computing the halo power spectrum. On the scales of interest, this power spectrum has a more direct and robust connection to the underlying linear, real space power spectrum than the power spectrum of the LRG galaxies themselves.

We calibrate our method using $N$-body simulations  with volume and resolution suitably tuned to trace the halo mass range relevant to LRGs, and provide several consistency checks between the observed and mock galaxy density fields to support our approach to model the LRG sample's clustering properties.
In particular,  we demonstrate the validity of our modeling of the small-scale clustering and FOG features by  matching the observed and mock catalogue higher-order statistics probed by the Counts-in-Cylinders group multiplicity function as well as the relative line of sight velocities between galaxies occupying the same halo.  We discuss and quantify  the sources of systematic error remaining in our modeling. For the LRG sample, with $\bar{n} P \sim 1$, both the shot noise subtraction and the large velocity dispersions of their host haloes can introduce uncertainty.  We identify the largest source of systematic uncertainty to be the velocity dispersion of central LRGs within their host haloes, and find its effects on cosmological parameters to be safely smaller than the size of the statistical errors.
We are able to derive quantitative bounds on our
model uncertainties and propagate these through the cosmological analysis by introducing nuisance parameters with tightly controlled allowed ranges, based on our understanding of the sources
of non-linearity in the spectrum.

Based on our  modeling of the LRG sample, we are able to extend our 
 model for $\hat{P}_{halo}(k)$ to $k=0.2 \hompc$, increasing
 the number of available modes by a factor of $\sim 8$ over an
 analysis restricted to $k_{max}=0.1 \hompc$, as was the case
 in the SDSS team's DR4 analysis \citep{tegmark/etal:2006}.  This allows us to simultaneously
 constrain the broadband shape of the underlying linear power spectrum 
 and detect the BAO signal with $\Delta \chi^2 = 8.9$, though most of the 
 shape information is confined to $k < 0.1 \hompc$.

If we fix $n_s$ and 
 $\Omega_b h^2$, $\hat{P}_{halo}(k)$
 alone constrains both $\Omega_m h^2 = 0.141^{+0.010}_{-0.012}$
 and $D_V(0.35) = 1380^{+60}_{-73}$.  
The agreement of our constraint on $\Omega_m h^2$ at $z_{eff} \sim 0.31$ with the one derived from the CMB at $z \sim 1000$ provides a remarkable consistency check for the standard cosmological model.
When $\hat{P}_{halo}(k)$ is combined with WMAP5,
 the error on $\Omega_c h^2$ is reduced by $\sim 15\%$, and the
 constraint on $D_V(0.35)$ allows us to place tight constraints
 on both $\Omega_m$ and $H_0$, as well as $\Omega_k$ or $w$.  If we also include
 the Union Supernova Sample, all four parameters can be tightly constrained:
 $\Omega_m = 0.312 \pm 0.022$, $H_0 = 65.6 \pm 2.5$ km s$^{-1}$ Mpc$^{-1}$, 
 $\Omega_k = -0.0109 \pm 0.008$, and $w = -0.99 \pm 0.11$, which is consistent
 with $\Lambda$CDM at the $\sim 1\sigma$ level.  In fact, in the spirit of Occam's razor,
these constraints can be taken as evidence against both $\Omega_k \neq 0$ and $w \neq -1$, since
their values must conspire to match the observed angular diameter distance at recombination as well
as $D_V(0.35)$; this can be seen from Fig.~\ref{fig:prettypic}.
 
 Finally, we show that the shape information in $\hat{P}_{halo}(k)$
can improve constraints on both massive neutrinos and the
number of relativistic species $N_{eff}$ in a $\Lambda$CDM model.
In combination with WMAP5 we find $\sum m_{\nu} < 0.62$ eV
at the 95\% confidence level and $N_{eff} = 4.8^{+1.8}_{-1.7}$.  
These represent 16\% (30\%) improvements over using the
WMAP5+BAO likelihood from P09.
 
 This paper represents a
first attempt to analyse the LRG redshift survey with a model that
accounts for the non-linear galaxy bias and non-linear redshft space distortions introduced by the so-called one-halo term, and to propagate the uncertainty in the modeling through the cosmological constraints.  We expect that the technique introduced here to estimate the halo density field will
be useful to further refinements such as reconstruction of the baryon acoustic peak \citep{eisenstein/etal:2007} and measurement of $\beta$ from redshift space distortions.  The modeling efforts presented in this paper are rather specific to the SDSS LRG sample.  However, similar techniques to probe the relation between the galaxy and underlying matter density fields as well as to quantify its uncertainty will be required in the analysis of larger data sets from future galaxy surveys.
 
\section*{Acknowledgements}

BAR gratefully acknowledges support from NSF Grant OISE/0530095 and
 FP7-PEOPLE-2007-4-3-IRG
while this work was being completed, and thanks Raul Jimenez for useful discussions and computing power.
WJP is grateful for support from the UK Science and Technology
Facilities Council, the Leverhulme trust and the European Research
Council. 
LV acknowledges support from FP7-PEOPLE-2007-4-3-IRG  n 20218 
and MICINN grant AYA2008- 03531.
DNS acknowledges NSF grants AST-0707731
and OISE05-30095.
DJE was supported by National Science Foundation grant AST-0707225
and NASA grant NNX07AC51G.
Simulated catalogues were calculated and analysed using the
COSMOS Altix 3700 supercomputer, a UK-CCC facility supported by HEFCE
and STFC in cooperation with CGI/Intel.  We acknowledge the use of the Legacy Archive for Microwave Background Data Analysis (LAMBDA).  Support for LAMBDA is provided by the NASA Office of Space Science.
We also acknowledge the use of the CAMB, CMBFAST,
CosmoMC, and WMAP5 likelihood codes.

Funding for the SDSS and SDSS-II has been provided by the Alfred
P. Sloan Foundation, the Participating Institutions, the National
Science Foundation, the U.S. Department of Energy, the National
Aeronautics and Space Administration, the Japanese Monbukagakusho, the
Max Planck Society, and the Higher Education Funding Council for
England. The SDSS Web Site is {\tt http://www.sdss.org/}.

The SDSS is managed by the Astrophysical Research Consortium for the
Participating Institutions. The Participating Institutions are the
American Museum of Natural History, Astrophysical Institute Potsdam,
University of Basel, Cambridge University, Case Western Reserve
University, University of Chicago, Drexel University, Fermilab, the
Institute for Advanced Study, the Japan Participation Group, Johns
Hopkins University, the Joint Institute for Nuclear Astrophysics, the
Kavli Institute for Particle Astrophysics and Cosmology, the Korean
Scientist Group, the Chinese Academy of Sciences (LAMOST), Los Alamos
National Laboratory, the Max-Planck-Institute for Astronomy (MPIA),
the Max-Planck-Institute for Astrophysics (MPA), New Mexico State
University, Ohio State University, University of Pittsburgh,
University of Portsmouth, Princeton University, the United States
Naval Observatory, and the University of Washington.

\setlength{\bibhang}{2.0em}

\appendix
\
\section{Testing Model Approximations}
In this Appendix we present tests to demonstrate the validity of several assumptions of our model $P_{halo}(k,{\bf p})$.
\subsection{Isotropy tests}
\begin{figure}
  \centering
  \resizebox{0.9\columnwidth}{!}
  {\includegraphics{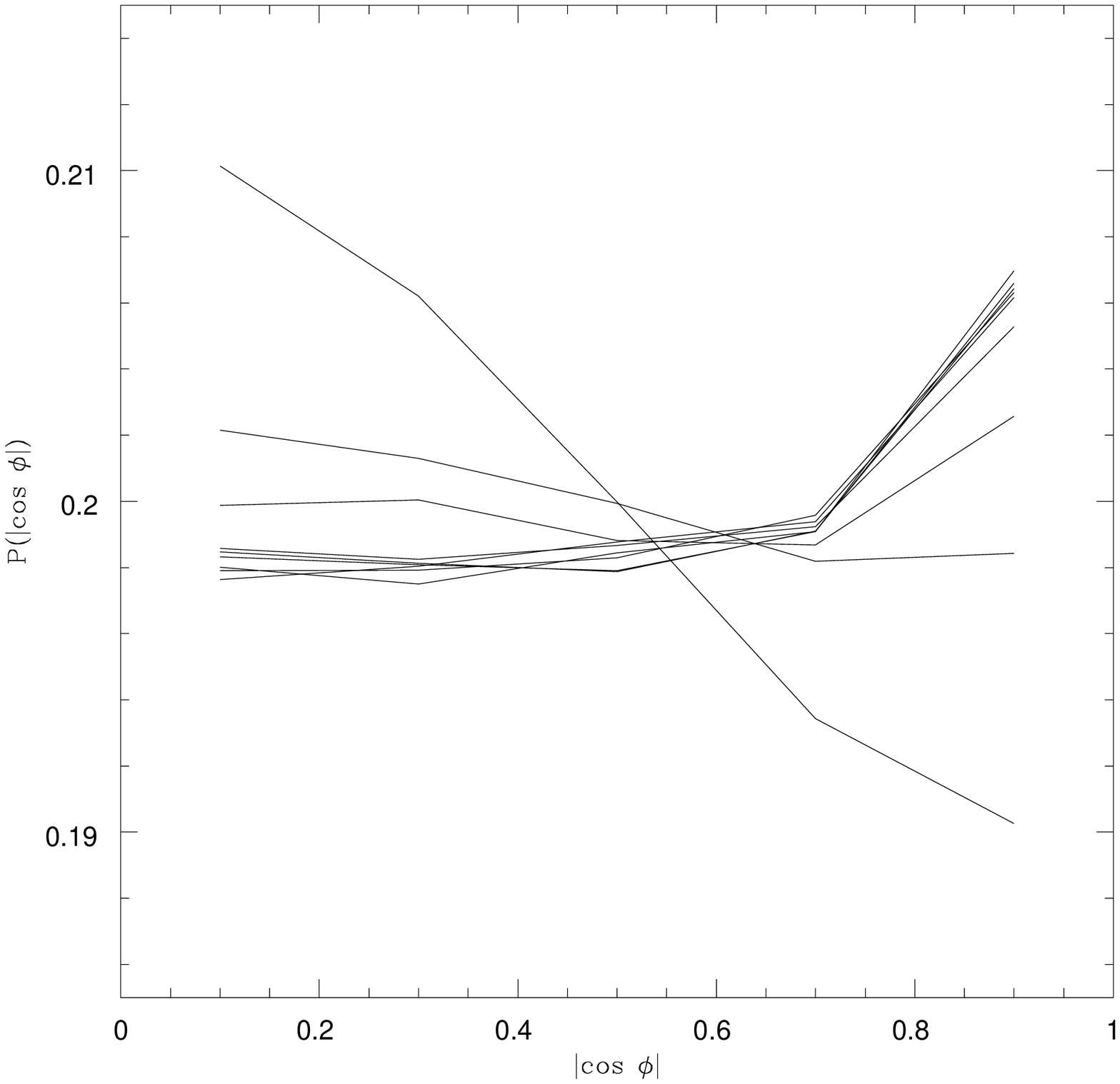}}
  \caption{\label{fig:pairiso2} $P(|\cos \phi|)$ vs $\cos \phi$ where
    $\phi$ is the angle between the galaxy pair separation vector and
    the line of sight defined by the observer and one of the galaxies in the
    pair (see text).  The smallest separation bin (15 $h^{-1}$ Mpc $<
    \Delta r < $ 30 $h^{-1}$ Mpc) shows the largest deviation from
    isotropy, with a $\sim 5\%$ preference for pairs perpendicular to
    the line of sight compared to along the line of sight due to FOGs.  
    The larger separation bins
    extend to 150 $h^{-1}$ Mpc and are nearly isotropic, but with a
    few percent excess of pairs directly along the line of sight.}
\end{figure}

Both our $P_{halo}(k, {\bf p})$ model (Eqn.~\ref{finalmodel}) and the $a_{scl}$ approximation
(Eqn.~\ref{asclcorr}) assume
 that the power spectrum modes are distributed isotropically with
respect to the line of sight.  We check this assumption in the
SDSS DR7 LRG galaxy sample using pairs of galaxies separated by
$\Delta r_{min} =15 \; h^{-1}$ Mpc to $\Delta r_{max} =150 \; h^{-1}$
Mpc, binned into nine equal bins in $\Delta r$ of width 15 $h^{-1}$ Mpc.
We consider the two angles in the triangle defined by the observer and
galaxy pair which give the angle between the galaxy pair separation
vector and the local line of sight vector defined between the observer and one
of the galaxies in the pair.  These two angles will be equal in the limit of a pair with $\Delta r
\ll max(\chi_1, \chi_2)$ where $\chi_1$ and $\chi_2$ are the distances to the two galaxies and $\Delta r$ is the separation between them.  We find
$\left<\cos^2 \phi \right> - 0.333$ is $-0.01$ for the smallest
separation bin (15 $h^{-1}$ Mpc $< \Delta r < $ 30 $h^{-1}$ Mpc) and
$+0.005$ in the largest separation bin.  Figure~\ref{fig:pairiso2}
shows the full distribution versus $|\cos \phi|$.  The small increase for
pairs perpendicular to the line of sight for the smallest separation bin is due to
non-linear redshift space distortions (FOGs), 
inducing a potentially large separation 
in redshift space between nearby pairs of galaxies in real space.
The few percent
deviations from isotropy will induce negligible variations in the
shape of the angle-averaged $P_{halo}(k,{\bf p})$, since the lower left panel of
Figure 7 in \citet{reid/spergel/bode:2008} indicates only a $\sim 5\%$
change to the power spectrum shape between real and redshift space at
$k=0.2 \hompc$.

\subsection{$D_V$ approximation}
\label{dvapprox}
As in Section \ref{modelz}, we use the approximation that pairs of 
galaxies contributing to $\hat{P}_{halo}(k)$ in the $k$-range of interest
are located at the same redshift to compute the effective survey redshift:
\begin{equation}
\label{zeff}
  z_{eff} = \frac{\int z n^2(z) \frac{w^2(z)}{b^2(z)} \frac{dV}{dz} dz}{\int n^2(z) \frac{w^2(z)}{b^2(z)} \frac{dV}{dz} dz}\,,
\end{equation}
where $n(z)$, $b(z)$, and $w(z)$ specify the average number density,
bias, and weight of the sample at redshift $z$ 
as defined in \citet{percival/verde/peacock:2004}.
We find $z_{eff} = 0.313$, and use this redshift to 
evaluate $a_{scl}$ in Eqn.~\ref{asclcorr}.
The effective redshift changes by
only $\Delta z = 0.004$ if one instead weights by the expected number
of galaxies at redshift $z$.  Given the distribution of pairs in the
small separation limit (Eqn.~\ref{pairwgt}) we estimate the fractional
bias remaining after the correction in Eqn.~\ref{asclcorr} is applied
as
\begin{equation}
  \label{dverr}
  \frac{\delta D_{V}}{D_{V}} \approx
  \frac{\int \left(\frac{D_V(z)}{D_V(z_{eff})} 
      \frac{D^{\rm fiducial}_V(z_{eff})}{D^{\rm fiducial}_V(z)}
      - 1\right) n^2(z) \frac{w^2(z)}{b^2(z)} 
    \frac{dV}{dz} dz}{\int n^2(z) \frac{w^2(z)}{b^2(z)} \frac{dV}{dz} dz}
\end{equation}
For a $\Lambda$CDM model,
the fractional bias on the distance scale is $<0.1\%$ in the range $\Omega_m =
0.2 - 0.4$ and the rms change is $<1.2\%$.  
This additional variance
about the peak is negligible for the BAO scale $\sim 100 \; h^{-1}$
Mpc since this is much smaller than the damping scale $\sigma_{BAO}
\sim 9 \; h^{-1}$ Mpc.  We find very similar results for the bias and rms
damping if we instead integrate over the full distribution of
isotropic pairs instead of using the $D_V$ approximation in
Eqn.~\ref{dverr}.

Testing this approximation in more general models is more subtle,
since $D_V(z)$ depends on $H_0$, $\Omega_m$, $\Omega_k$, and $w$.
We instead do a consistency check: for $\Omega_m h^2$ constrained by 
WMAP5, $D_V(z_{eff})$ constrained by WMAP5+$\hat{P}_{halo}(k)$, $\Omega_k = 0$, and
$-2 < w < -0.5$, the maximum fractional bias is $\sim 0.5\%$, and the maximum
rms change is 3.5\%; a similar analysis for $-0.025 < \Omega_k < 0.025$ and $w=-1$
 shows much smaller deviations.  We therefore conclude that in the range of models
considered here, a single scale factor $a_{scl}$ can accurately account for the effects of 
the model redshift distance relation on the interpretation of the measured power spectrum.

\subsection{Comparing $P_{nw}$ approximations}
\label{pnwapprox}
\begin{figure}
  \centering
  \resizebox{0.9\columnwidth}{!}
  {\includegraphics{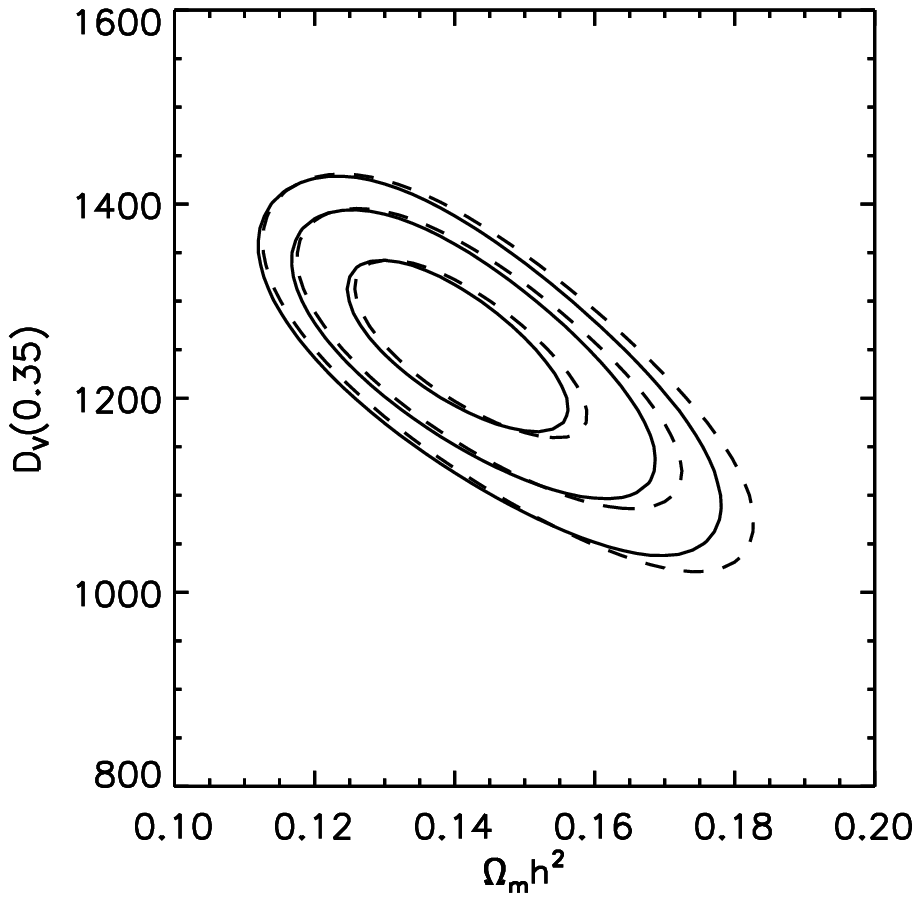}}
  \caption{\label{fig:pnwcompare} Comparison of the LRG-only
  likelihood surface computed with the analytic approximation 
  of $P_{\rm nw}$ in Eqn. 29 of \citet{eisenstein/hu:1998} (solid, 
  as in Fig.~\ref{fig:dr7onlyonepanel}) compared
  with the result when using the b-spline fit described in Appendix \ref{pnwapprox} (dashed).}
\end{figure}

In the models without massive neutrinos, we have used the 
\citet{eisenstein/hu:1998} formula (Eqn. 29) to compute $P_{nw}$,
which enters our model in Eqn.~\ref{eq:Pdamp}.  However, for more 
general models such as those containing massive neutrinos or 
which vary the number of relativistic species, it
is more convenient to use a spline to obtain a smooth version of
$P_{\rm lin}$ without BAO features.  We fit a cubic b-spline to 
$P_{\rm lin} k^{1.5}$ in order to minimize the slope in the $k$ region
of interest.  There are eight equally spaced nodes starting at 
$k=0.0175$ Mpc$^{-1}$
and ending at $k=0.262$ Mpc$^{-1}$, and an additional node at
$k=0.0007$ Mpc$^{-1}$.  Note we fix the location of the nodes in
units of Mpc$^{-1}$ since the linear power spectrum is fixed in those
units for fixed $\Omega_m h^2$ and $\Omega_b h^2$.
Fig.~\ref{fig:pnwcompare} shows that the LRG-only likelihood
surfaces computed with these two approximations agree well in
the region preferred by WMAP5: $\Omega_m h^2 = 0.133 \pm 0.0063$.
\section{QUANTIFYING MODEL UNCERTAINTIES AND CHECKS FOR SYSTEMATICS: DETAILS}
\label{quantnuisdetails}
In this Appendix we aim to quantify the sources of systematic uncertainty in the model $P_{halo}(k)$.  The model is calibrated on the mock catalogues of \citet{reid/spergel/bode:2008}.  In Appendix \ref{haloparam} and Appendix \ref{galaxiesinhaloes} we present the detailed assumptions we have made to produce the mock catalogues from the $N$-body simulation halo catalogues, and discuss the expected impact of these assumptions on the predicted relation between the reconstructed halo and matter density fields.  Appendix \ref{cicstats} through \ref{wmap3cats} present consistency checks between the observed and mock catalogue LRG density fields that address the modeling uncertainties.  In Section \ref{getmaxnuisvals}, the results of these tests are used to establish quantitative bounds on the nuisance parameters in Eqn.~\ref{eq:nuis} to be used in our cosmological parameter analysis.
\subsection{Halo model parametrization}
\label{haloparam}
In \citet{reid/spergel/bode:2008} we adopt the following
parametrization for the average number of LRGs in a halo of mass $M$
\citep{zheng/etal:2005}:
\begin{eqnarray}
  \left<N(M)\right> 
  = \left<N_{cen}\right>(1 + \left<N_{sat}\right>) \label{censatsum}\\
  \left<N_{cen}\right> = \frac{1}{2} \left[ 1 + {\rm erf} 
    \left ( \frac{\log_{10} M - \log_{10} M_{min}}{\sigma_{log M}}\right)
  \right]\label{NcenM}\\
  \left<N_{sat}\right> 
  = \left(\frac{M - M_{cut}}{M_1}\right)^{\alpha}\label{NsatM}.
\end{eqnarray}
For our adopted fiducial cosmological model, we find $\sigma_{log M} \sim
0.6-0.9$ in order to match the amplitude of the observed large scale
clustering of the LRGs; the exact parameter values used to generate the mock catalogues
are given in \citet{reid/spergel/bode:2008}.
  Since the scale dependence of halo bias varies
with halo mass at the $\sim 10\%$ level at $k=0.15 \hompc$
\citep{smith/scoccimarro/sheth:2007}, changes in the distribution of
LRGs with halo mass that preserve the large scale clustering amplitude
could result in few percent changes in non-linear bias of the haloes
traced by the LRGs.  Changes in the distribution of halo biases traced
by the LRGs could also alter the relation between the CiC and true
group multiplicity function, which would introduce further uncertainty
in the relation between the reconstructed and underlying halo density
fields.

\subsection{Distribution of mock galaxies within haloes}
\label{galaxiesinhaloes}
In the mock catalogues of \citet{reid/spergel/bode:2008} used to calibrate our
model $P_{halo}(k, {\bf p})$, we have assumed a sharp distinction between
`central' and `satellite' galaxies.  The first or `central' LRG in
each halo is assumed to sit at the halo centre and move with the mean
velocity of the halo dark matter; roughly 94\% of the LRGs in our
sample are central galaxies \citep{zheng/etal:2008,reid/spergel/bode:2008}.  For the $\sim 6\%$ of LRGs that are
`satellites', we assume that they trace the phase space distribution
of the halo dark matter, so that their positions and velocities are
assigned to be that of a random dark matter particle in the halo.

We do not evaluate the impact of errors in our assumed real space
distribution of galaxies in their haloes on the fidelity of the halo density field
reconstruction; the impact will be negligible in the case where
there is a single LRG per halo.  However, if the observed galaxies
have a significantly different real space distribution in their haloes than we have
assumed, the relationship between the reconstructed halo density
field and underlying matter density field will be different in the observed
and mock galaxy catalogues.  We test our
assumed spatial distribution in Appendix \ref{cicstats} by checking for consistency between
the observed and mock catalogues for CiC group multiplicity functions,
measured with two distinct sets of cylinder parameters.  
Furthermore, we can use Eqn.~\ref{onehaloterm} (where the measured
CiC group multiplicity function specifies $\left<N_{gal} (N_{gal} - 1)\right>$) as
an upper limit on the error on the shot noise term due to differences
between the model and observed reconstructed halo density fields.

\begin{figure}
  \centering
  \resizebox{0.9\columnwidth}{!}{\includegraphics{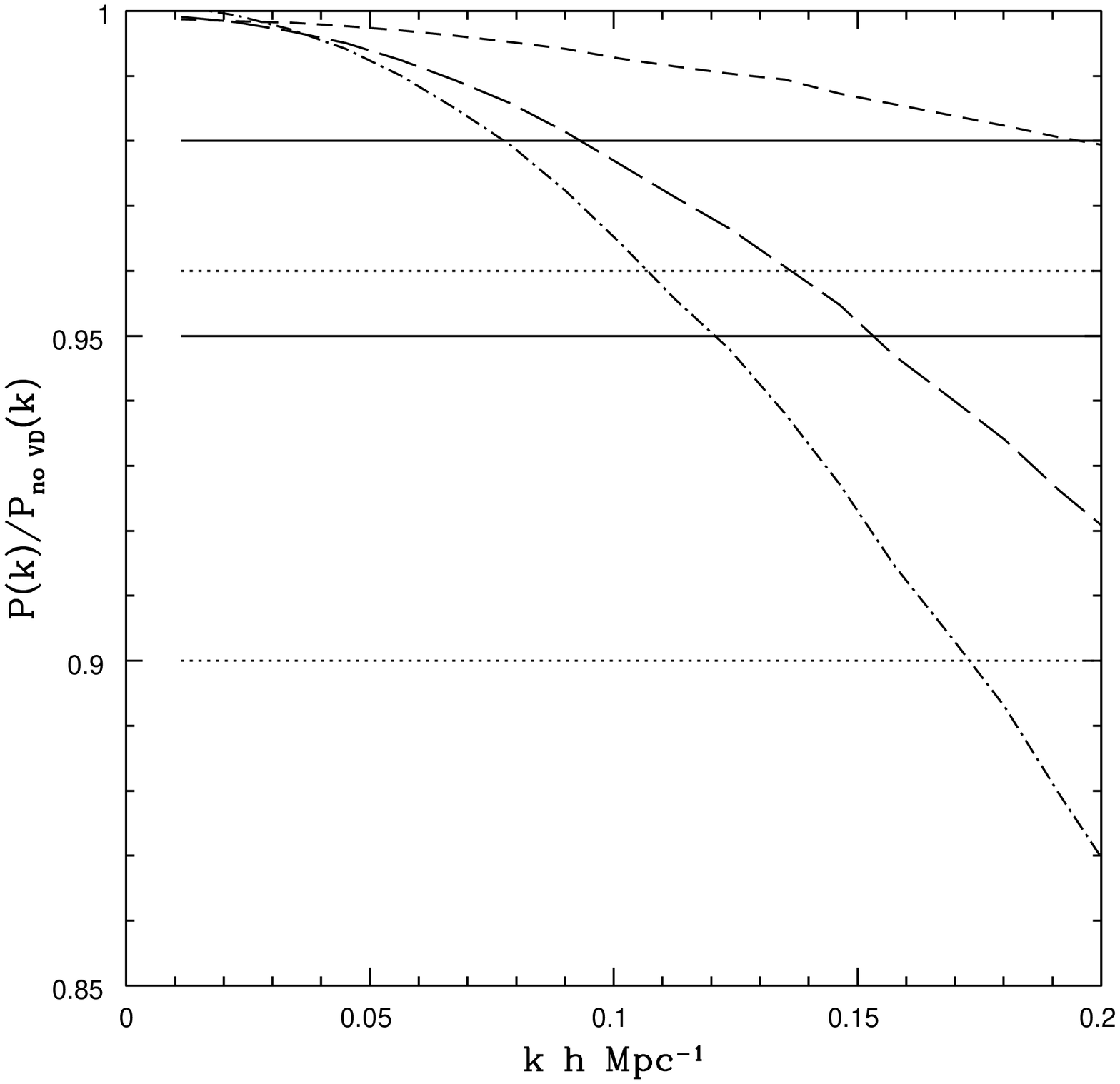}}
  \caption{\label{fig:veldispPk} We compare three models including 
  central galaxy velocity dispersion to our fiducial model with no central galaxy velocity
  dispersion ($f_{cen,err} = b_{vel}
    = 0$) by showing the ratio of $P_{halo}(k, {\bf p_{fid}})$ for the models.
  The dashed curve has $f_{cen,err} = 0.2$, $b_{vel} = 0$ (`optimistic');
    the long dashed curve has $f_{cen,err} = 0.4$, $b_{vel} = 0$ (`conservative'); the
    dash-dot curve has $f_{cen,err} = 0.2$, $b_{vel} = 0.6$ (`extreme').  The
    straight lines show our fiducial (solid) and conservative (dotted)
    nuisance parameter constraints determined in
    Section~\ref{getmaxnuisvals}.}
\end{figure}

We consider two possible sources of deviation from our assumed galaxy
distribution within haloes.  The first is that on occasion an isolated
LRG in our sample is not the `central' galaxy in its halo, but a
satellite galaxy, while the `central' galaxy in that halo is not
selected by our sample cuts.  We call this situation `central
misidentification', and denote its probability $f_{cen,err}$, assumed
independent of halo mass for simplicity.  The brightest LRGs are
indeed centrally concentrated, with $\sim 80\%$ of them within $\sim
0.2 r_{vir}$ of the X-ray peak \citep{ho/etal:2009}.
\citet{lin/mohr:2004} similarly find that 80\% of the BCGs in their
X-ray selected cluster sample are within $\sim 0.1 r_{vir}$, and in
the $\sim 8\%$ of cases where the BCG is outside $0.5 r_{vir}$, the
second ranking galaxy in the group is within $0.1 r_{vir}$.  In some
of these cases, both the first and second brightest galaxies would be
identified as LRGs; \citet{vandenbosch/etal:2007} showed that the luminosity
difference between first and second brightest galaxies in massive groups is
typically small.  In this situation there would be no error in our catalogues
since we are not assigning luminosities to our mock LRGs.
From these studies
we would expect $f_{cen,err} <0.2$ for the halo mass scales probed by these
studies, $M>10^{14} M_{\sun}$, and it is reasonable to assume this
holds at lower masses where there are fewer massive galaxies per halo.
We therefore choose $f_{cen,err} = 0.2$ as our `optimistic' value in the cases 
we consider in Fig.~\ref{fig:veldispPk}.
Using a galaxy group and cluster catalogue from SDSS \citep{yang/etal:2007},
 \citet{skibba/etal:prep} find that the fraction of clusters
  in which the central galaxy is fainter than the brightest satellite is 
  $\approx 30\%$ in the mass range $M \sim 10^{13} - 10^{14} M_{\sun}$ and
  $\approx 40\%$ for $M \sim 10^{14} - 10^{15} M_{\sun}$.  It is not clear
  what these results imply for the LRG galaxy sample, but the parameter
  $f_{cen,err}$ aims to encompass this case.  We choose $f_{cen,err} = 0.4$
  as our  `conservative' estimate for the cases we consider in 
  Fig.~\ref{fig:veldispPk}.

The second situation we consider is the breakdown of our assumption
that the central galaxy has no peculiar motion with respect to the
mean velocity of the halo dark matter.  Any offset with respect to the
halo centre implies that central galaxies are moving with respect to
the halo centre \citep{vandenbosch/etal:2005,skibba/etal:prep}.  We
call this situation central--halo velocity bias and parametrize the amplitude as
$b^2_{vel} = \sigma_{cen}^2/\sigma_{DM}^2$, the ratio of the mean
square velocity of the central galaxy to the halo dark matter.    
 \citet{skibba/etal:prep} find $b_{vel} \sim 0.1$ once central
 misidentification has been accounted for.  This small value is negligible
 for our purposes, so we set $b_{vel} = 0$ in the `optimistic' and `conservative'
 cases we consider in  Fig.~\ref{fig:veldispPk}.
 However, \citet{coziol/etal:2009}
 find $b_{vel} \sim 0.3$ for brightest cluster members.  This quantity
 is difficult to extract from observations, and it is not clear how the literature
 results apply to the LRG sample because of the color-magnitude cuts defining
 the LRG selection.  We set $b_{vel} = 0.6$ in 
 the `extreme' case we consider in Fig.~\ref{fig:veldispPk}.

On the large scales of interest, the effect of nonzero $f_{cen,err}$ or $b_{vel}$ is to 
give the mock galaxies a velocity with respect to the halo centre.
In Figure~\ref{fig:veldispPk} we show the impact of nonzero central
galaxy velocities on the recovered $P_{halo}(k,{\bf p})$ for the three cases
we described above.  In the `optimistic' case, we set 
$(f_{cen,err}, b_{vel}) = (0.2, 0)$; in the `conservative' case, we set
$(f_{cen,err}, b_{vel}) = (0.4, 0)$; and in the `extreme' case, we set
$(f_{cen,err}, b_{vel}) = (0.2, 0.6)$.  To construct mock catalogues in each
of these cases we leave the real space distribution of
galaxies fixed.  To mimic central misidentification, we replace the
central galaxy's velocity with the velocity of a randomly selected
dark matter particle halo member.  For central--halo velocity bias, we replace the
central galaxy velocity with $b_{vel} v_{ran}$, where $v_{ran}$ is the
velocity of a randomly selected dark matter particle halo member.  
For comparison, we also outline both our fiducial nuisance function constraints
(2\% deviation at $k=0.1 \hompc$ and 5\% at $k=0.2 \hompc$) and the conservative nuisance
function constraints (4\% deviation at $k=0.1 \hompc$ and 10\% at $k=0.2 \hompc$) established
in Section~\ref{getmaxnuisvals}.  The `optimistic' case is well within
the fiducial nuisance constraints, and the `conservative' case is well
within the conservative nuisance constraints.  The `extreme' case,
however, exceeds the conservative nuisance constraints for $k > 0.17 \hompc$.
In Section~\ref{cosmoconstraints} we also evaluate the cosmological parameter
constraints when $P_{halo}(k,{\bf p})$ is calibrated using the power spectrum of
 the `extreme' velocity dispersion model in order to
derive a limit on the systematic errors on our final results.
 
\subsection{Comparison of mock and observed CiC group statistics}
\label{cicstats}
In Table~\ref{table:ciccompare} we present CiC group multiplicity
functions normalized by the number of galaxies per sample for two sets
of cylinder parameters: $\Delta r_{\perp} \leq 0.8$ $h^{-1}$ Mpc,
$\Delta v_{p} = 1800$ km s$^{-1}$ (these are the parameters used to
define our CiC groups and reconstucted halo density field for
$P_{halo}(k,{\bf p_{fid}})$) and $\Delta r_{\perp} \leq 1.2$ $h^{-1}$ Mpc, $\Delta
v_{p} = 1800$ km s$^{-1}$.  The second CiC multiplicity function
computed with larger $\Delta r_{\perp}$ 
is used to demonstrate consistency between the mock and observed
catalogues.  If the observed satellite galaxies were significantly less
concentrated than in our mock catalogues, we would detect these
galaxies when $\Delta r_{\perp}$ is increased.

The observed groups contain 2158 LRGs that
were assigned redshifts by the fiber collision correction.  
According to \citet{reid/spergel:2009}, where colors are used as a redshift
indicator, up to $\sim 36\%$ of these may be
erroneous assignments; correcting this 
would remove $\sim 780$ galaxies from the
observed groups.  We find 6.2\% of the observed galaxies are
`satellite' galaxies using the reconstructed haloes, or 5.5\% if we
apply a correction for erroneous fiber collision assignments, while
our mock catalogues have 5.9\%.  The structures of the multiplicity
functions are generally similar.  Since our mock catalogues were
designed to match this measurement but for LRGs selected as in
\citet{zehavi/etal:2005a}, the level of agreement is as expected.  We
verify that the agreement extends to the multiplicity function when we
adjust the group finding parameter $\Delta r_{\perp}$ to be 50\%
larger.  Accounting for the possible contamination from fiber collision
corrections, which is likely to manifest mostly at $n_{group}=2$, we see that in general
the observed distribution is smaller than in the mock catalogues at all
multiplicities and for both values of $\Delta r_{\perp}$.  This result may be
understood as one or more of three possibilities: 
the mocks having too many satellites altogether,
different amounts of contamination from interlopers due to errors in
the small-scale two-halo redshift space correlation function, or a
tighter distribution of satellite galaxies about the central one in the mocks.  An
error of the first kind would result in no error in the reconstructed
density field; errors of the other kinds would result in small changes
to the effective shot noise or FOG features in the density field.  The
last line in Table~\ref{table:ciccompare} shows that the difference in
the effective one halo term derived from the mock and observed
catalogues is $< 2\%$ of the total shot noise correction.
Since the difference between $P^{1h} \bar{n}_{gal}$ measured at $\Delta
r_{\perp} \leq 0.8$ $h^{-1}$ Mpc and $\Delta r_{\perp} \leq 1.2$ $h^{-1}$ Mpc
is less for the observed catalogues compared with the mocks, we cannot be missing
significant contributions to $P^{1h}$ due to a less concentrated
distribution of the satellite galaxies in the observed haloes 
compared with the simulated ones; rather, the
increase in the number of groups comes from the increase in
contamination from galaxies residing in nearby haloes.  Our final conservative nuisance
parameter bounds, discussed in Section~\ref{getmaxnuisvals},
allow an error of the order of 40\% in the one-halo shot noise
subtraction.  Also note that because the maximum line of sight separation
($\Delta v_{p} = 1800$ km s$^{-1}$ or $\sim 20 \; h^{-1}$ Mpc) is so
large, the model CiC multiplicity functions are nearly identical when
we consider the model with `extreme' central galaxy velocity dispersion.
Finally, adding some spatial dispersion of the central galaxies would
slightly reduce the number of CiC groups for an otherwise fixed
catalogue; this may bring the models and observations into even closer
agreement.

\begin{figure}
  \centering
  \resizebox{0.9\columnwidth}{!}{\includegraphics{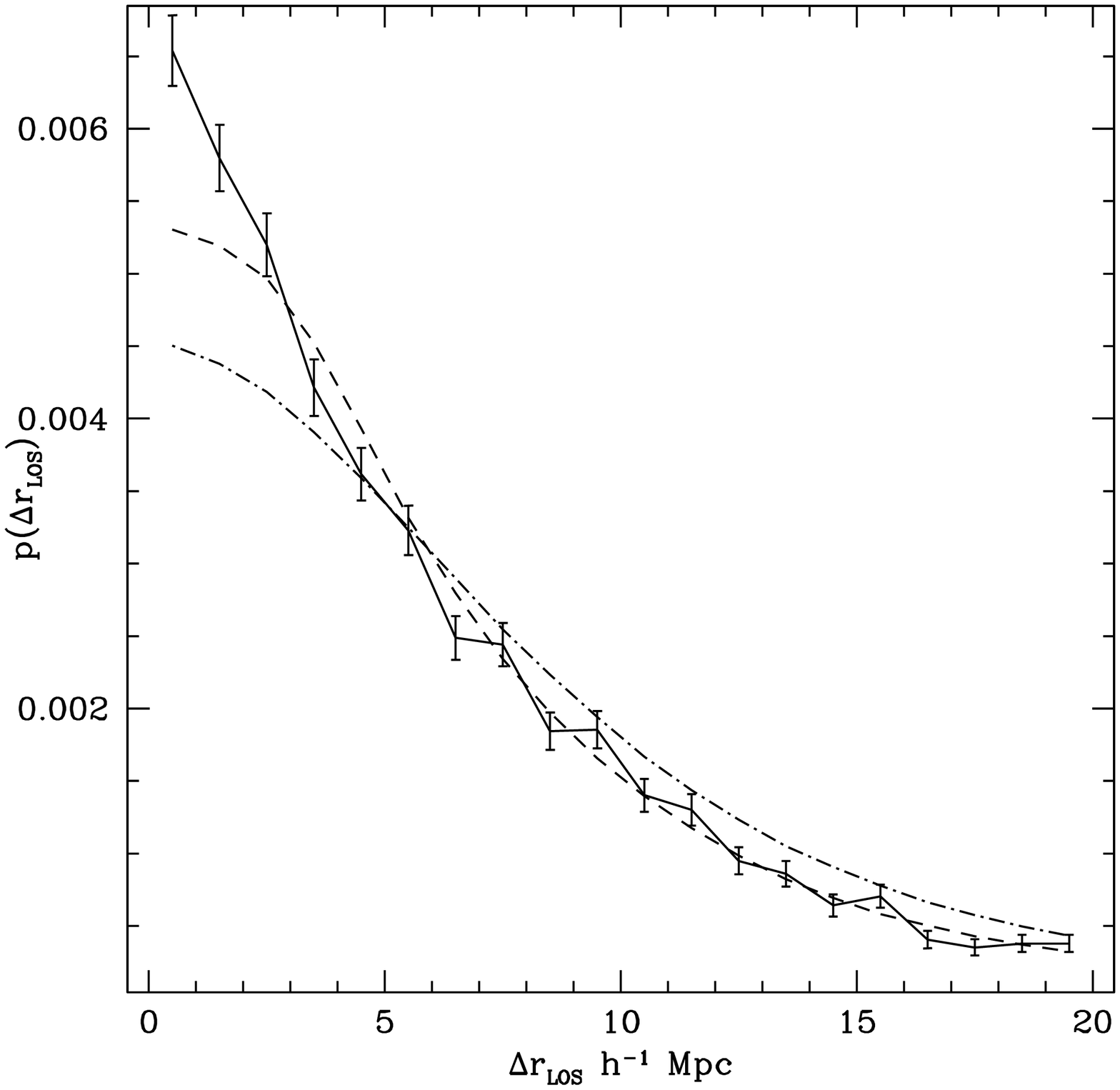}}
  \caption{\label{fig:FOGhist} Solid line with error bars is the
    observed probability that a galaxy has a member of its CiC group
    with a separation $\Delta r_{LOS}$ along the line of sight
     for
    pairs of galaxies identified as pairs by the CiC criteria, once all groups
    containing a fiber collision galaxy are removed.  Error bars
    indicate fractional errors of $1/\sqrt N(\Delta(r_{LOS})$, giving
    a sense of the Poisson level of uncertainty in the measurement
    without considering the contribution from cosmic variance.  The
    dashed line is the expected distribution for our model with no
    central galaxy velocity dispersion, and the dot-dashed line is for
    the model with central galaxy velocity dispersion.  
    Note that $\Delta r_{LOS} = 1 \; h^{-1}$ Mpc corresponds 
    to $\Delta v \approx 115$ km s$^{-1}$ for the redshift distribution of
    our sample.}
\end{figure}

\begin{table}
\begin{center}
\begin{tabular}{lllll}
  $n_{group}$ & $N_{\rm CiC,obs}(n)$ & $N_{\rm CiC,mock}(n)$ & 
  $N_{\rm big,obs}(n)$ & $N_{\rm big,mock}(n)$ \\
  2 & 5283 & 4717 & 6432 & 6280 \\
  3 & 539 & 658 & 899 & 1076 \\
  4 & 110 & 124 & 198 & 252 \\
  5 & 26 & 28.2 & 39 & 71.4\\
  6 & 7 & 7.68 & 27 & 22.9 \\
  7 & 1 & 2.32 & 5 & 8.65 \\
  8 & 3 & 0.78 & 5 & 3.34 \\
  9 & 0 & 0.30 & 0 & 1.39 \\
  10 & 0 & 0.10 & 0 & 0.66 \\
  $P^{1h} \bar{n}_{gal}$ & 0.144 & 0.143 & 0.205 & 0.225
\end{tabular}
\caption{\label{table:ciccompare} The observed and mock catalogue CiC
  group multiplicity functions of groups with $n_{group}$ galaxies
  for our fiducial group finding parameters $\Delta r_{\perp,max} =
  0.8 \; h^{-1}$ Mpc, $\Delta v_p = 1800$ km s$^{-1}$ and for a
  bigger $\Delta r_{\perp,max} = 1.2 \; h^{-1}$ Mpc.  The final row
  shows the ratio of the one-halo shot noise $P^{1h}_{LRG}$
  (Eqn.~\ref{onehaloterm}) to the standard shot noise correction
  $1/n_{LRG}$.}
\end{center}
\end{table}
We compute the line of sight separation of galaxies in the same CiC group as a probe
of the accuracy of our model galaxy velocities at the high halo mass
end, where there is more than one LRG per halo.  The comparison is
complicated by the presence of fiber collision corrected galaxies,
since their redshifts are artificially set to that of another galaxy in
their group.  We discard all such groups, and discard an equal
fraction at each $n_{group}$ in our mock sample.  The
resulting distributions are shown in Fig.~\ref{fig:FOGhist}.  The
fiducial mocks with no central galaxy velocity dispersion appear to
fit the data better, though neither matches the observed sharpness of the rise at small
separations.  Note that the fiducial mock catalogues with no velocity dispersion are determined only by the observed $N_{CiC}(n_{group})$; no free parameters have been adjusted to match the observed velocity distribution.  This comparison indicates that the residual FOG features
in the reconstructed observed and mock halo density fields will be in
satisfactory agreement.

\subsection{Comparison of $\hat{P}_{halo}(k)$ and $\hat{P}_{LRG}(k)$}
\label{cicvsnotcompare}
In this subsection we examine the difference between the observed redshift space
monopole spectrum $\hat{P}_{LRG}(k)$ (no density field preprocessing of
FOG features) and the power spectrum of the reconstructed halo density
field, $\hat{P}_{halo}(k)$, and compare with our mock galaxy catalogues.
This comparison provides an additional consistency check between 
the mock and observed LRG catalogues, and quantifies the effect of the halo 
density field reconstruction step on the measured power spectrum shape.

\begin{figure}
  \centering
  \resizebox{0.9\columnwidth}{!}{\includegraphics{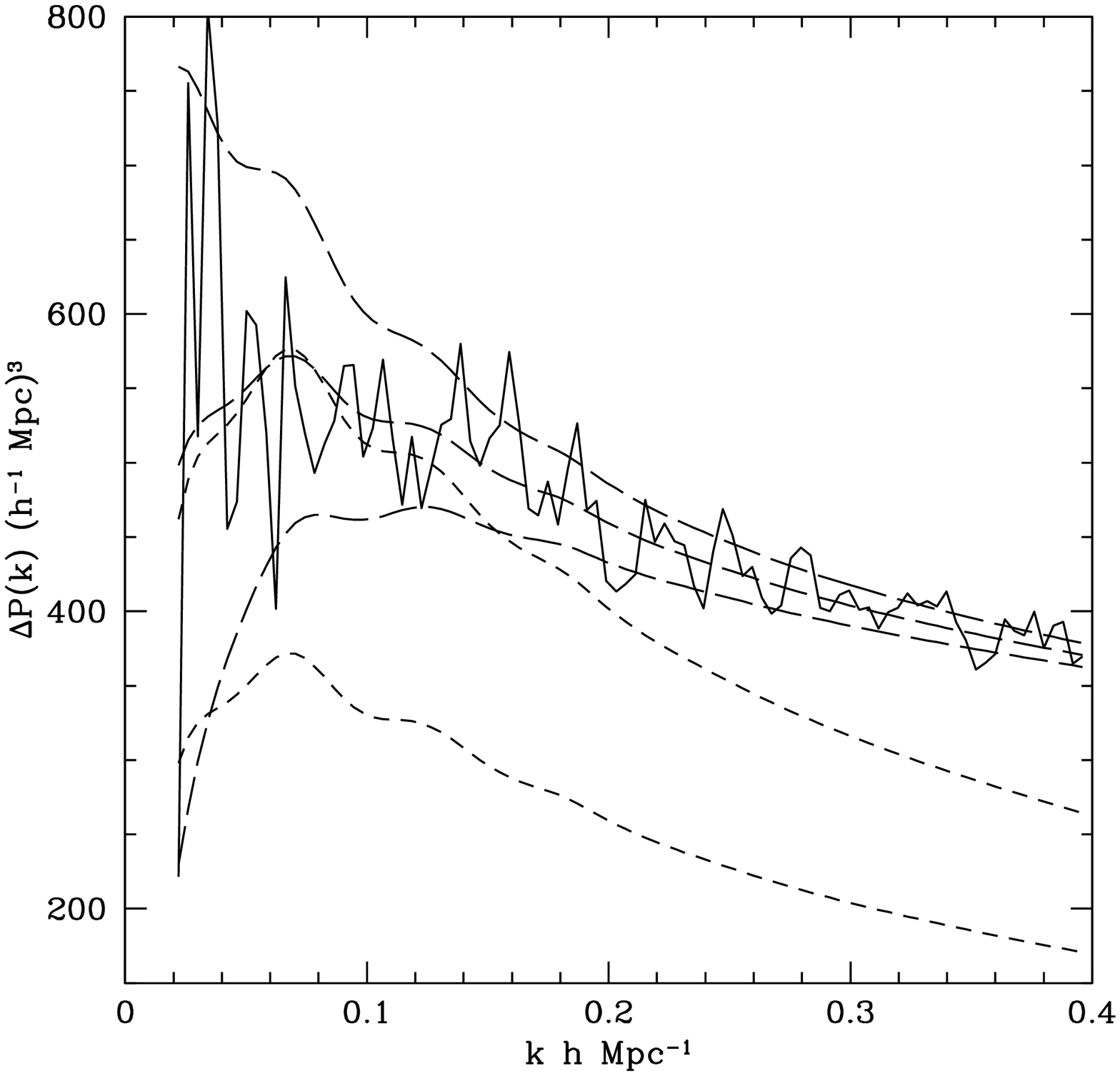}}
  \caption{\label{fig:cccompare} The solid curve is the difference
    between the observed spectra $\hat{P}_{halo}(k)$ and 
    $\hat{P}_{LRG}(k)$, the lower short-dashed
    curve is the predicted difference from our
    simulated catalogues, and the upper short-dashed curve is the same
    curve but scaled by a factor of 1.5.  The scale
    dependence of $\Delta P(k)$ is smaller for the observed spectra
    than for the simulation results.  Furthermore, there is some
    uncertainty in the appropriate value of $b_{rel}$, which changes
    the shape of $\Delta P(k)$.  However, at high $k$, the prediction
    is robust to changes in $b_{rel}$ since $P(k)$ is small.  The
    long-dashed curves show $\Delta P_{mock}(k) + 200 (h^{-1}$
    Mpc)$^3$ for several values of $b_{rel}$.  This demonstrates that
    the difference between $\hat{P}_{halo}(k)$ and 
    $\hat{P}_{LRG}(k)$ is
    consistent with the difference measured in the simulated catalogues
    if the excess shot noise from fiber collisions is accounted for.
    Moreover, the difference between the observed halo and LRG
    spectra is large compared with the statistical errors on the
    bandpowers.}
\end{figure}

We consider
\begin{equation}
\Delta P(k) = P_{LRG}(k) - b_{rel}^2 P_{halo}(k)
\end{equation}
where $b_{rel}$ is a constant that parametrizes the enhancement of the
overall bias when satellite galaxies are included, since they occupy
the most highly biased regions.  In real space on large scales,
$\Delta P(k)$ would be a simple shot noise, but in redshift space we
expect the detailed $\Delta P(k)$ to result from the transfer of power
between scales caused by the FOGs, making $\Delta P(k)$ dependent on
the underlying power spectrum shape.  We will ignore this possible
$<10\%$ level modification to the expected $\Delta P(k)$ since
we have demonstrated good agreement between the shape of the mock and
observed halo power spectra.  The lower short
dashed curve in Fig.~\ref{fig:cccompare} shows the predicted
$\Delta P(k)$ from our mock catalogues, the upper
short dashed curve shows the predicted $\Delta P(k)$ scaled by a factor of 1.5,
and the solid curve shows $\Delta P(k)$ for the observed spectra.  The observed
$\Delta P(k)$ is clearly shallower than the predicted shape.

A crucial difference between the observed and mock LRG density fields
is the application of fiber collision corrections discussed in
Section~\ref{sampleselection} in the observed density field.  2158 galaxies
without spectra were added to the LRG sample and assigned the redshift
of the nearest LRG, while the CiC group multiplicity results in
Table~\ref{table:ciccompare} indicate that 6857 galaxies are
`satellite' galaxies.  
First, since 
$\sim 36\%$ of the fiber-collision corrections are erroneous \citep{reid/spergel:2009},
we expect an additional shot noise of $\sim 125 (h^{-1}$ Mpc)$^3$ from these
galaxies, which are not represented in our mock catalogues.  Second, the
fiber collision corrected galaxies that are physically associated with a neighboring LRG
will change the distribution of $\Delta P(k)$ relative to the mocks because their
line of sight separation from the neighboring galaxy has been eliminated.  
The long dashed curves in Fig.~\ref{fig:cccompare} shows that we can
match the observed $\Delta P(k)$ as a sum of the mock catalogue $\Delta P(k)$
and a shot noise of $200 (h^{-1}$ Mpc)$^3$.  The $\Delta P(k)$ 
for the observed spectra is
consistent with a constant power for $k<0.2 \hompc$ and amounts to a
significant difference between the two spectra: $\sim 8\%$ at $k=0.1 \hompc$
and $\sim 18\%$ at $k=0.2 \hompc$.  Therefore, 
differences in the preprocessing of the LRG density field can lead to
changes in $P(k)$ much larger than the statistical errors on the
measurements, which could then be propagated to errors in the derived
cosmological parameters.
Note that the
reconstructed halo density field is basically unaffected by
errors in the close-pair fiber collision correction applied to the data, 
since these galaxies are all assigned to haloes containing other LRGs already.

In summary, the difference between $\hat{P}_{halo}(k)$ and $\hat{P}_{LRG}(k)$ 
can be understood once we account for the effects of fiber collisions, and the model 
predictions $P_{halo}(k,{\bf p})$ are robust to any uncertainty associated with these effects.

\subsection{The effect of luminosity-weighting on $\hat{P}_{halo}(k)$}
\label{lumweight}
\begin{figure}
  \centering
  \resizebox{0.9\columnwidth}{!}{\includegraphics{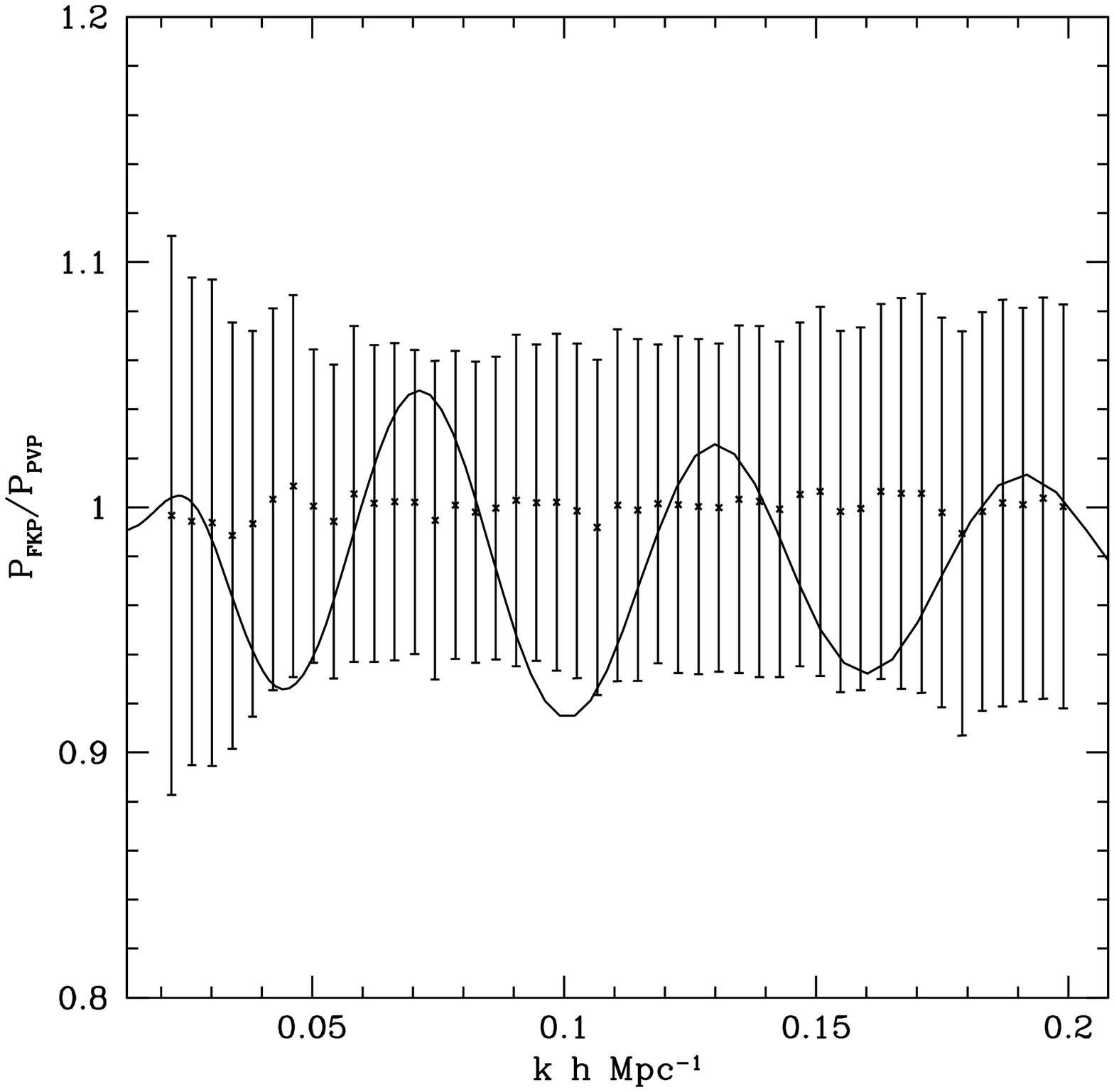}}
  \caption{\label{fig:nolumbiasrat} Ratio of the power spectra
    computed using the weights in Eqn.~\ref{nobiaswgts} to the
    standard \citet{percival/verde/peacock:2004} weighting scheme after rescaling the overall
    normalization.  We also overplot $P_{lin}(k)/P_{nw}(k)$ for our
    fiducial model to demonstrate no correlation between the small
    shifts in the measured power spectrum and expected BAO feature.
    Errors show the fractional errors on $\hat{P}_{halo}(k)$, 
    $\sqrt{C_{ii}}/\hat{P}_{halo}(k)$.}
\end{figure}

A further subtle difference between the mock and observed halo power spectrum
is that the mock catalogues were evaluated using a redshift snapshot
with constant $\bar{n}_{LRG}$, and luminosities were not assigned to
the mock LRGs; each reconstructed halo is weighted equally when
computing the overdensity field.  To verify that the luminosity
weighting used to compute the $\hat{P}_{halo}(k)$ does not
significantly alter the relative amplitude of the shot noise to total
power compared with our mock catalogues, we recompute $\hat{P}_{halo}(k)$ from
the data with luminosity-independent weights from 
\citet{feldman/kaiser/peacock:1994}:
\begin{eqnarray}
  b(L) & = & 1 \label{bis1} \\
  w(r,L) & = & \frac{1}{1 +P_o \bar{n}_{LRG}} \label{nobiaswgts}
\end{eqnarray}
where $P_o = 10000 (h^{-1} \; {\rm Mpc})^3$.
Fig.~\ref{fig:nolumbiasrat} shows the ratio of the observed spectra
with our fiducial weights compared with the luminosity-independent
weights.  The good agreement even at large $k$ where the power is small
indicates there is no significant difference from the shot noise
subtraction between these two weightings; we find no statistically
significant change in the power spectrum shape.  Moreover, the change in
the windowed theory power spectrum due to the change in weights is
negligible ($<0.1\%$), indicating that the window function
will not be sensitive to the particular weighting choices of
Section~\ref{data} for reconstructed haloes containing more than one galaxy.
While the luminosity-weighting is critical for the SDSS main 
sample \citep{tegmark/etal:2004a:la},
Fig.~\ref{fig:nzcompare} shows that the LRGs are close to volume-limited
over much of the redshift range of the sample; it is therefore unsurprising
that the \citet{feldman/kaiser/peacock:1994} and \citet{percival/verde/peacock:2004} weighting schemes produce nearly identical power spectra
for the LRG sample.

\subsection{Checking the cosmological dependence of the model}
\label{wmap3cats}
Our model uses {\sc halofit}  
to describe the cosmological parameter dependence
of the non-linearity in the matter power spectrum,
and is calibrated from $N$-body simulations
at the fiducial cosmology (Eqn.~\ref{eq:DM_model}).
Below $k=0.1 \hompc$, the dark matter power spectrum
is linear at the 1\% level, apart from the BAO damping, 
and it is only $\sim 15\%$ larger than the linear one at $k=0.2 \hompc$.
Using the publicly available WMAP5 $\Lambda$CDM MCMC
chain, we find $P_{\rm halofit}(k)/P_{lin}(k)$ changes by $\sim \pm
2\%$ for $k \leq 0.2$ in the space of cosmologies allowed
by the WMAP5 data alone; the error on this small correction
will therefore be well below 1\%.  
Therefore we expect the model of the 
non-linear matter power spectrum to be accurate at the $< 1\%$ level
at $k=0.1 \hompc$ and $\sim 1\%$ at $k=0.2 \hompc$.

\begin{figure}
  \centering
  \resizebox{0.9\columnwidth}{!}{\includegraphics{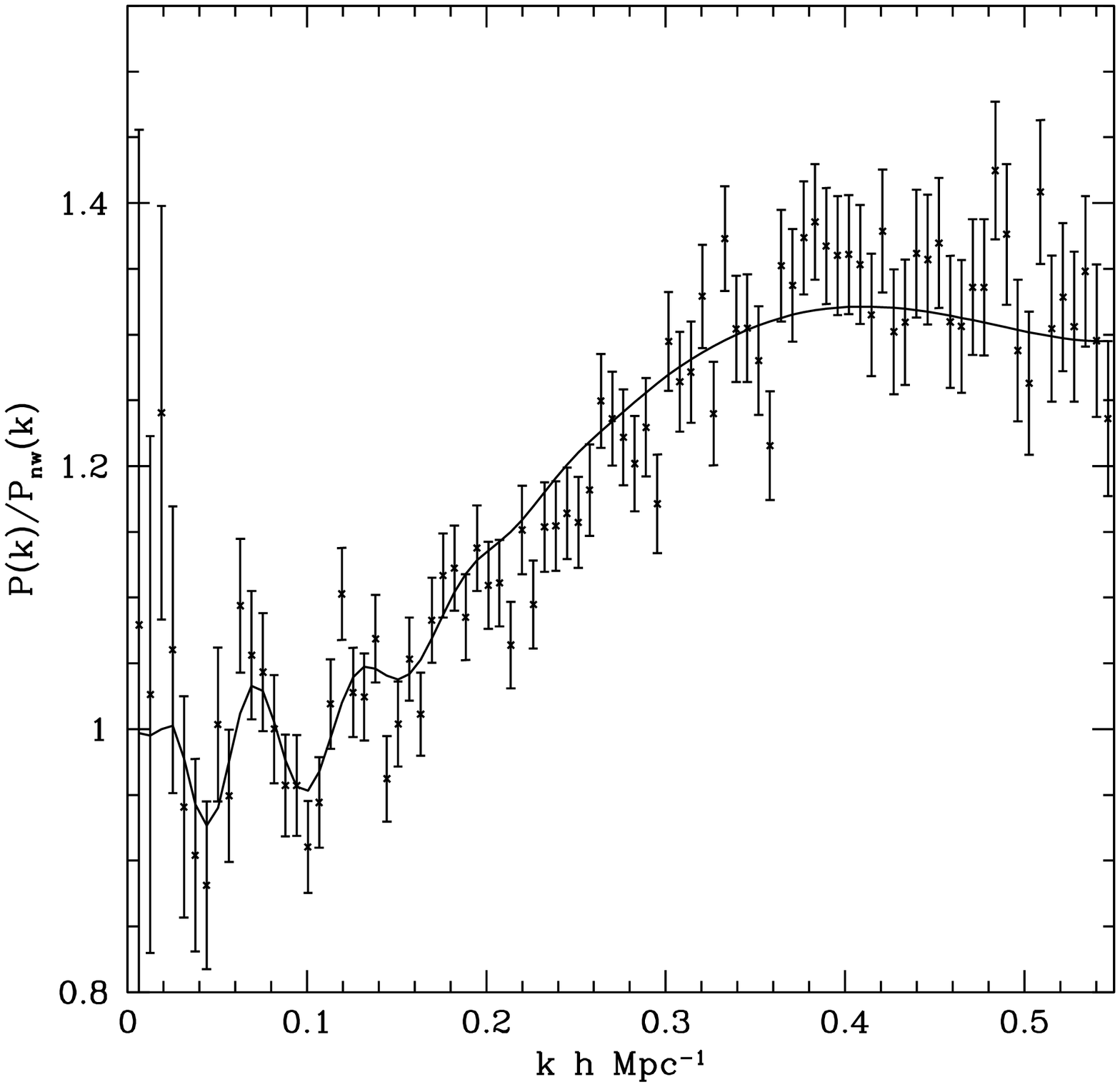}}
  \caption{\label{fig:wmap3fit} Agreement between
    $P_{halo,WMAP3}(k)/P_{nw}(k,{\bf p}_{WMAP3})$ measured from the
    catalogues in \citet{reid/spergel:2009} based on an $N$-body
    simulation $z=0.2$ snapshot with WMAP3 cosmological parameters (points with error bars)
    vs. the model prediction from Eqn~\ref{finalmodel} at $z_{NEAR} = 0.235$.}
\end{figure}

We use the LRG catalogues from \citet{reid/spergel:2009} evaluated at
the WMAP3 preferred cosmological parameters $(\Omega_m, \Omega_b,
\Omega_\Lambda, n_s, \sigma_8, h) = (0.26, 0.044, 0.74, 0.95, 0.77,
0.72)$ at $z=0.2$ with $L_{box} = 1 \; h^{-1}$ Gpc to test the
cosmological dependence of our model $P_{halo}(k,{\bf p})$
 in Eqn.~\ref{eq:DM_model}.  We
plot a mock catalogue power spectrum $P_{halo,WMAP3}(k)/P_{nw}(k,{\bf
  p}_{WMAP3})$ against our $P_{halo}(k,{\bf p})$ 
  model predictions for a NEAR subsample in
Fig.~\ref{fig:wmap3fit} to demonstrate the agreement both in the BAO
features and overall shape of the deviation out to $k=0.55 \hompc$.  $\chi^2
= 96.6$ for 86 DOF ($k \leq 0.55$) and $\chi^2 = 29.1$ for 31 DOF ($k
\leq 0.2$).  This provides further evidence that the cosmological
dependence of our model $P_{halo}(k,{\bf p})$ is sufficiently accurate for the
SDSS DR7 data, which probe a somewhat smaller volume.

\section{Effects of central galaxy velocity dispersion and nuisance parameters}
\label{nuiseffects}

\begin{figure*}
  \centering
  \resizebox{0.9\textwidth}{!}{\includegraphics{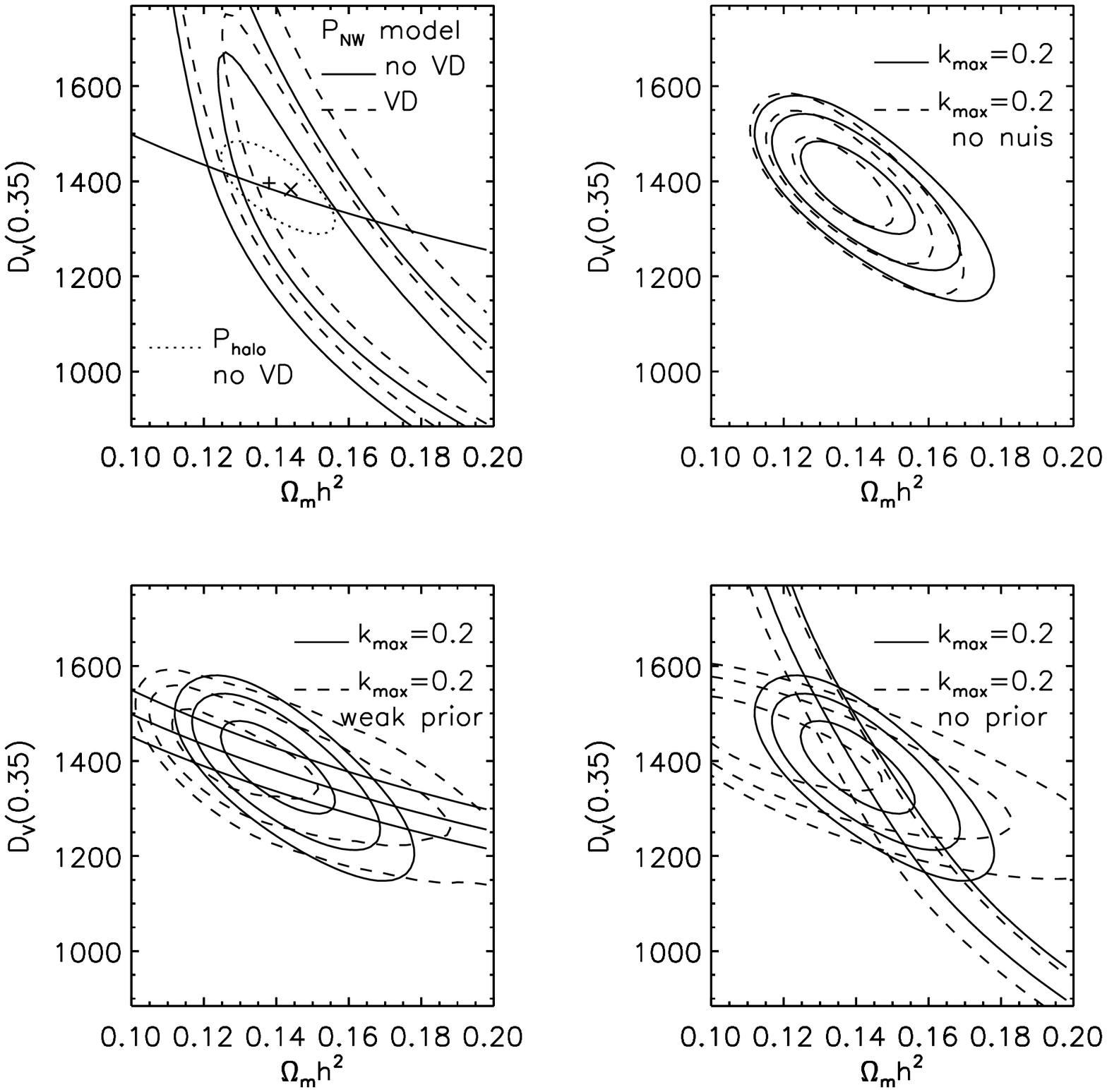}}
  \caption{\label{fig:dr7onlynuis} Effects of velocity dispersion and nuisance parameters on constraints from the LRG DR7 $\hat{P}_{halo}(k)$ for a $\Lambda$CDM model.  In each panel we hold
    $\Omega_b h^2 = 0.02265$ and $n_s = 0.960$ fixed.  {\em Upper left
      panel}: $\Delta \chi^2 = 2.3$ and 6.0 contours for the
    $\hat{P}_{halo}(k)$ fit to a no-wiggles model with no central velocity
    dispersion (solid) and extreme velocity dispersion
    (dashed).  The $\Delta \chi^2 = 2.3$ for the fiducial model with BAO features
    is shown for comparison by the dotted line.
    The cross shows the maximum likelihood point
    for our fiducial model, while the 'X' shows it for the extreme
    velocity dispersion model.  The solid line indicates $r_s/D_V(0.35) = 0.1097$,
    demonstrating that adopting the velocity dispersion model shifts the likelihood surface
    along constant $r_s/D_V(0.35)$.
    {\em Upper right panel}: $\Delta
    \chi^2 = 2.3, 6.0, $ and 9.3 contours.  The
    solid contours use our fiducial marginalization over
    $b_o^2$, $a_1$, and $a_2$ (as in Fig.~\ref{fig:dr7onlyonepanel}), while in the dotted contours
    fix $a_1 = a_2 = 0$ and $b_o^2$ to the value which minimizes
    $\chi^2$.  {\em Lower left panel}: The
    solid contours as in Fig.~\ref{fig:dr7onlyonepanel}, while the dashed contours
    take the minimum $\chi^2$ value for which $|F(k=0.1 \hompc)|/b_0^2 < 0.2$
    and $|F(k=0.2 \hompc)|/b_0^2 < 0.5$.  {\em Lower right panel}: The
    solid contours as in Fig.~\ref{fig:dr7onlyonepanel}, while the dashed contours
    minimize $\chi^2$ with no restrictions on $a_1$ and $a_2$.
    For comparison with the fiducial nuisance restrictions, the solid 
    lines enclose the region where for the best-fitting $\chi^2$, 
    $|F(k=0.1 \hompc)|/b_0^2 < 0.04$ and the dashed lines enclose $|F(k=0.2 \hompc)|/b_0^2 < 0.1$.}
\end{figure*}

In Section \ref{quantnuis} we established that the largest remaining known source of systematic uncertainty is the central galaxy velocity dispersion.  To test the impact of this uncertainty on the cosmological constraints, we reevaluate the $\hat{P}_{halo}(k)$ likelihood surface using the `extreme' velocity dispersion model in Appendix \ref{galaxiesinhaloes} to calibrate the model $P_{halo}(k, {\bf p})$.  The maximum likelihood points for the fiducial, no velocity dispersion model (cross) and the `extreme' velocity dispersion model ('X') are shown in the upper left panel of Fig.~\ref{fig:dr7onlynuis}.  
The systematic shift in the contours between the zero
and extreme central velocity
dispersion model is small compared to the width of the $\Delta \chi^2
= 2.3$ constraint (dotted curve).
When we marginalize over nuisance parameters $b_o^2,
a_1$, and $a_2$, $\Delta \chi^2$ between the maximum likelihood model
values for the zero and extreme velocity dispersion models is $\sim 0.3$.  If one
instead adopts the $a_1$, $a_2$, and $b_o^2$ values which minimize
$\chi^2$, the shift decreases to $\Delta \chi^2 \sim 0.1$; the
difference is because the preferred nuisance parameters $a_1$ and
$a_2$ in the no velocity dispersion model are closer to the boundary of
the allowed values.  These $\Delta \chi^2$ values are approximately the same
when considering a fit to the model with or without BAO wiggles. 
This shift is small compared to the statistical errors, and since the velocity
dispersion model considered is extreme compared with the available
estimates in the literature \citep{skibba/etal:prep,coziol/etal:2009}, we can
safely neglect this systematic uncertainty in the present analysis.

Within our fiducial nuisance parameter bounds and using our
fiducial model with no central galaxy velocity dispersion, we have verified
 that the effect of the nuisance parameters in Eqn.~\ref{eq:nuis} 
 is small on the $\hat{P}_{halo}(k)$  cosmological parameter
constraints.  The preferred nuisance
parameters are off-center in the allowed $a_1-a_2$ space, although not at the boundary:
$\left<F_{nuis}(0.1 \hompc)/b_0^2 \right> = 0.016$ and 
$\left<\left(F_{nuis}(0.1 \hompc) - F_{nuis}(0.2 \hompc)\right)/b_0^2\right> = 0.060$
, where we have 
computed a likelihood-weighted average over the DR7-only constraints.  
The upper right
panel of Fig.~\ref{fig:dr7onlynuis} shows $\Delta \chi^2 = 2.3, 6.0, $ and
9.3 contours where $a_1 = a_2 = 0$ and $b_o^2$ is varied to
minimize $\chi^2$ (dashed contours) compared to our fiducial
marginalization over $b_o^2$, $a_1$ and $a_2$ (solid
contours).  Allowing nuisance parameters to account for our 
imperfect modeling induces both a small shift and widening of
the likelihood surface.  The difference in the contours is negligible 
when $\chi^2$ is evaluated instead
at the values $a_1$ and $a_2$ that minimize $\chi^2$.  Therefore the
hard boundary we impose in $a_1-a_2$ space does not seriously affect
the likelihood contours, and $a_1$ and $a_2$ are not strongly degenerate
with the cosmological parameters constrained by $\hat{P}_{halo}(k)$
when $a_1$ and $a_2$ are tightly constrained by the arguments in 
Section \ref{getmaxnuisvals}.

However, when one substantially relaxes the constraints on the nuisance 
function, the constraints from the power spectrum shape degrade.  The lower right
panel of Fig.~\ref{fig:dr7onlynuis} shows how the $\chi^2 = 2.3, 6.0$, and 9.3 
constraints relax when $a_1$ and $a_2$ are chosen to minimize $\chi^2$
such that $F_{nuis}(k=0.1 \hompc)/b_0^2 < 0.2$ and $F_{nuis}(k=0.2 \hompc)/b_0^2 < 0.5$.  While the constraints on 
$r_s/D_V(0.35)$ are unchanged, the shape information is degraded.  The effects of scale 
dependent halo bias are well below these allowed deviations
\citep{smith/scoccimarro/sheth:2007}, and we have argued that our reconstruction
of the halo density field should leave much smaller uncertainties as well.
The dashed contours in the lower right panel of Fig.~\ref{fig:dr7onlynuis} show
a further broadening of the constraints when $a_1$ and $a_2$ are varied without restriction to minimize $\chi^2$.
For comparison with the adopted nuisance 
restrictions, the bottom right panel of Fig.~\ref{fig:dr7onlynuis} also shows the regions
where the best-fitting nuisance parameters satisfy $|F_{nuis}(k=0.1 \hompc)|/b_0^2 < 0.04$ (solid lines)
and $|F_{nuis}(k=0.2 \hompc)|/b_0^2 < 0.1$ (dashed lines).  
The width of this region is smaller than the statistical errors derived from the
shape constraint, which are shown in the
upper left panel. Consequently, it is
unsurprising that our marginalized likelihood contours with the
fiducial nuisance restrictions deviate only slightly from the contours
where $a_1 = a_2 = 0$.
Finally we note that for the models with and without velocity dispersion,
the likelihood-weighted best-fitting nuisance
functions have small deviations from one at $k=0.1 \hompc$ ($< 2\%$), the region
containing most of the shape information.  The two models differ in the
quasi-linear regime: 
 $\left<\left(F_{nuis}(0.1 \hompc) - F_{nuis}(0.2 \hompc)\right)/b_0^2\right> = -0.033$
  for the velocity dispersion model and 0.060 without
 velocity dispersion.  However, we cannot distinguish between velocity
dispersion and other modeling uncertainties to explain the shape of
the nuisance function preferred by the data.  Moreover, using the 
velocity dispersion model does not improve the overall $\chi^2$ of
the fit.

We conclude that, for this data set,
the statistical errors are comfortably larger than the errors from modeling
uncertainties.
\end{document}